\documentclass[12pt]{article}
\usepackage[utf8]{inputenc}
\usepackage[sectionbib]{natbib}
\usepackage{setspace}

\title{Bayesian Transfer Learning Using Muti-Source Summary-level Data}

\usepackage{mylatexstyle,comment}
\usepackage{float}
\usepackage[margin=1in]{geometry}
\usepackage{graphicx, color}
\usepackage{booktabs}
\usepackage{xr}
\usepackage{pifont}
\usepackage{titlesec}
\usepackage{multirow}
\usepackage{multicol}
\usepackage{amsfonts}
\usepackage{textcomp}
\usepackage{amsthm}
\usepackage{mathtools}
\usepackage{algorithm}
\usepackage{algorithmic}
\usepackage{amsmath}
\usepackage{amssymb}
\usepackage[toc,page]{appendix}
\usepackage[mathscr]{euscript}

\let\counterwithin\relax
\usepackage{chngcntr}
\usepackage{apptools}
\usepackage{caption}
\usepackage{tabularx} %
\usepackage{array} %
\usepackage{soul}
\usepackage{tikz}
\usetikzlibrary{shapes.misc}
\usetikzlibrary{shapes}
\usepackage{underscore} 
\AtAppendix{\counterwithin{lemma}{section}}
\usepackage{indentfirst}
\usepackage{epstopdf}
\usepackage{diagbox}
\usepackage{lipsum}
\usepackage{subcaption}
\usepackage{bbm}
\usepackage{adjustbox}
\usepackage[affil-it]{authblk}

\newtheorem{proposition}{Proposition}

\newcolumntype{P}[1]{>{\centering\arraybackslash}p{#1}} %
\usepackage{xcolor}
\usepackage[mathscr]{euscript}
\usepackage{mathtools}
\usepackage{amsthm}
\usepackage{amsfonts}
\usepackage{amssymb}
\usepackage{color}
\usepackage{multirow}
\usepackage{makecell}
\usepackage{booktabs}
\newcommand{\bm}{\boldsymbol}

\makeatletter
\newcommand*{\addFileDependency}[1]{
\typeout{(#1)}
\@addtofilelist{#1}
\IfFileExists{#1}{}{\typeout{No file #1.}}
}\makeatother

\definecolor{blue}{rgb}{0.0,0.0,1.0}
\definecolor{green}{rgb}{0.0,0.5,0.0}
\definecolor{green}{rgb}{1.0,0.0,0.0}
\definecolor{cyan}{rgb}{0.0,1.0,1.0}

\usepackage{lipsum}
\usepackage[colorlinks,linkcolor=blue,citecolor=blue]{hyperref}
\newcommand{\bet}{\begin{theorem}}
	\newcommand{\eet}{\end{theorem}}
\newcommand{\bel}{\begin{lemma}}
	\newcommand{\eel}{\end{lemma}}
\newcommand{\bep}{\begin{proposition}}
	\newcommand{\eep}{\end{proposition}}
\newcommand{\bed}{\begin{definition}}
	\newcommand{\eed}{\end{definition}}
\newcommand{\bec}{\begin{corollary}}
	\newcommand{\eec}{\end{corollary}}
\newcommand{\bex}{\begin{example}}
	\newcommand{\eex}{\end{example}}
\newtheorem{theorem}{Theorem}
\newtheorem{corollary}{Corollary}
\newtheorem{remark}{Remark}
\newtheorem{assumption}{Assumption}

\title{Bayesian Transfer Learning for Enhanced Estimation and Inference}

\author[1,2]{Daoyuan Lai}
\author[3]{Oscar Hernan Madrid Padilla}
\author[4]{Xiang Li}
\author[2,\footnote{Corresponding author. Email address: \texttt{tg2880@cumc.columbia.edu}}]{Tian Gu}
\affil[1]{\small Division of Nephrology, Department of Medicine, Vagelos College of Physicians and Surgeons, Columbia University, New York, NY}
\affil[2]{Department of Biostatistics, Columbia University Mailman School of Public Health, New York, NY}
\affil[3]{Department of Statistics and Data Science, University of California, Los Angeles, Los Angeles, CA}
\affil[4]{Department of Biostatistics, The University of Texas MD Anderson Cancer Center, Houston, TX}
\date{}

\begin{document}
{
\cleardoublepage
\pagenumbering{arabic}
\singlespacing
\maketitle
\thispagestyle{empty} 

\begin{abstract}
Transfer learning enhances model performance in a target population with limited samples by leveraging knowledge from related studies. While many works focus on improving predictive performance, challenges persist in statistical inference. Bayesian approaches naturally provide uncertainty quantification for parameter estimates; however, existing Bayesian transfer learning methods are typically limited to single-source scenarios or require individual-level data. We introduce TRansfer leArning via guideD horseshoE prioR (TRADER), a novel approach enabling multi-source transfer through pre-trained models in high-dimensional linear regression. TRADER shrinks target parameters toward an adaptively weighted average of source estimates, while remaining robust to differences in source scale and correlation. Theoretical investigation shows that TRADER achieves faster posterior contraction rates than standard continuous shrinkage priors when sources are well-aligned with the target, while preventing negative transfer from heterogeneous sources. The finite-sample marginal posterior behavior of TRADER is established. Extensive simulations show that TRADER achieves inference performance no worse than using target data alone, performs comparably to a competing frequentist method despite using only summary-level source data, and offers substantial computational advantages. Application to a high-dimensional genetic dataset further shows TRADER's effectiveness in inference under strong multicollinearity. Supplementary materials for this article, including a standardized reproducibility guide detailing all materials required to replicate the results, are available online.
\end{abstract}

}

\newpage
\section{Introduction}

In the era of big data, many advanced statistical methods rely on large, diverse datasets to perform well. However, in high-dimensional settings--especially when studying populations with limited data--data scarcity remains a major challenge. For instance, fine-mapping in statistical genetics aims to identify genetic variants that have non-zero effects on diseases but often struggles with small sample sizes, weak signal strength, and low allele frequencies in such populations \citep{schaid2018genome, li2024bayesian}. These challenges are compounded by complex genetic structures and limited linkage disequilibrium (LD) information. To improve power, researchers have explored integrating data from larger cohorts with more comprehensive genomic profiles \citep{gao2024mesusie}, motivating the use of transfer learning. As a broader data integration strategy \citep{chatterjee2016constrained, han2019empirical}, transfer learning enables models to leverage information from richer datasets (\textit{source}) and improve performance in a target population (\textit{target}) with limited data. Its success has been demonstrated across applications in risk prediction \citep{cheng2019informing, gu2022transfer, li2023targeting, gu2023commute, lu2024enhancing, cai2024semi}, classification \citep{al2016transfer}, and other areas including generative modeling and meta-learning \citep{gu2019synthetic, jin2021unit, gu2023meta, gu2023synthetic, xu2025representation, sui2025multi}.

Recently, transfer learning has gained attention in high-dimensional linear regression. \cite{bastani2021predicting} studied prediction and estimation with a single source, while \cite{li2022transfer} extended this to multiple sources with varying similarity. Extensions to generalized linear models have also been proposed \citep{tian2023transfer, li2023estimation}. These methods typically assume (i) access to individual-level data across studies, and (ii) that the difference between source and target models can be captured via their $\ell_1$ distance. To address data-sharing constraints, federated learning algorithms have emerged \citep{duan2018odal, duan2020learning, duan2022heterogeneity, cai2022individual, li2023targeting, chen2024distributed, lu2024enhancing}, enabling collaboration without centralized data access. However, these methods rely on robust infrastructure and synchronous communication. In contrast, approaches that use pre-trained models, such as estimated coefficients from source studies, offer a more flexible and communication-efficient alternative \citep{zhai2022data, taylor2023data, gu2023commute, han2023integrating}. Still, relying on distance-based assumptions can be problematic when source and target models differ in scale or structure, as shown in \cite{gu2022robust}, which is a common issue in mixed-type outcome integration \citep{miglioretti2003latent} and cross-population studies \citep{need2009next}. These limitations motivate the development of more adaptive and robust transfer learning frameworks.

The Bayesian framework is naturally suited for transfer learning, as it allows prior information from external sources to be seamlessly integrated into inference \citep{suder2023bayesian}. Although Bayesian methods are often perceived as computationally intensive relative to frequentist approaches, this view is not always accurate. For uncertainty quantification across all $  p  $ parameters, frequentist debiasing or desparsification methods require solving $  p  $ optimization problems to approximate the inverse covariance matrix, incurring high costs as $  p  $ increases \citep{boot2017scalable}. For instance, \cite{javanmard2014confidence} employs per-variable numerical optimization, while \cite{van2014asymptotically} solves $  p  $ LASSO problems for inverse covariance estimation. Both yielding $  O(p^4)  $ complexity. By contrast, our proposed Bayesian method uses Markov Chain Monte Carlo (MCMC), avoiding large-matrix inversion and achieving $  O(n_{\text{iter}} p^2)  $ complexity, where $  n_\text{iter}  $ is the number of MCMC iterations. This illustrates the computational advantage of Bayesian approaches when quantifying uncertainty for all parameters is the primary concern.

Despite the attractiveness of the Bayesian paradigm, existing Bayesian transfer learning methods encounter significant limitations. \cite{hickey2022transfer} proposed a random effect calibration method for calibrating target posterior predictive credible regions, leveraging a single informative source with individual-level, low-dimensional data. \cite{abba2024bayesian} used a horseshoe (HS) prior on the $\ell_2$-distance between target and source estimates for prediction, but their method is limited to single-source settings, requires individual-level data, and is restricted to the normal-means model \citep{song2023nearly}. \cite{zhang2024concert} introduced a conditional spike-and-slab (SSL) prior for partial information sharing when only some covariates are shared between source and target datasets. However, their approach also requires individual-level source data, lacks statistical inference procedures, and is highly sensitive to the specification of the number of non-zero coefficients \citep{zhang2022bayesian,li2024bayesian}. These issues motivate the use of continuous (global-local) shrinkage priors \citep[e.g.,][]{carvalho2010horseshoe, polson2011shrink, bhattacharya2015dirichlet, Bhadra2017thehorseshoe, Veronika2019Bayesian}, which are more robust and scalable. Although Bayesian transfer learning remains an emerging field, prior work on distributed Bayesian inference for independent non-identically distributed data \citep{srivastava2015wasp} and applications in spatio-temporal models \citep{guhaniyogi2018meta,guhaniyogi2022distributed,guhaniyogi2023distributed} provides essential foundational contributions. Recently, \cite{presicce2024bayesian} proposed improving inference on the current dataset by partitioning it into smaller subsets, performing distributed inference on each subset, and aggregating the posteriors to enhance the overall inference.  In contrast, we focus on enhancing inference in a small target dataset using external source estimates from independent datasets, without joint modeling or access to individual-level source data.

To address these gaps, we propose a novel Bayesian transfer learning method, which builds on the continuous shrinkage priors and leverages multiple source estimates from pre-trained models to improve estimation and inference in a target population with limited samples. By incorporating robust priors, our approach accommodates differences in scale between source and target parameters and mitigates the risk of performance degradation when source information is less relevant. Our key contributions are fourfold. \textbf{(1) Communication- and computation-efficient multi-source transfer learning:} our approach enhances estimation in a target population with limited samples by utilizing only summary-level information from multiple pre-trained source models, achieving comparable performance to competing methods that require individual-level source data, while offering substantially faster computation, enabling scalability to high-dimensional settings. \textbf{(2) Robust Bayesian transfer learning with theoretical guarantees:} We establish posterior contraction rates and analyze finite-sample properties, demonstrating that the proposed method improves both estimation accuracy and uncertainty quantification over target-only baselines. \textbf{(3) Adaptivity to heterogeneous sources:} Our approach adaptively borrows strength from informative sources while ensuring performance no worse than the target-only baseline. It remains robust when source and target parameters differ in scale and correlation. \textbf{(4) Application to challenging variable selection scenarios:} In a real genetic data analysis, our approach demonstrates effectiveness in performing variable selection under severe multicollinearity, where frequentist methods typically struggle.

The structure of the paper is as follows. After introducing the proposed method in Section \ref{section-method} and discussing its theoretical properties in Section \ref{section-theory}, we show the results of extensive simulation studies to validate its performance in Section \ref{section-simulation}. In Section \ref{section-real-data}, we apply the proposed method to identify the genetic variants that regulate the gene expression on a high-dimensional multi-source genetic dataset, followed by discussions in Section \ref{section-discussion}.

\section{Methodology} \label{section-method}
Suppose we have a target dataset $\mathcal{D}^{(0)}=(\Xb^{(0)},\yb^{(0)})$, where $\Xb^{(0)}\in\mathbb{R}^{n\times p}$ is the $p$-dimensional covariate of size $n$ and $\yb^{(0)}\in\mathbb{R}^{n}$ is the outcome of interest. The target model can be written as
\setlength{\abovedisplayskip}{1pt}
\setlength{\belowdisplayskip}{1pt}
\begin{equation}\label{eq:target}
y_i^{(0)}=\xb_i^{(0)\top}\bbeta+\epsilon^{(0)}_i,
\end{equation}
where $(\xb_i^{(0)},y_i^{(0)}), i=1,\cdots,n$, are independent samples, and the error term $\epsilon_i^{(0)}$ are independently normal distributed with zero mean and $\sigma^{2}$ variance. We assume $\bbeta=(\beta_1,\cdots,\beta_p)^\top\in\mathbb{R}^p$ has at most $s$ non-zero entries, with $s\ll p$. Without relying on external source, we let $\widehat{ {\bm{\beta}}}_0$ denote a target-only estimate of $\bm{\beta}$ obtained from $\mathcal{D}^{(0)}$ alone. By default, we take $\widehat{ {\bm{\beta}}}_0$ to be the posterior mean of a standard HS prior applied to each regression coefficient $\beta_j$, for $j \in \{1,\cdots, p\}$:
\begin{equation}
\vspace{2mm}
\begin{split}
    \label{eq:HS}
    &\beta_j \mid \lambda_j, \tau, \sigma \sim \mathcal{N} \left(0, \sigma^2 \lambda_j^2 \tau^2 \right), \\
    &\lambda_j, \tau \sim \text{Cauchy}^+(0,1), \quad \sigma^2 \sim \text{Inverse-Gamma}(\nu, \nu),
\end{split}
\vspace{2mm}
\end{equation}
where $\text{Cauchy}^+(0,1)$ denotes a standard half-Cauchy distribution assigned to the local shrinkage parameter $\lambda_j$ and the global shrinkage parameter $\tau$, and $\text{Inverse-Gamma}(\nu, \nu)$ represents an inverse Gamma distribution governing the variance parameter $\sigma^2$.

Suppose there are $K$ independent source datasets $\mathcal{D}^{(k)}=(\Xb^{(k)},\yb^{(k)}), k=1,\cdots,K$, which may provide useful information for estimating $\bbeta$, where $\Xb^{(k)}\in\mathbb{R}^{n_k\times p}$ and $\yb^{(k)}\in\mathbb{R}^{n_k}$, $\Xb^{(k)}$ represents the same covariates as $\Xb^{(0)}$, while $\yb^{(k)}$ may correspond to the same or to a related outcome of $\yb^{(0)}$. To support a more practical and privacy-preserving framework, we assume access only to source parameter estimates $\widehat{\bomega}^{(k)} \in \mathbb{R}^p$ from each pre-trained source model $\yb^{(k)} | \Xb^{(k)}; \bomega^{(k)}$, rather than the full source datasets $\mathcal{D}^{(k)}$'s. When $\bbeta$ and $\bomega^{(k)}$ share certain similarities, we want to transfer useful information from $\widehat{\bomega}^{(k)}$ to guide the estimation of $\bbeta$ with appropriate uncertainty quantification. 

In what follows, a superscript * in the parameters denotes their true values in the frequentist sense. A true coefficient $\beta_j^* \neq 0$ is referred to as a signal, while $\beta_j^* = 0$ is referred to as a noise. The notations $\norm{\cdot}_1$ and $\norm{\cdot}$ represent the $\ell_1$- and $\ell_2$-norm, respectively, where the norm calculations exclude the intercept term. For two positive sequences $a$ and $b$, $a \prec b$ indicates $\lim a/b = 0$, $a \asymp b$ represents $0 < \lim\inf a/b \leq \lim\sup a/b < \infty$, and $a \preccurlyeq b$ denotes either $a \prec b$ or $a \asymp b$.

\vspace{-1cm}
\subsection{A robust two-step method for leveraging multiple unequal-scaled source estimates} 
TRADER consists of a robust two-step procedure to handle cases where source and target estimates are concordant but differ in scale. As illustrated in Figure~\ref{fig:angle-fig}A, traditional distance-based transfer learning methods employ the $\ell_2$-norm to measure the distance between the target parameter $\bbeta$ and source parameters $\bomega^{(k)}$. These methods assume that $\bomega^{(k)}$ defines a ball-shaped region, with radius $h$, centered at $\bomega^{(k)}$, which guides the estimation of $\bbeta$. When the distance, $h$, between the two vectors becomes large, the ball-space may not provide sufficient guidance for learning the target $\bbeta$, which is common when $\bomega^{(k)}$ is of unequal scale from $\bbeta$ \citep{gu2022robust}. A two-step procedure is proposed to address this problem (Figure \ref{fig:angle-fig}B):

\begin{figure}[!htb]
	\centering
	\includegraphics[width=1\textwidth]{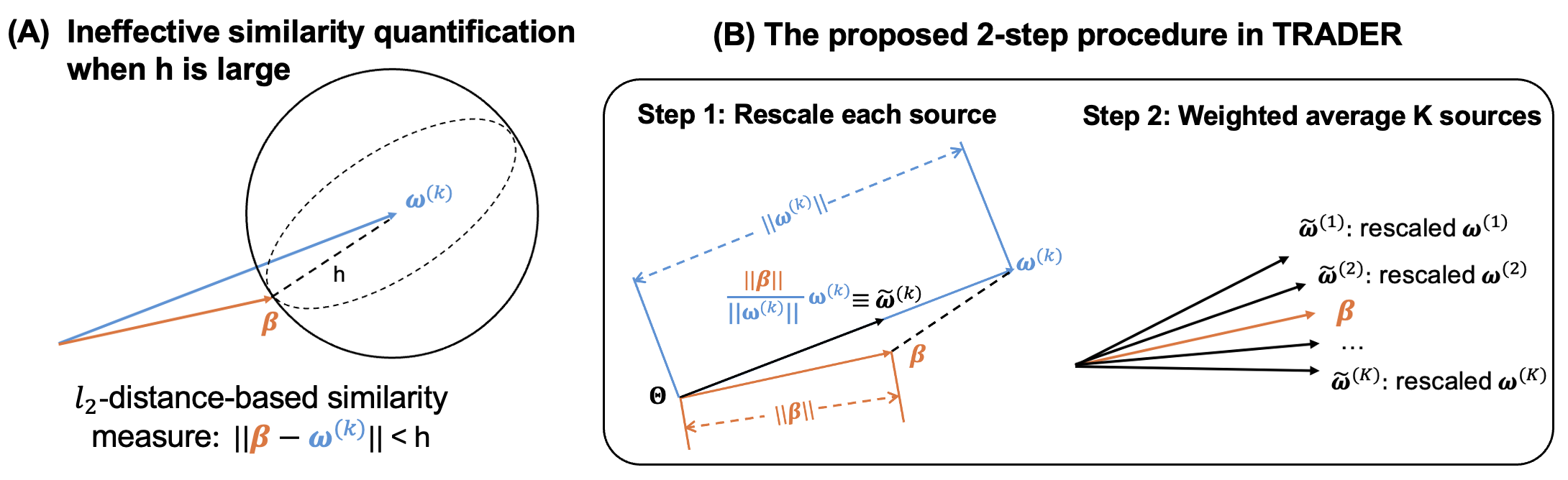}
	\caption{Schematic illustration of an existing challenge and the proposed two-step procedure. (A) When the scale of $\bbeta$ and $\bomega^{(k)}$ are different, the $\ell_2$-distance-based similarity measure spans a large ball space, which may be ineffective in leveraging the source information to estimate $\bbeta$; and (B) the proposed two-step procedure to combine multiple unequal-scaled source estimates in TRADER.}
	\label{fig:angle-fig}
\end{figure}

\noindent\textbf{Step 1: Re-scaling each source estimate.} Conceptually, each source estimate $\widehat{\bomega}^{(k)}$ is rescaled to have the same Euclidean norm as the target $\bbeta$, as illustrated in Figure \ref{fig:angle-fig}(B). Since $\bbeta$ is unknown in practice, we replace $\|\bbeta\|$ by the norm of the target-only estimate $\|\widehat{\bbeta}_0\|$, and define $\widetilde{\bm{\omega}}^{(k)} = (\norm{\widehat{\bbeta}_0}/\norm{\widehat{\bomega}^{(k)}})\widehat{\bm{\omega}}^{(k)}$ as the $k$-th scaled source estimate.

\noindent\textbf{Step 2: Adaptive weighting to combine multiple sources.} To leverage information from multiple source estimates, we compute a weighted average of all $K$ rescaled source estimates, $\sum_{k=1}^K\eta_k \widetilde{\bomega}^{(k)}$, where the weights $\eta_k$ are adaptively assigned based on the similarity between each source and the target.

Steps 1--2 are implemented by specifying the following prior on $\beta_j$:
\begin{equation}
\vspace{2mm}
\begin{aligned}
    \label{eq:TRADER}
    \beta_j & \sim \mathcal{N} \left( \sum_{k=1}^{K}\eta_k \widetilde{\omega}^{(k)}_j +  \eta_{K+1} \cdot 0, \quad \sigma^2\lambda_j^2\tau^2 \right),\\
    \left(\eta_1,\cdots,\eta_K,\eta_{K+1}\right) &\sim \text{Dirichlet}\left(\theta_1,\cdots,\theta_K,\zeta\right),\\
    \lambda_j &\sim \text{Cauchy}^+\left(0,1\right),\\
    \sigma^2 &\sim\text{Inverse-Gamma}(\nu,\nu).
    \end{aligned}
    \vspace{2mm}
\end{equation}

We refer to Equation~\eqref{eq:TRADER} as a \textit{source-guided HS prior}. Unlike the standard HS prior in Equation \eqref{eq:HS}, which is centered at zero, our proposed prior has a mean that is adaptively informed by source datasets. Specifically, the prior mean is a weighted sum of two components: a weighted average of rescaled source estimates and a zero mean weighted by $\eta_{K+1}$. To encourage borrowing from informative sources, we place a Dirichlet prior on the weights $\left(\eta_1, \cdots, \eta_K, \eta_{K+1}\right)$, where each $\theta_k$ is set by the cosine similarity between the $k$-th source and the target: $\theta_k = \widehat{\bomega}^{(k)\top}\widehat{\bbeta}_0 / (\norm{\widehat{\bomega}^{(k)}}\cdot\norm{\widehat{\bbeta}_0})$, and $\zeta=1$. This design assigns greater weight to sources more aligned with the target, enabling TRADER to leverage informative sources while downweighting others. In the extreme case where all sources are uninformative or highly heterogeneous, TRADER reduces to the standard HS prior (i.e., target-only HS) by concentrating prior mass on $\eta_{K+1}$--thus avoiding negative transfer.

The following proposition formalizes this adaptive behavior and highlights the advantage of our similarity-informed prior over a naive $\text{Dirichlet}(1,\cdots,1)$ prior, which treats all sources equally.
\begin{proposition}
    \label{prop_prior}
    Consider the prior defined in Equation~(\ref{eq:TRADER}) and denote the expectation with respect to this prior as $\,\mathbb{E}_{ \theta_1,..,\theta_K}(\cdot)$, where $\theta_1,\ldots,\theta_K$ are the hyperparameters of the Dirichlet prior. Then
   \begin{equation}
       \label{eqn:prior1}
         \mathbb{E}_{ \theta_1,..,\theta_K}(\beta_j)\,=\,   \sum_{k=1}^{K}\text{\small$\widetilde{\omega}^{(k)}_j$}\cdot  \frac{\theta_k}{ \theta_1 + \ldots +\theta_K+1  }.
        \vspace{1mm}
   \end{equation}
In particular, for $K=1 $ and $\theta_1 = \widehat{\bomega}^{(1)\top}\widehat{\bbeta}_0 / (\norm{\widehat{\bomega}^{(1)}}\cdot\norm{\widehat{\bbeta}_0})$ we have that 
\vspace{1mm}
\begin{equation}
   \label{eqn:prior2}
\frac{  \mathbb{E}_{ \theta_1}(\beta_j) }{ \mathbb{E}_{ 1}(\beta_j)}\,=\,     \frac{2\theta_1}{1+\theta_1}.   
\vspace{1mm}
\end{equation}
\end{proposition}

Proposition~\ref{prop_prior} (proof in Supplementary Section~\ref{S-supp:prop1}) shows that TRADER's adaptive prior, with $\theta_k = \widehat{\bomega}^{(k)\top}\widehat{\bbeta}_0 / (\norm{\widehat{\bomega}^{(k)}}\cdot\norm{\widehat{\bbeta}_0})$ effectively downweights uninformative sources. Unlike the naive $\text{Dirichlet}(1,\ldots,1)$ prior, which assigns equal weights to all sources, TRADER's prior assigns weights based on $\theta_k$. Informative sources (high similarity) gain larger $\theta_k$, while noninformative ones (low similarity) approach zero influence as $\theta_k \to 0$, minimizing bias from irrelevant sources. For example, in the case of one noninformative source ($K=1$ and small $\theta_1$), Equation~(\ref{eqn:prior2}) yields $\vert \mathbb{E}_{\theta_1}(\beta_j) \vert \ll \vert \mathbb{E}_1(\beta_j) \vert$, correctly shrinking the prior mean toward zero.

We follow the approach of \cite{vanderpas2017adaptive,vanderpas2017uncertainty} to determine the global parameter $\tau$. Specifically, we put a half-Cauchy $\mathcal{C}^+(0,1)$ prior on $\tau$ and restrict it in $(1/p,1)$. This restriction reflects the interpretation of $\tau$ as the proportion of non-zero parameters, implying that at least one and at most all $\beta_j$'s are non-zero \citep{vanderpas2017adaptive,vanderpas2017uncertainty}. The lower bound mitigates computational issues associated with extremely small values of $\tau$ \citep{vanderpas2014thehorseshoe}. Alternatively, users may choose to fix $\tau$ as a prespecified constant. For other parameters, we set a standard half-Cauchy prior for $\lambda_j$, an inverse Gamma prior with $\nu=0.001$ for $\sigma^2$.

We implement an efficient MCMC algorithm to sample from the posterior distribution of TRADER, with details provided in Supplementary Section \ref{S-supp:mcmc}. We find that our algorithm offers a significant computational advantage over competing frequentist methods, especially when $p$ increases, which will be discussed further in Section \ref{sec:high-d-sim}.	

\subsection{Posterior distribution of TRADER}
We show that the posterior mean of TRADER is a weighted average between the target and source estimates. To illustrate, we first consider the case of a single source estimator $\widehat{\bomega}^{(1)}$. Given the target data $\mathcal{D}^{(0)}$, $\widehat{\bomega}^{(1)}$, and hyperparameters, the posterior distribution of TRADER can be expressed as $\pi \left(\bbeta\mid \tau,\bLambda,\widehat{\bomega},\mathcal{D}^{(0)}\right) \sim \mathcal{N}\left(\widehat{\bbeta},\widehat{\bSigma} \right)$, where
\vspace{1mm}
\begin{equation}
\begin{aligned}
\label{eq:normal-means}
 &\begin{cases} 
	\widehat{\bbeta}=\tau^2\bLambda[\tau^2\bLambda+(\Xb^{(0)\top}\Xb^{(0)})^{-1}]^{-1}\widehat{\bbeta}^\text{OLS}+\left(\tau^2\bLambda\Xb^{(0)\top}\Xb^{(0)}+\Ib\right)^{-1}\widetilde{\bomega}^{(1)}\\
	\widehat{\bSigma}=\sigma^2\left(\Xb^{(0)\top}\Xb^{(0)}+\tau^{-2}\bLambda^{-1}\right)^{-1}
\end{cases}
\end{aligned}
\end{equation}
\vspace{1mm}
with $\bLambda=\text{diag}(\lambda^2_1,\cdots,\lambda_p^2)$, $\widehat{\bbeta}^\text{OLS}=(\mathbf{\Xb}^{{(0)}\top} \mathbf{\Xb}^{(0)})^{-1} \mathbf{\Xb}^{(0)\top} \yb^{(0)}$ as the ordinary least squared (OLS) estimate if $(\mathbf{\Xb}^{(0)\top} \mathbf{\Xb}^{(0)})^{-1}$ exists, and the scaled source parameter \\ $\widetilde{\bomega}^{(1)}=(\norm{\widehat{\bbeta}_0}/\norm{\widehat{\bomega}^{(1)}})\widehat{\bomega}^{(1)}$. When the columns of $\mathbf{\Xb}^{(0)}$ are uncorrelated with zero mean and unit variance, we have $\mathbf{\Xb}^{(0)\top} \mathbf{\Xb}^{(0)} \approx n\mathbf{I}$, and we can approximate the $j$-th element of $\widehat{\bbeta}$ as a weighted average of the $j$-th element of $\widehat{\bbeta}^\text{OLS}$ and $\widetilde{\bomega}^{(1)}$
\begin{equation}
\vspace{1mm}
\label{eq:trajectory}
\widehat{\beta}_j=\left(1-\kappa_j\right){\widehat{\beta}^\text{OLS}}_j+\kappa_j \widetilde{\omega}^{(1)}_j,
\vspace{1mm}
\end{equation}
where the weight of the source is $\kappa_j=1/(1+\tau^2\lambda_j^2n)$. 

Equation \eqref{eq:trajectory} indicates that the posterior of TRADER is a weighted average of $\widehat{\bbeta}^\text{OLS}$ and $\widetilde{\bomega}^{(1)}$, where the degree of proximity towards source is determined by $\kappa_j$. For multiple source estimates, the results extend by replacing the single source estimate with a weighted average of multiple scaled source estimates, $\sum_{k=1}^K\eta_k\widetilde{\bomega}^{(k)}$. The weighted average form of the TRADER estimator is consistent with findings in other transfer learning methods, where the target-source similarity is measured by various distance-based similarity metrics, such as $\ell_2$-norm \citep{gu2024equivalence} or Kullback-Leibler divergence \citep{hector2024turning}.

\section{Theoretical properties} \label{section-theory}
In this section, we study the theoretical properties of TRADER. We demonstrate that by incorporating informative source information, TRADER achieves more accurate posterior concentration around the true coefficients than standard continuous shrinkage priors. Moreover, when the sources are uninformative, TRADER can prevent negative transfer and covers the standard continuous shrinkage priors as a special case. We first impose some common assumptions that follow \cite{song2023nearly}:
\begin{assumption}
    \label{as1}
    All the covariates are uniformly bounded. For simplicity, we assume that $x_{ij} \in [-1,1]$ where $x_{ij}$ denotes the element in the $i$-th row and the $j$-th column of $\Xb^{(0)}$.
\end{assumption}

\begin{assumption}
\label{as2}
    We require that the dimension be high: $p \succcurlyeq n$.
\end{assumption}

\begin{assumption}
	\label{as3}
	We assume that $s\log p \prec n$. 
\end{assumption}

\begin{assumption}
\label{as4}
	There exists an integer $\overline{p}$ such that $\overline{p}> s$, $\overline{p} \asymp s$, and a constant $\lambda_0 >0$ satisfying $\lambda_{\min}(  \Xb_{\xi}^{(0)\top}\Xb_{\xi}^{(0)} ) \geq n \lambda_0$ for all $\xi$ with $\vert \xi\vert \leq \overline{p}$.
\end{assumption}

Assumption \ref{as1} puts a minor condition on the covariates, while Assumptions \ref{as2}-\ref{as4} allow us to have identifiability of the model and develop theories. Although Assumption~\ref{as3} puts a condition on the target sample size $n$, we will show in Corollary \ref{cor:n0assumption} that TRADER remains valid even as $n$ increases. Besides these assumptions, for readability, we omit the superscript when referring to a single scaled source and denote it as $\widetilde{\bomega} = (\norm{\widehat{\bbeta}_0}/\norm{\widehat{\bomega}})\widehat{\bomega}$ in this section. 

\subsection{Posterior contraction rate}
The first result below considers the single source setting, establishing conditions under which the source is helpful, and providing the posterior consistency of TRADER.
\begin{theorem}[\textbf{Posterior contraction oracle inequality in single-source}]
	\label{thm1}
	Consider the target model \eqref{eq:target}, where the design matrix $\Xb^{(0)}$ and the true $\bbeta^*$ satisfy the Assumptions \ref{as1}-\ref{as4}. The prior for $\bbeta$ and $\sigma^{2}$ is given by Equation~\eqref{eq:TRADER} with fixed $\eta_1 =1 $. Let 
    \vspace{2mm}
	\[
    \begin{split}
	\mathcal{A}  \,:=\,\Bigg\{  r \geq0:\, \, r \leq C\min\left\{\frac{1}{\sqrt{s} }, \sigma^*\left(\frac{ \widetilde{s}(r) \log p }{s^2 n }\right)^{1/4}    \right\}   \\  \,\text{and}\, \sum_{j\,:\,\vert\beta^*_j - \widetilde{\omega}_j\vert\leq r   }\vert\beta^*_j - \widetilde{\omega}_j\vert \,\leq \, C \sqrt{\frac{  \widetilde{s}(r) \log p}{n}}   \Bigg\}
    \end{split}
	\]
    \vspace{3mm}
	for a large enough constant $C>0$ and where $
	\widetilde{s}(r) \,=\, \vert\{ j\,:\, \vert\beta^*_j - \widetilde{\omega}_j\vert > r    \}\vert$.
	Let $r_{\sup} = \sup\{ r\,:\, r\in \mathcal{A}\}$.  Suppose that there exists $\gamma>0$ such that $ \| (\bbeta^*-\widetilde{\bomega})/\sigma^*\|_{\infty} \leq E:= p^{\gamma}$ for some $\gamma>0$.
	If $\tau $ satisfies $\tau \asymp p^{-u}$ for some appropriate $u>0$, then for $\epsilon_{n} =  M  \sqrt{\widetilde{s}(r_{\sup}) \log p / n}$  for a large enough constant $M$, it holds that 
	\[
	\mathbb{P}^*\left(  \pi( \| \bbeta -\bbeta^*\| \geq   c_1\sigma^* \epsilon_{n} | \mathcal{D}^{(0)}) \geq   e^{-c_2 n \epsilon_{n}^2 }\right)   \leq  e^{-c_3 n \epsilon_{n}^2},
	\]
	for some positive constants $c_1$, $c_2$, and $c_3$. 
\end{theorem}
Theorem~\ref{thm1} states that, with high probability under the true data-generating process, the posterior assigns exponentially small mass outside a shrinking ball around $\bbeta^*$. While the inequality may appear nonstandard in form, it follows the convention used in posterior contraction theory \citep{song2023nearly}. The outer probability bounds the chance that the posterior assigns non-negligible mass outside a ball of radius $\epsilon_n$ around $\bbeta^*$, and the inner term measures the tail probability under the posterior. Together, they quantify posterior concentration at rate $\epsilon_n$ with high probability.
\begin{remark}
In Theorem~\ref{thm1}, $r_{\sup}$ represents a threshold that provides relaxation to the usual notion of sparsity. The function $\widetilde{s}(r)$ counts the number of coefficients $\widetilde{\omega}_j$ whose deviation from $\beta^*_j$ is greater than $r$, which measures how well $\widetilde{\bomega}$ approximates $\bbeta^*$, with a larger $r$ corresponding to fewer coefficients contributing to the sum. We choose $r_{\sup}$ to maximize $r$ within $\mathcal{A}$ because $\widetilde{s}(r)$ decreases as $r$ increases, meaning fewer coefficients violate the threshold as $r$ grows. The constraint $r \in \mathcal{A} $ balances the approximation quality of $\widetilde{\bomega}$ to $\bbeta$ with sparsity, based solely on differences $\widetilde{\omega}_j -\beta_j$. The small $r$ requirement arises from the proof,  and $0 \in \mathcal{A}$ ensures $\widetilde{s}(r_{\sup}) \lesssim s + \text{sparsity of}\,\, \widetilde{\bomega}$. For settings where $\widetilde{\bomega}$ has $O(s)$ nonzero coefficients, we obtain $\widetilde{s}(r) \lesssim s$. Thus, if $\widetilde{\bomega}$ reasonably estimates the sparsity of  $\bbeta^*$, the posterior contraction rate is no worse than when $\widetilde{\bomega} = 0$. 
\end{remark}

\begin{corollary}
	\label{cor1}
	Under the conditions of Theorem \ref{thm1}, if $\bbeta^* -\widetilde{\bomega}$  is sparse in the sense that $s(\bbeta^* -\widetilde{\bomega}) \,:=\, \vert \{ j\,:\, \vert \beta^*_j -\widetilde{\omega}_j \vert \neq 0 \}\vert \ll  s$, then
	\begin{equation}
	\widetilde{s}(r_{\sup}) \,\leq \, s(\bbeta^* -\widetilde{\bomega}) \,\ll\, s.
	\end{equation}
	Hence,  the posterior contraction rate is much faster than the original rate of target-only HS (i.e., no source is used), thus
	\[
    \vspace{2mm}
	\sqrt{\frac{ \widetilde{s}(r_{\sup})\log p} {n}} \ll \sqrt{\frac{ s\log p} {n}}. 
    \vspace{2mm}
	\]
	If, in addition, many entries of $\bbeta^* -\widetilde{\bomega}$  are small such that $\widetilde{s}(r_{\sup}) \ll s(\bbeta^* -\widetilde{\bomega}) $, then the rate in Theorem \ref{thm1} satisfies
	\[
    \vspace{2mm}
	\sqrt{\frac{ \widetilde{s}(r_{\sup})\log p} {n}} \ll \sqrt{\frac{ s(\bbeta^* -\widetilde{\bomega})\log p} {n}},
     \vspace{2mm}
	\]
    highlighting the usefulness of the oracle inequality established in Theorem \ref{thm1}.
\end{corollary}
\begin{remark}
Corollary~\ref{cor1} demonstrates the helpfulness of leveraging rescaled source estimate $\widetilde{\bomega}$ in improving the posterior contraction rate, provided that $\widetilde{\bomega}$ aligns well with the true parameter $\bbeta^*$. It indicates that the closer $\widetilde{\bomega}$ is to $\bbeta^*$, both sparsely and in magnitude, the more significant the gains in posterior contraction rate, reinforcing the practical value of the oracle inequality in Theorem~\ref{thm1}. Specifically, (1) If the differences between $\bbeta^*$ and $\widetilde{\bomega}$ are sparse, meaning that only a small number of entries in $\bbeta^* - \widetilde{\bomega}$ are non-zero, the effective sparsity parameter $\widetilde{s}(r_{\sup})$ is much smaller than $s$, leading to a faster posterior contraction rate; (2) When these differences are not only sparse but many are small in magnitude, the effective sparsity $\widetilde{s}(r_{\sup})$ becomes even smaller compared to $s(\bbeta^* - \widetilde{\bomega})$. This further enhances the posterior contraction rate, making it significantly faster than the rate when no source is used; and (3) These results highlight the utility of incorporating well-aligned source information. When $\widetilde{\bomega}$ closely approximates $\bbeta^*$, both in terms of sparsity and magnitude of differences, the model achieves much faster learning rates. This underscores the importance of selecting or constructing high-quality source information.
\end{remark}

\begin{corollary}
\label{cor:n0assumption}
   TRADER is particularly effective when the target sample size $n$ is relatively small, which is a common scenario in transfer learning. However, even when $n$ increases and the source data remain fixed, TRADER retains consistency. In some cases, it can still attain the optimal convergence rate, possibly with larger constants.
\end{corollary}
We prove Corollary \ref{cor:n0assumption} in the Supplementary Section \ref{S-cor:n0assumption-proof}. Our next result outlines conditions for ensuring posterior consistency in the presence of multiple sources, where the key condition~(\ref{eqn:extra-cond}) regulates the support magnitude when aggregating information across sources.

\begin{theorem}[\textbf{Posterior consistency with multiple sources that underestimate the support}]
	\label{thm2}
	Let $\widetilde{\bomega}^{(1)},\ldots,\widetilde{\bomega}^{(K)} \in \mathbb{R} ^p$ be vectors learned from $K$ source data, respectively, independent of the target data. Suppose that
	\[
	\max\{\|\bb^*\|_{\infty}, K\underset{k=1,\ldots,K}{\max}\| \widetilde{\bomega}^{(k)}\|_{\infty} \}/\sigma^* \,\leq\, p^{\gamma}
	\]
	for a positive constant $\gamma$. Consider the prior for $\bbeta$ and $\sigma^{2}$ defined by Equation~\eqref{eq:TRADER} with $\tau >0$.  Next, let $\widetilde{\bomega} \,=\, \sum_{k=1}^{K}  \eta^*_k \widetilde{\bomega}^{(k)}$
	for arbitrary $\eta_1^*,\ldots,\eta_k^* \in [0,1]$ such that $\sum_{k=1}^K \eta^*_k \leq 1$.
	Let $r \in \mathcal{A}$, where $\mathcal{A}$ is as defined in Theorem \ref{thm1},  and suppose that for an appropriate constant $c>0$, 
	\begin{equation}
		\label{eqn:extra-cond}
		\vert   \{  j\in \{1,\ldots,p\}\,:\, \underset{k=1,\ldots,K}{\max  } \, \vert \widetilde{\bomega}_j^{(k)} \vert \,>0\,   \}\vert \,\leq \, c \widetilde{s}(r).
	\end{equation}
	If $\tau $ satisfies $\tau \asymp n^{-u}$ for some appropriate $u>0$, then for $\epsilon_n =  M  \sqrt{\widetilde{s}(r) \log p / n } $  and a large enough positive constant $M$, it holds that
	\[
	\mathbb{P}^*\left(  \pi( \| \bbeta -\bbeta^*\| \geq   c_1\sigma^* \epsilon_n | \mathcal{D}^{(0)}) \geq  e^{-c_2 n \epsilon_n^2 }\right)   \leq  e^{-c_3 n \epsilon_n^2}
	\]
	for some positive constants $c_1$, $c_2$, and $c_3$.	
\end{theorem}
\begin{remark}
	Theorem~\ref{thm2} highlights the flexibility of combining multiple sources, even when each individually underestimates the support of $\bbeta^*$. The weights $\eta_1^*, \ldots, \eta_K^*$, determine how the rescaled sources $\widetilde{\bomega}^{(1)}, \ldots, \widetilde{\bomega}^{(K)}$ are combined to approximate $\bbeta^*$. While these weights are unknown, they can be arbitrary as long as the theorem's conditions are met, and an oracle with perfect knowledge of $\bbeta^*$ could select the optimal weights to achieve the sharpest posterior contraction rate. Furthermore, Theorem~\ref{thm2} includes realistic scenarios where each source provides a partial view of the true support, capturing only a subset of the relevant nonzero coefficients in $\bbeta^*$. By combining these sources with appropriately chosen weights, the model leverages their collective strength to better approximate $\bbeta^*$, even when individual sources are noisy or incomplete. Moreover, the sparsity condition (\ref{eqn:extra-cond}) ensures that the combined source information retains a sparse structure, enabling efficient and robust posterior contraction. This makes the framework well-suited for high-dimensional settings, where reliable information is often spread across multiple sources. In the next result, we give more intuition about the actual upper bound. 
\end{remark}
Another interesting observation from Theorem \ref{thm2} is that TRADER can account for the scenarios of partial information sharing discussed in \cite{zhang2024concert}, where sources underestimate the true support. Specifically, they assumed that $\widetilde{\bomega}^{(k)}_\xi$ closely resembles $\bbeta_\xi$ only for some $\xi \in \{1,\dots,p\}$, while other covariates are uninformative. Under condition (\ref{eqn:extra-cond}), Theorem~\ref{thm2} demonstrates TRADER's ability to leverage source information even when individual $\widetilde{\bomega}^{(k)}$ underestimates the true parameter support in this scenario.

\begin{corollary} \label{cor2}
	Using the notation from Theorem \ref{thm2}, if we take $r=0$, then $r\in \mathcal{A}$, and condition (\ref{eqn:extra-cond}) becomes 
	\begin{equation}
		\label{eqn:cond}
		\vert   \{  j\in \{1,\ldots,p\}\,:\, \underset{k=1,\ldots,K}{\min  } \, \vert \widetilde{\omega}_j^{(k)} \vert \,>0\,   \}\vert \,\leq \, c  \vert\{ j\,:\, \vert\beta^*_j - \widetilde{\omega}_j\vert > 0    \}\vert\,:=\,s(\eta^*),
	\end{equation}
	since the weights $\eta_1^*,\ldots,\eta_K^*$ determine $\widetilde{\bomega}$. Then the contraction rate can be expressed as
	\[
    \vspace{2mm}
	\sqrt{\frac{\log p}{n}} \cdot \underset{  \eta_1^*,\ldots,\eta_K^* \in [0,1],\, \sum_{k=1}^K \eta_k^* \leq 1,\, \eta^* \,\text{satisfies} \, (\ref{eqn:cond})}{\inf}  \,\sqrt{s(\eta^*) }.
    \vspace{2mm}
	\]
	If we take $\eta^*=0$, then condition (\ref{eqn:cond}) becomes 
	\begin{equation}\label{eqn:cond2}
		\vert   \{  j\in \{1,\ldots,p\}\,:\, \underset{k=1,\ldots,K}{\max } \, \vert \widetilde{\omega}_j^{(k)} \vert \,>0\,   \}\vert \,\leq \, c  s,
	\end{equation}
	and the corresponding contraction rate is $\sqrt{s\log p/n}$. Thus, under condition (\ref{eqn:cond2}), the worst-case contraction error rate is the same as if no sources were used, i.e., target-only HS. 
\end{corollary}

\begin{remark}
	Corollary~\ref{cor2} highlights the role of weights $\eta_1^*, \ldots, \eta_K^*$ in determining how the sources contribute to the contraction rate. When $r = 0$, the effective sparsity $s(\eta^*)$ reflects the overlap between the true support of $\bbeta^*$ and the combined source information. A smaller $s(\eta^*)$ results in a faster contraction rate, showing the benefit of well-aligned sources. In the worst-case scenario where $\eta^* = 0$, i.e., no contribution from sources, the contraction rate reduces to the baseline $\sqrt{s \log p / n}$ \citep{song2023nearly}, equivalent to not using source information in the target-only estimate.
\end{remark}

\subsection{Finite-sample marginal posterior behavior}
\label{sec:fintesample}
The following result establishes TRADER's finite-sample marginal posterior behavior. We will show that TRADER achieves reliable inference by balancing the contributions of source and target information, even in scenarios with large biases or misaligned sources.

To start with, we express model \eqref{eq:target} as 
\begin{equation*}
    \yb^{(0)}=\xb_j^{(0)}\beta_{j}+\Xb_{-j}^{(0)}\bbeta_{-j}+\bepsilon^{(0)},
\end{equation*}
where $\Xb_{-j}^{(0)}$ denotes $(\xb_k^{(0)}, k\neq j)$ and $\bbeta_{-j}=\left(\beta_1,\cdots,\beta_{j-1},\beta_{j+1},\cdots,\beta_p\right)^\top$. Based on Equation \eqref{eq:normal-means}, the conditional distribution of $\bbeta_{-j}$ given all other parameters is
\begin{equation}
	\label{eq:con-posterior}
	\pi\left(\beta_j\mid\bbeta_{-j},\tau,\bLambda, \xb^{(0)}_j, \Xb^{(0)}_{-j},\yb^{(0)}\right)\propto \mathcal{N}\left(g_j\left(\widehat{\bbeta}_{-j}\right),\sigma^2\left(\xb^{(0)\top}_{j}\xb_{j}^{(0)}\right)^{-1}\right)\pi_{\beta_{j}}(\beta_{j}),
\end{equation}
where $\pi_{\beta_{j}}(\beta_{j})$ denotes a general continuous shrinkage prior on $\beta_j$. Because of the Bayesian estimation consistency, $\bbeta_{-j}$ will converge to $\bbeta_{-j}^*$. The center of the marginal posterior of $\beta_j$'s will then approximately be $g_j(\widehat{\bbeta}_{-j})$, where $\widehat{\bbeta}_{-j}$ is the posterior of $\bbeta_{-j}$ and 
\begin{equation}
\label{eq:marginal-posterior}
g_j\left(\widehat{\bbeta}_{-j}\right)=\underbrace{\beta^*_{j}+\left(\xb^{(0)\top}_j\xb_j^{(0)}\right)^{-1}\xb_j^{(0)\top}\bepsilon^{(0)}}_{\text{MLE of }\beta_j}+\underbrace{\left(\xb_j^{(0)\top}\xb_j^{(0)}\right)^{-1}\xb_j^{(0)\top}\Xb^{(0)}_{-j}\left(\bbeta_{-j}^*-\widehat{\bbeta}_{-j}\right)}_{\text{bias term}}.
\end{equation}
When $\widehat{\bbeta}_{-j}=\bbeta_{-j}^*$, the first two terms of Equation \eqref{eq:marginal-posterior} represent the maximum likelihood estimate (MLE) of $\bbeta_j$ with the last term being a bias term.

\begin{theorem} [\textbf{Finite-sample marginal posterior behavior under single-source}]
    \label{thm3}
  Suppose that $\sigma^{*2} =\sigma^2$ is known and consider the setting of Theorem \ref{thm1} with no prior on $\sigma^2$.  Let $j \in \{1\ldots,p\}$ and 
  $\sigma_j^2 =  \sigma^2  ( \xb_j^{(0) \top}  \xb_j^{(0) }   )^{-1} $. Assume that $\| \xb_j^{(0) } \|\asymp \sqrt{n}$, and  there exists $\overline{p}> \max\{\widetilde{s}(r_{\sup}),s\}$ and a constant  $a_0>0$ such that 
  \[
  \vspace{2mm}
    \lambda_{\max}\,:=\,\Lambda_{\max}\left( \frac{ \Xb_{\xi}^{(0)\top} \Xb_{\xi}^{(0)}  }{n }\right) < a_0
    \vspace{1mm}
  \]
  for all $\xi \subset \{1,\ldots,p\}$ with $\vert \xi\vert < \overline{p}$. Let  us  also assume that $\vert \beta_j^* -\widetilde{w}_j\vert^2 = o( \sigma p^2  )$. Define
\begin{equation}
\vspace{2mm}
    \label{set}
     \mathcal{B}\,:=\,\left\{ \bbeta_{-j}\,:\, \bigg\vert\underbrace{\left(\xb_j^{(0)\top}\xb_j^{(0)}\right)^{-1}\xb_j^{(0)\top}\Xb^{(0)}_{-j}\left(\bbeta_{-j}^*-\bbeta_{-j}\right)}_{\text{bias term}}\bigg\vert \leq C_0 \sqrt{\frac{  \tilde{s}({r_{\sup}})\log p }{n}}  \right \}
     \vspace{2mm}
\end{equation}
 for some appropriate constant $C_0>0$. The following holds:\\
\textbf{(i)} Suppose  $\vert \widetilde{\omega}_j -  \beta_j^*\vert  > C \sigma \epsilon_n$ for some constant $C>0$. 
  Then, there exists a distribution $F_j $ with support contained in $\mathcal{B}$ such that the random variable 
 $\beta_j$ generated as 
  \begin{equation}
      \label{eqn:post_approx}
          \begin{array}{lll}
            \beta_j \,|\,  \bbeta_{-j} &\sim   &  \mathcal{N}\left( g_{j}\left(\bbeta_{-j}\right),\sigma_j^2\right),\\
            \bbeta_{-j}  & \sim &  F_j,
          \end{array}
  \end{equation}
has a marginal distribution $f_j$ that satisfies
\begin{equation}
       \label{eqn:e77}
       \| f_j -   \pi(\beta_j \,|\, \mathcal{D}^{(0)}, \sigma^2  )\|_{ \mathrm{TV} }  \,=\,  O\left( \frac{ \vert \beta_j^* -\widetilde{\omega}_j\vert^2  }{\sigma  p^2  }  \,+\,  e^{- c_2n \epsilon_n^2}     \,+\, \frac{ \sigma\epsilon_n}{\vert \beta_j^* - \widetilde{\omega}_j \vert  }  \right)
   \end{equation}
    with probability approaching one.\\
 \textbf{(ii)} In contrast, if $\vert \widetilde{w}_j -  \beta_j^*\vert  \leq C \sigma \epsilon_n$, then in the construction of $\beta_j  \,|\,   \bbeta_{-j} $ in (\ref{eqn:post_approx}), we have 
\[
p(\beta_j  \,|\,   \bbeta_{-j} \,) \,\propto   \, \mathcal{I}(\beta_j \in R(\bbeta_{-j}) ) \,\mathcal{N}(\beta_j; g_j(\bbeta_{-j} ),\sigma_j^2   ) \,\cdot \,h(\beta_j ),
 \]
where $\mathcal{I}(\beta_j \in R(\bbeta_{-j}))$ is the indicator for the interval $R(\bbeta_{-j}) \,=\, [g_j(\bbeta_{-j})- \widetilde{C} \sigma\epsilon_n,  g_j(\bbeta_{-j})+ \widetilde{C} \sigma\epsilon_n]$, with the constant $\widetilde{C}>0$ large enough such that $\widetilde{\omega}_j \in R(\bbeta_{-j})$. Moreover,   the function $h$ satisfies $\underset{\beta_j \rightarrow \widetilde{\omega}_j}{\lim  } h(\beta_j)  \,=\,\infty$. Then, for the corresponding $f_j$, with probability approaching one, it holds that
  \begin{equation}
       \label{eqn:e955}
       \| f_j -   \pi(\beta_j \,|\,\mathcal{D}^{(0)}, \sigma^2  )\|_{ \mathrm{TV} }  \,=\,  o\left(1   \right).
   \end{equation}
\end{theorem}

\begin{remark}
Theorem~\ref{thm3} describes the behavior of the posterior mean for each coefficient $\beta_j$ under TRADER. When \(|\widetilde{\omega}_j - \beta_j^*| > C\sigma\epsilon_n\), we obtain a finite-sample result establishing that the marginal posterior for $\beta_j$ can be approximated as $\mathcal{N}(\widehat{\beta}_j,\sigma_j^2 ) + \text{bias}$, where  $\widehat{\beta}_j:= \beta^*_{j}+\left(\xb^{(0)\top}_j\xb_j^{(0)}\right)^{-1}\xb_j^{(0)\top}\bepsilon^{(0)}$ is the MLE for $\beta^*_j$ and the bias term is guaranteed to satisfy $O(\sqrt{\tilde{s}(r_{\sup})\log p/n})$. Thus, 
if the sparsity level $s(r_{\sup})$ is oracle-like, as stated in Theorem~\ref{thm1}, TRADER's posterior centers around MLE with a diminishing bias that can be substantially smaller than $O(\sqrt{s\log p/n})$. This result is consistent with the shape approximation guarantees established in Theorem 2.7 and Corollary 2.8 of \cite{song2023nearly}. In fact, under the more restrictive beta-min condition specified in Corollary 2.8, our result reduces precisely to the one presented there. Our finding demonstrates TRADER's ability to leverage the overall informativeness of the source $\widetilde{\bomega}$ to improve the estimation of $\beta_j$, even when $\widetilde{\omega}_j$ deviates by at least $C\sigma\epsilon_n$ from the true coefficient $\beta_j^*$. 

In contrast, when \(|\widetilde{\omega}_j - \beta_j^*| < C\sigma\epsilon_n\), indicating that the source $\widetilde{\omega}_j$ is informative for $\beta_j$, Theorem \ref{thm3} indicates that the marginal posterior concentrates tightly around the true coefficient $\beta_j^*$, even if the bias remains at the order of $\sqrt{s\log p/n}$. However, in practice, our results do not directly describe how coverage behaves when $n$ is moderate or the signal is weak. Consequently, TRADER, similar to standard HS priors, may exhibit undercoverage, as noted in \cite{vanderpas2017uncertainty}.
\end{remark}
\begin{remark}
In TRADER, if at least one source estimate is informative, Theorems \ref{thm1} and \ref{thm2} establish conditions under which the fitting error of $\beta_{j}$ converges faster than the usual $O(\sqrt{s\log{p}/n})$. Building on Theorem~\ref{thm3}, TRADER achieves faster convergence through its adaptive guided structure, which prioritizes target-only information in the presence of large biases, i.e., $|\widetilde{\omega}_j - \beta_j^*| \gg C\sigma \epsilon_n$. Thus, TRADER balances contributions from source and target information, ensuring robust coverage and reliable inference.
\end{remark}

\section{Simulations}\label{section-simulation}
We conduct extensive simulation studies to evaluate the performance of TRADER. We compare TRADER to the target-only HS and the competing frequentist method, TransGLM \citep{tian2023transfer}, which also aims to enhance the target data performance by borrowing information from the source. The two transfer learning methods differ in their use of source data: TransGLM requires full individual-level data $\mathcal{D}^{(k)}$ from each source, whereas TRADER only uses source parameter estimates $\widehat{\bomega}^{(k)}$. Our objective is to assess whether TRADER can outperform the target-only HS by leveraging source information while achieving estimation accuracy comparable to TransGLM. The source estimates $\widehat{\bomega}^{(k)}$ and the target-only estimates $\widehat{\bbeta}_0$ are obtained by fitting a standard SSL prior to each $\mathcal{D}^{(k)}, k=0,1,\cdots, K$, and only these fitted values are used in TRADER. We evaluate model performance using several metrics, with detailed definitions in Supplementary Section \ref{S-supp:simu-metric}. Unless otherwise noted, each simulation is repeated 100 times. HS is implemented using the \texttt{R} package \texttt{horseshoe} and TransGLM via \texttt{glmtrans}. For all Bayesian methods, we run 10,000 MCMC iterations and discard the first 5,000 as burn-in. Additional simulation results are reported in the Supplementary Material.

\subsection{Setting I: Some but unknown sources are informative}
We consider the case where a subset of $K = 10$ sources is informative, i.e., their true parameter vectors are concordant with the target, while the remaining sources are uninformative. In practice, the informativeness of sources is often unknown. As TRADER is designed to adaptively downweight uninformative sources, this simulation evaluates its robustness in such heterogeneous scenarios.

The target dataset has $n = 200$ samples and $p = 500$ dimensions. Covariates $\xb_i^{(0)}$ are generated from a multivariate normal distribution $\mathcal{N}(\mathbf{0}_p, \Ab)$, where $\Ab = \left[A_{jj'}\right]_{p \times p}$ and $A_{jj'} = 0.5^{\left|j-j'\right|}$ for $i = 1, \ldots, n$. The true target coefficients are $\bbeta = (0.5 \times \bm{1}_s^\top, \bm{0}_{p-s}^\top)^\top$ with $s = 5$ non-zero entries, where $\bm{1}_s$ is a length-$s$ vector of ones and $\bm{0}_{p-s}$ is a length-$(p - s)$ vector of zeros.

Each of the $K = 10$ sources has $n_k = 200$ samples. Covariates $\xb_i^{(k)}$ are drawn independently from a Student's $t$-distribution with four degrees of freedom.  Let $\mathcal{I}_p^{(k)} \in \{-1, 1\}^p$ be a length-$p$ vector of i.i.d. random variables, independent across $k$. For informative sources ($k = 1, \ldots, K_a$), we set $\omega^{(k)}_j = \beta_j + (h/p)\cdot \mathcal{I}_p^{(k)}$. For the remaining $K-K_a$ uninformative sources ($k = K_a + 1, \ldots, K$), we define a subset $\mathcal{S}^{(k)} \subset \{2s+1, \ldots, p\}$ of size $s$ sampled at random, and set:
$$
\vspace{1mm}
\omega_j^{(k)}= \begin{cases}0.5+2 h \mathcal{I}_j^{(k)} / p, & j \in\{s+1, \ldots, 2 s\} \cup \mathcal{S}^{(k)}, \\ 2 h \mathcal{I}_j^{(k)} / p, & \text {otherwise.}\end{cases}
$$
Source datasets are generated independently, and neither TRADER nor TransGLM knows which sources are informative a priori.

We assess model performance based on their ability to detect true signals (\(\beta^*_j \neq 0\)). For Bayesian methods (HS and TRADER), a signal is detected when the credible level (CL) for $\beta_j$ exceeds 0.95, where CL is defined as
\begin{equation}
\label{eq:CL}
\mathrm{CL}_j = \left|\, \widehat{\mathbb{P}}(\beta_j > 0 \mid \mathcal{D}) - \widehat{\mathbb{P}}(\beta_j < 0 \mid \mathcal{D}) \,\right| \in [0,1],
\end{equation}
and the posterior probabilities are estimated using MCMC samples. Equation~\eqref{eq:CL}, introduced in \cite{li2024bayesian}, is called CL for $\beta_j$ as it can be understood as the highest value at which the corresponding equal-tailed credible interval of $\beta_j$ does not include zero. For the frequentist method (TransGLM), a signal is identified when the $p$-value for $\beta_j$ is below 0.05, computed based on Corollary 2.1 of \citet{van2014asymptotically} due to the absence of built-in $p$-value computation in the \texttt{glmtrans} \texttt{R} package. We compare the average estimation error, power, false discovery rate (FDR), and computation time of the three methods.

\begin{figure}[ht]
	\centering
	\includegraphics[width=0.8\textwidth]{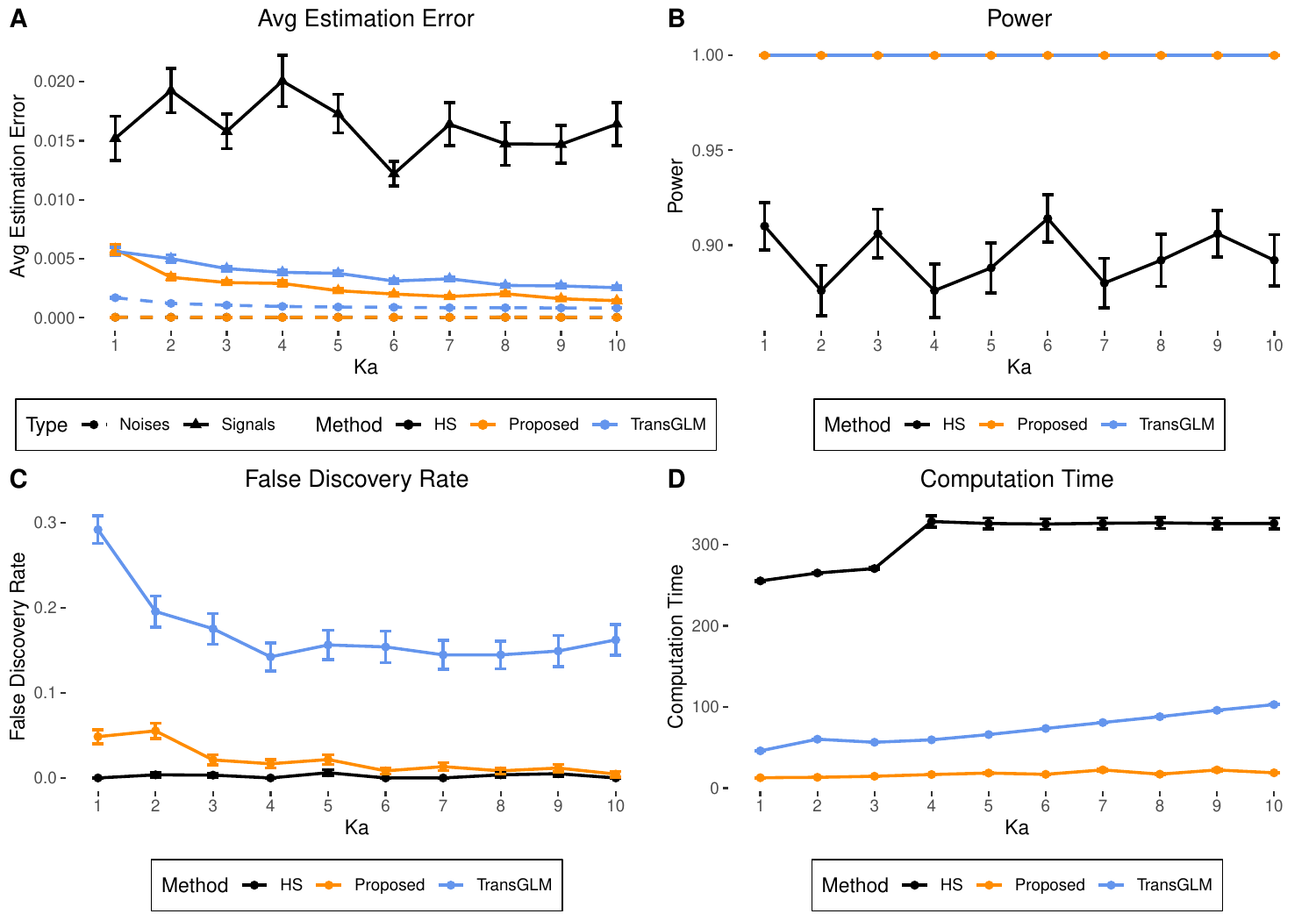}
        \caption{Average estimation error, power, false discovery rate (FDR), and computational time of proposed TRADER (orange), TransGLM (blue; implemented with 20 additional CPU cores per simulation), and target-only HS (black) over 100 simulations under Simulation Setting I. $K = 10$ sources, with a varying number $K_a$ of informative sources; $h = 5$, $n = 200$, $n_k = 200$ for all $k = 1, \ldots, 10$, $p = 500$, $s = 5$. Results are separated by true noise (dots and dashed line) and signal (triangles and solid line), with error bars denoting the standard deviations.}
	\label{fig:stratified-covtype2}
\end{figure}

Figure~\ref{fig:stratified-covtype2}A shows that TRADER consistently achieves lower average estimation errors for signal variables than the target-only HS approach, even when source data are uninformative. In panel B, TRADER attains statistical power comparable to TransGLM, while HS lags noticeably behind both methods. Panel C reveals that TRADER controls the FDR more effectively than TransGLM. Taken together, these results indicate that TRADER makes efficient use of informative source data to compensate for the limited target sample size, delivering estimation accuracy for signals similar to TransGLM, despite TransGLM's access to individual-level source data. Moreover, Figure~\ref{fig:stratified-covtype2}D highlights TRADER's clear advantage in computational efficiency over TransGLM, even though the latter employs 20 additional CPUs per simulation. Further discussions on run times appear in Section~\ref{sec:high-d-sim}.

As a sensitivity analysis, we varied the degree of heterogeneity $h$ between source and target. Figures~\ref{S-fig:simu-setting-I-h-10} and \ref{S-fig:simu-setting-I-h-15} show that TRADER maintains estimation errors for signals comparable to TransGLM across increasing $h$, while all methods control noise variable errors effectively. TRADER continues to match TransGLM in power, yet consistently achieves a lower FDR. Overall, the sensitivity analysis confirms TRADER's robustness to target-source heterogeneity, exhibiting comparable estimation accuracy and power to TransGLM together with improved FDR control.

\subsection{Setting II: Sources with varying correlations and scales}
In Setting II, we test TRADER's robustness when sources vary in scale and correlation with the target. This setup captures different levels of source usefulness and scale differences. We explore two cases: (1) sources with strong or weak correlations to the target but consistent scales, and (2) sources with mismatched scales but constant correlations.

The target dataset consists of $n=100$ samples and $p=200$ parameters. For the source datasets, we consider $  K \in \{1,\ldots,10\}  $ sources, each with the same number of parameters as the target and $  n_k=100  $ samples per source. Covariates for both source and target datasets are generated in the same approach as Simulation Setting I.

We then generate the target and source parameters as follows. A random support set $  S \subset \{1,\ldots,p\}  $ is selected uniformly at random from all subsets of cardinality $  |S| = s  $. For each $  j \in S  $, we independently sample

\[
\resizebox{.8\hsize}{!}{$
(\beta_j,\omega_j^{(1)}, \ldots, \omega_j^{(K)})^\top \overset{\mathrm{iid}}{\sim} \mathcal{N}\left(\mathbf{0}_{K+1}, \begin{pmatrix}
\alpha_\text{tgt}^2 & \rho_1\alpha_\text{tgt}\alpha_{\text{src,}1} & \cdots & \rho_K\alpha_\text{tgt}\alpha_{\text{src},K} \\
\rho_1\alpha_\text{tgt}\alpha_{\text{src},1} & \alpha_{\text{src},1}^2 & \cdots & 0 \\
\vdots & \vdots & \ddots & \vdots \\
\rho_K\alpha_\text{tgt}\alpha_{\text{src},K} & 0 & \cdots & \alpha_{\text{src,}K}^2
\end{pmatrix}\right)$},
\vspace{1mm}
\]
where $  \alpha_\text{tgt}^2 = \mathbb{E}\|\bbeta\|^2  $, $  \alpha_{\text{src,}k}^2 = \mathbb{E}\|\bomega^{(k)}\|^2  $, and $  \rho_k  $ represents the correlation between $  \bbeta  $ and $  \bomega^{(k)}  $. For all $  j \notin S  $, we set $  \beta_j = 0  $ and $  \omega_j^{(k)} = 0  $ for $  k = 1,\ldots,K  $. We set $s=5$ across all experiments in Simulation Setting II.

\begin{figure}[ht]
	\centering
    \includegraphics[width=0.8\textwidth]{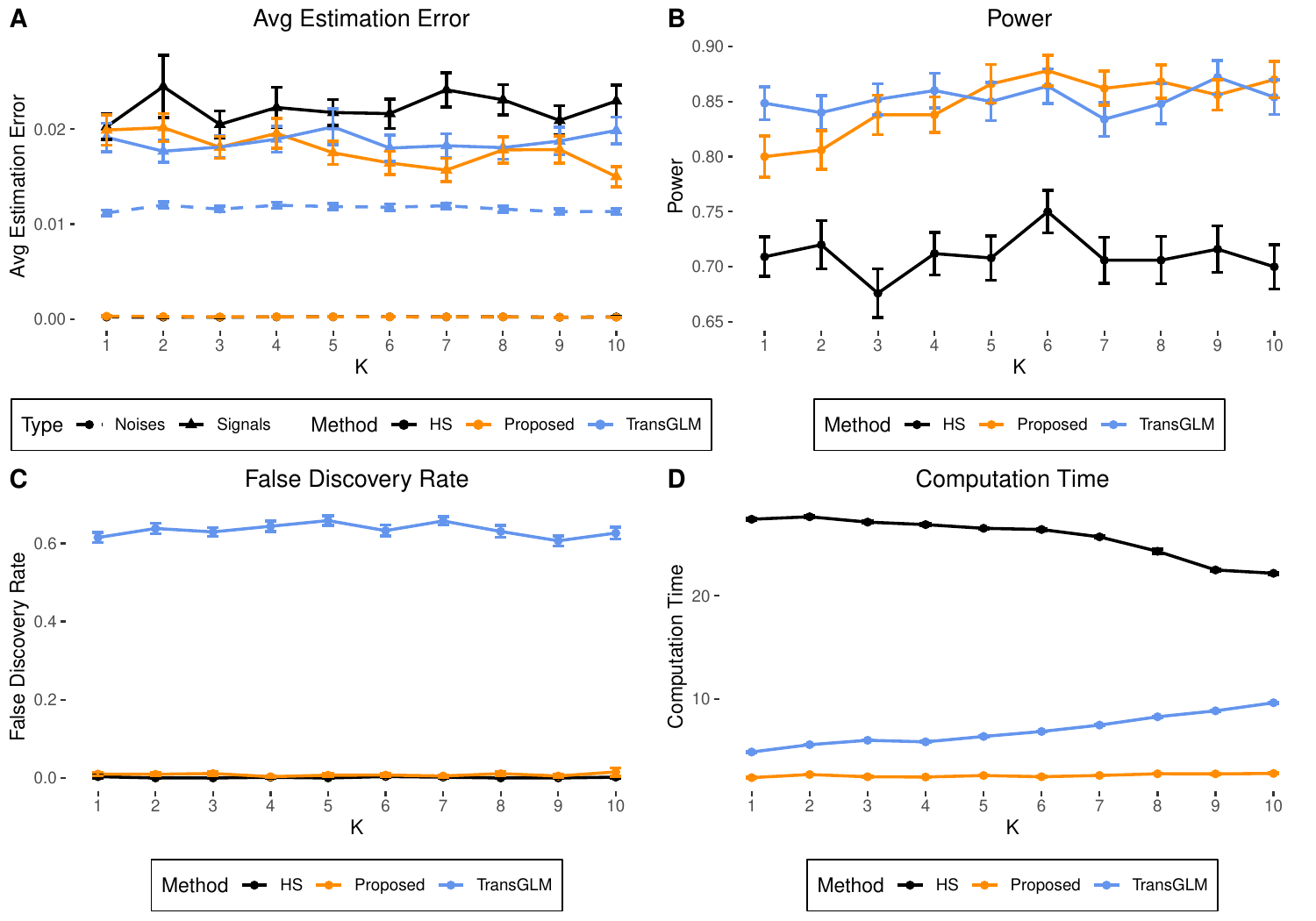}
	\caption{Average estimation error, power, false discovery rate (FDR), and computational time of proposed TRADER (orange), TransGLM (blue; implemented with 20 additional CPU cores per simulation), and target-only HS (black) over 100 simulations under Simulation setting II. $K = 1,\cdots,10$, $p = 200$, fixed scale ratios $\alpha_\text{tgt}/\alpha_{\text{src},k} = 1
    $ with fixed correlation at $\rho_k = 0.8$ for $k = 1, \ldots, 10$, with error bars denoting the standard deviations.}
	\label{fig:angle-simu}
\end{figure}

We first examine sources with varying correlations but fixed scale ratios of 1. As shown in Figure~\ref{fig:angle-simu}, TRADER achieves average estimation errors below HS levels for both signals and noise at $\rho_k=0.8$; TransGLM exhibits slightly higher errors than TRADER with comparable power, but a markedly larger FDR than both HS and TRADER. Even at weaker correlations like $\rho_k=0.3$, Figure~\ref{S-fig:additional-setting-II-rho-0.3-alpha-1} highlights substantial improvements in estimation accuracy and power over the target-only HS baseline, demonstrating TRADER's capacity to prevent negative transfer and exceed baseline performance.

We next fix correlations at $\rho_k=0.8$ and vary scale ratios $\alpha_{\text{tgt}}/\alpha_{\text{src},k}$ to assess performance under heterogeneous source estimator lengths. As shown in Figure~\ref{S-fig:additional-setting-II-rho-0.8-alpha-2}, TRADER yields estimation errors comparable to or slightly below those of TransGLM and HS, while attaining similar power to TransGLM and markedly superior FDR control.

Overall, results from Setting II affirm TRADER's robustness across multiple sources, effectively mitigating negative transfer and surpassing competitors even amid weak correlations or scale mismatches. TransGLM, by contrast, consistently struggles with elevated FDR and higher computational costs, despite additional CPUs.

\subsection{Setting III: A higher-dimensional scenario}
\label{sec:high-d-sim}
To evaluate the scalability of TRADER, we consider a challenging high-dimensional setting with $  p = 2,000  $, target sample size $  n = 100  $, and source sample sizes $  n_k = 500  $ for $  k = 1, \ldots, 10$.This setup mimics real-world scenarios where covariates greatly exceed the number of observations, complicating inference and computation. We set $  s = 20  $, $  h = 10  $, and retain other parameters from Setting I.

\begin{figure}[ht]
    \centering
    \includegraphics[width=0.8\textwidth]{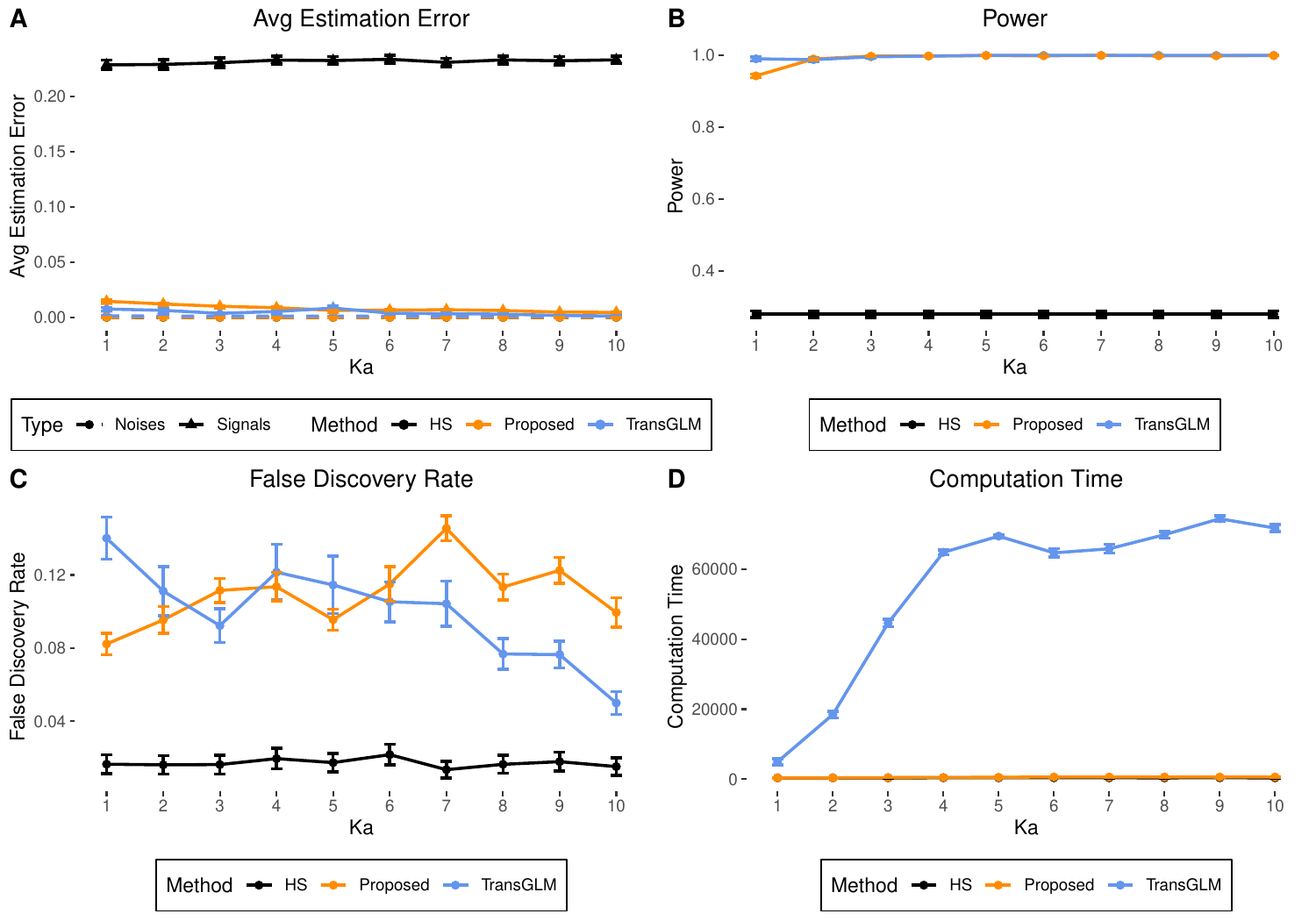}
    \caption{Average estimation error, power, false discovery rate (FDR), and computational time of proposed TRADER (orange), TransGLM (blue; implemented with 20 additional CPU cores per simulation), and target-only HS (black) over 100 simulations under Simulation setting III. Higher dimension of $p = 2,000$, with $K = 10$ sources and a varying number $K_a$ of informative sources; $h = 10$, $n = 100$, $n_k = 500$ for all $k = 1, \ldots, 10$, and $s=20$ with error bars denoting the standard deviations.}
    \label{fig:setting-III}
\end{figure}

Figure \ref{fig:setting-III} shows that TRADER achieves estimation performance comparable to TransGLM with substantial computational gains. Panel A indicates similar average estimation errors for signals between the two transfer learning methods, with TRADER yielding lower errors for noise; both outperform HS on signals due to limited target data.

Panels B--C assess signal detection. TRADER attains power comparable to TransGLM (Panel B), while HS shows lower power. TRADER's FDR is slightly higher than TransGLM's (Panel C), possibly due to its reliance on summary statistics rather than individual-level source data, yet it remains controlled. HS exhibits the lowest FDR with reduced power.

Panel D highlights TRADER's computational advantages, which are also seen in the previous two simulations. Despite using only a single core, TRADER runs significantly faster than TransGLM, which uses 20 additional CPU cores per simulation. This discrepancy is expected, as TransGLM relies on the desparsified LASSO framework \citep{van2014asymptotically}, which requires solving $p$ penalized regressions, each involving $p - 1$ predictors, to approximate the inverse covariance matrix. The LARS algorithm \citep{efron2004least} provides an efficient solver with a complexity of $O(p^3)$ per column. Therefore, the total complexity of computing the approximate inverse covariance matrix is $O(p^4)$. In contrast, TRADER uses an efficient MCMC sampler with per-iteration cost $O(p^2)$, yielding total complexity $O(n_\text{iter} p^2)$ (see Supplementary Section \ref{S-supp:mcmc}). Because TRADER scales quadratically in $p$ per iteration, whereas TransGLM scales quartically in $p$, TRADER is substantially more scalable in high-dimensional regimes. This demonstrates TRADER's (and Bayesian approaches more broadly) scalability for inference with uncertainty quantification in high dimensions.

Supplementary Section~\ref{S-supp:ModelH} compares TRADER with a hierarchical model that uses individual-level source data. The results, presented in Figures~\ref{S-fig:AE-cov-type2} and~\ref{S-fig:AE-cov-type1}, indicate that TRADER substantially outperforms the hierarchical model when uninformative sources are present. When all sources are informative, TRADER achieves comparable performance while incurring significantly lower computational cost.

\section{Real-data application}\label{section-real-data}
We apply TRADER to the Genotype-Tissue Expression (GTEx) dataset (\url{https://gtexportal.org/}), which contains gene expression profiles from 49 tissues across 838 human donors, comprising over 1.2 million observations on 38,187 genes. Our objective is to identify causal variants that genuinely influence gene expression levels. Here, the term ``causal variants" represents variants that have non-zero effects on gene expression.

\subsection{Credible level and credible set in fine-mapping}
\label{explanation-finemapping}
This task is called \textit{fine-mapping} in genetic studies. It aims to identify causal variants despite high correlation among variants due to LD (that is, multicollinearity in the genotype matrix) \citep{wang2020susie,li2024bayesian}. Standard regression methods struggle to distinguish truly causal variants from highly correlated ones. These methods may also fail to provide valid uncertainty quantification. Additionally, discarding correlated variants to ensure identifiability, though common, risks eliminating true signals.

Frequentist approaches are generally unsuitable for fine-mapping tasks, as they struggle with the ambiguity introduced by highly correlated variants. To illustrate, consider the example from \cite{wang2020susie}: suppose $\xb_1$ and $\xb_4$ are true causal variants, and each is perfectly correlated with a noise variable—specifically, $\xb_1 = \xb_2$ and $\xb_3 = \xb_4$, with no other correlations. In this case, it is impossible to confidently select the correct effect variables, even with large sample sizes. Instead of selecting individual variants, \cite{wang2020susie} propose making the following inferential statement:
\begin{equation}
\label{eq:finemapping}
\beta_1 = 0 \quad \text{or} \quad \beta_2 = 0 \quad \text{and} \quad \beta_3 = 0 \quad \text{or} \quad \beta_4 = 0.
\end{equation}
This statement narrows associations to small sets of correlated variants without pinpointing individuals, acknowledging ambiguity to avoid erroneous selections. In fine-mapping, such ambiguity is critical, as scientific conclusions may depend on arbitrary choices. This explains why most penalized regression methods are not preferred in fine-mapping: \textit{they neither generate nor aim to produce statements like \eqref{eq:finemapping}}. Instead, they typically select a single ``optimal'' variable set, ignoring other plausible combinations. In the example above, the elastic net would select all four variables, including the noise variables $\xb_2$ and $\xb_3$, without acknowledging alternative plausible configurations.

Bayesian approaches are better equipped for this task. They naturally yield statements like \eqref{eq:finemapping} by constructing credible sets (CSs): a group of variants that together have high posterior probability of containing a true signal. For example, a 95\% CS includes at least one causal variant with at least 95\% posterior probability. \cite{wang2020susie,li2024bayesian} have shown that such a set can comprise multiple highly correlated potential causal variants, effectively addressing the ambiguity in the previous example. Accordingly, our goal is to report as many concise CSs as supported by the data.

We adapt the framework of \cite{li2024bayesian} to construct the CSs under the continuous shrinkage prior. Their approach begins by defining CLs for individual $\beta_j$ as in Equation~\eqref{eq:CL} and then extends the concept to groups of $\beta_j$ to form CSs. For a group of variants $\mathcal{C} = \{j_1, \ldots, j_k\}$, we test the joint null $H_{0\mathcal{C}}: \bbeta_\mathcal{C} = 0$. Following \cite{liu2020comparison}, we project $\bbeta_\mathcal{C}$ onto the leading principal components of the LD matrix (i.e., the correlation matrix of $\Xb^{(0)}$). Let $\ub_1$ be the eigenvector of the largest eigenvalue. The CL for set $\mathcal{C}$ is defined as:
\begin{equation}
\mathrm{CL}_\mathcal{C} := \left|\, \widehat{\mathbb{P}}(\ub_1^\top \bbeta_\mathcal{C} > 0 \mid \mathcal{D}) - \widehat{\mathbb{P}}(\ub_1^\top \bbeta_\mathcal{C} < 0 \mid \mathcal{D}) \,\right|.
\end{equation}
If $\mathrm{CL}_\mathcal{C} \geq 1 - \alpha$, we reject $H_{0\mathcal{C}}$ at level $\alpha$ and designate $\mathcal{C}$ as a level $1-\alpha$ CS. A greedy search algorithm is used to enumerate all CSs that reach the pre-specified level.

\subsection{Application to GTEx dataset}
\label{sec:applicationtogtex}
We apply TRADER to identify causal variants for gene \textit{FADS3}, which encodes an enzyme involved in long-chain polyunsaturated fatty acid (PUFA) metabolism. \cite{zhou2020unified} has shown that borrowing information across tissues can enhance understanding of \textit{FADS3} expression. The GTEx dataset contains varying sample sizes across tissues (Figure~\ref{S-fig:gtexsamplesize}); for example, liver has only 208 samples, limiting power for causal variant discovery \citep{wainberg2019opportunities}. Since \textit{FADS3} is expressed in multiple tissues, we aim to integrate multi-tissue data to improve discovery in liver tissue. 

Following preprocessing from \cite{lai2025transfertwas} (details in Section~\ref{S-sec:realdata}), we retain $p = 2,554$ variants. To preserve potential causal variants, we do not remove highly correlated variants, resulting in a more realistic but challenging setting than standard approaches. We compare TRADER and the target-only HS. TransGLM is excluded because its inference relies on desparsified LASSO, which is known to be unstable under extreme multicollinearity (see Section \ref{explanation-finemapping}). The target-only HS is applied to the $n=208$ target liver data, while TRADER also leverages information from $K=48$ source tissues.

\begin{figure}[ht]
	\centering
	\includegraphics[width=0.8\textwidth]{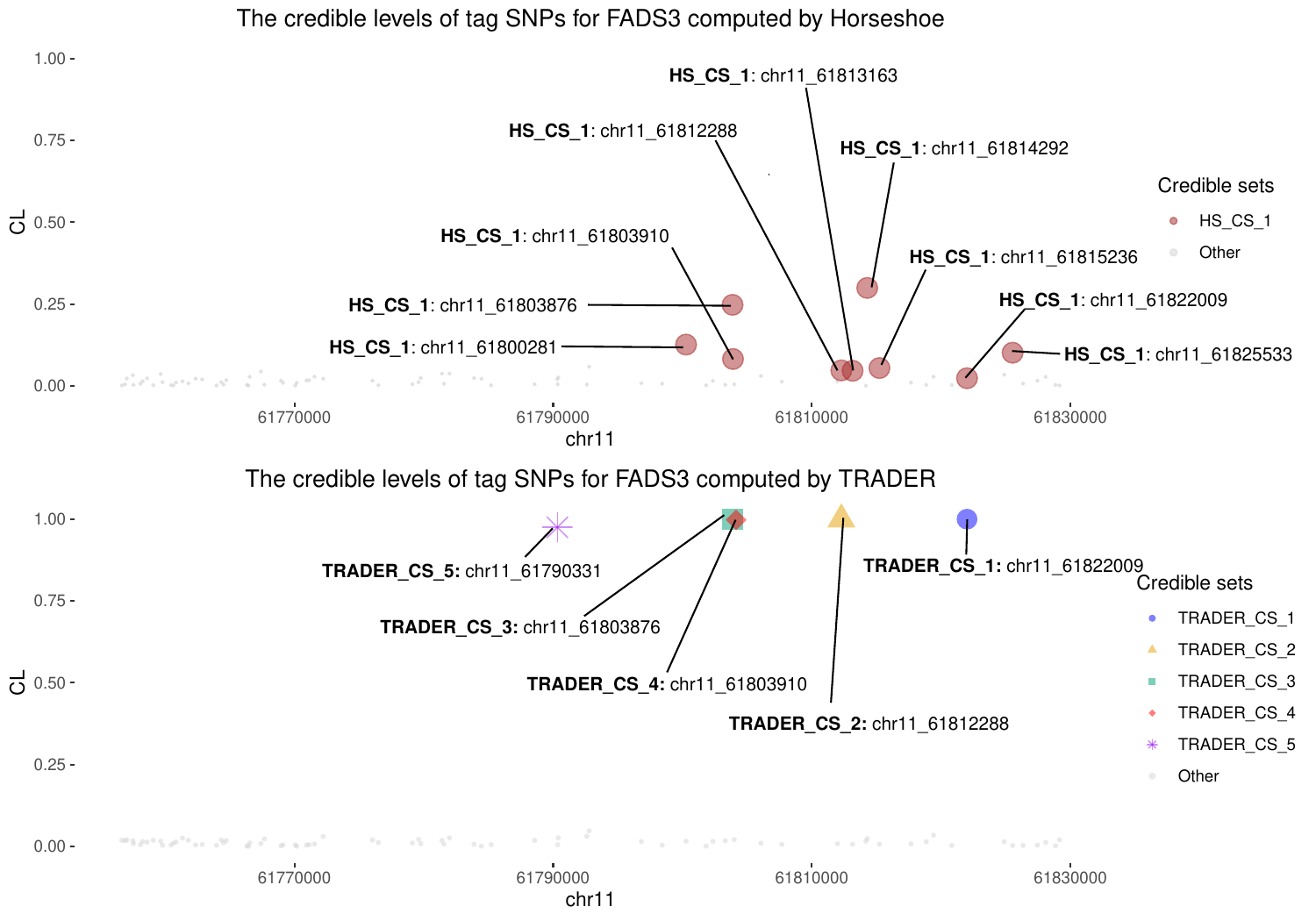}
	\caption{Zoomed-in view of the fine-mapping results for \textit{FADS3}; the full region is shown in the Supplementary Section \ref{S-supp:finemapping-example}. Semitransparent points represent genetic variants not included in any CS. The first and second panels display the CLs of tag genetic variants computed using the standard horseshoe prior and TRADER, respectively. In each panel, distinct colors denote different 95\% CSs. Each CS is labeled in the format \{\textbf{method name}\}\_\textbf{CS\_}\{\textbf{index}\}:\ \text{chr\_}\{\text{chromosome ID}\}\_\{\text{chromosome position}\}. }
	\label{fig:FADS3-zoom-in}
\end{figure}

Fine-mapping results are shown in Figures~\ref{fig:FADS3-zoom-in} and \ref{S-fig:FADS3}. HS identifies one CS of eight variants with a CL of 0.9510, suggesting at least one variant in the set regulates \textit{FADS3}'s expression. In contrast, TRADER identifies five CSs with CLs exceeding 95\%, each containing a single variant. We rank CSs by their CLs. The top CS (CL = 0.9996) contains the variant chr11:61822009 (rs28456), previously reported as a causal variant regulating the expression of the \textit{FADS} gene family \citep{tabassum2019genetic}. The third CS (CL = 0.9988) includes chr11:61803876 (rs174548), a variant strongly associated with reduced \textit{FADS3} expression and increased accumulation of pro-inflammatory eicosanoids due to delta-5 desaturase activity \citep{fadason2018chromatin}. The fourth CS (CL = 0.9976) consists of chr11:61803910 (rs174549), regulates gene expression at the \textit{FADS1-2-3} locus in liver tissue \citep{wang2020cross}, possibly explaining the similar CLs assigned by TRADER in Figure~\ref{fig:FADS3-zoom-in}. The final CS (CL = 0.9748) contains chr11:61790331 (rs102275), a variant previously shown to predict \textit{FADS3} expression in liver tissue \citep{kathiresan2009common}; notably, this variant is not identified by the target-only HS, underscoring the benefit of incorporating gene expression data and TRADER’s capacity to identify novel causal variants. Additionally, TRADER identifies a novel candidate causal variant in its second CS, chr11:61812288 (rs174555), recently found associated with dihomo-gamma-linolenic acid levels, a precursor to arachidonic acid and a long-chain omega-6 PUFA \citep{mathias2011impact}. Since \textit{FADS3} contributes to long-chain PUFA biosynthesis, these two findings suggest that rs174555 may be a novel causal variant for \textit{FADS3}.

In conclusion, TRADER provides higher resolution in fine-mapping by identifying five distinct CSs, each containing a single variant, compared to HS, which yields one CS with eight variants. Notably, the top four variants identified by TRADER are included within the broader CS of the target-only HS, reflecting strong concordance between the methods. These findings show TRADER's effectiveness in enhancing fine-mapping resolution through efficient cross-tissue information borrowing.

\section{Discussion}\label{section-discussion}
In this paper, we introduce TRADER, a Bayesian transfer learning method that adjusts the prior mean of the HS prior from zero to an adaptively weighted average of source estimates. This approach extends existing Bayesian transfer learning methods to high-dimensional linear regression model and only requires summary-level source information. Simulations show that TRADER outperforms target-only HS and achieves estimation and inference performance comparable to methods that require individual-level data, while remaining robust when sources differ in scale or correlation from the target. Additionally, TRADER offers substantial computational advantages over frequentist alternatives, particularly in high dimensions. A real-data application to identify genetic variants regulating gene expression demonstrates TRADER's effectiveness and shows a practical scenario where frequentist approaches struggle, highlighting the unique advantages of the Bayesian framework.

One of the key features of TRADER is using source estimates to define prior means in Bayesian models. It formalizes a flexible and generalizable framework for transfer learning in high dimensions. By centering the prior at an adaptively weighted average of source estimates, the approach enables robust and scalable borrowing across datasets and offers a foundation for devising new Bayesian methods tailored to various transfer learning tasks. 

While incorporating the uncertainty of source estimates could, in principle, improve the performance of TRADER, we note that in high-dimensional regimes, reliable estimation of standard errors or posterior variances is generally difficult and may introduce further instability. Moreover, modeling precision explicitly would substantially increase the computational burden. A potential extension is to explore heuristic or data-driven filtering strategies that down-weight poorly estimated sources or components prior to integration. Future work could also adapt TRADER to handle mismatched covariates between source and target datasets, a common challenge in multi-site studies where each site may collect different sets of variables. 

\begin{spacing}{1}
\section*{Supplementary Materials}
The Supplementary Materials contain technical proofs of all main theorems, together with additional discussions, numerical experiments, theoretical results, and further details of the real-data applications.

\section*{Acknowledgments}
We thank Drs. Yuxi Cai, Jeremy MG Taylor, Jin Jin, Huaqing Jin, and Chenyang Zhang for their insightful discussions and valuable feedback during the development of this paper. We are grateful to the Editor, Professor Annie Qu, the Associate Editor, and three anonymous reviewers for their constructive comments and valuable suggestions, which substantially improved the manuscript.

\section*{Disclosure Statement}
No potential conflict of interest was reported by the author(s).

\section*{Funding}
This study was funded by NIH R01CA296289.



\bibliography{ref}
\end{spacing}
\end{document}


\maketitle

\setcounter{equation}{0}
\setcounter{figure}{0}
\setcounter{table}{0}
\setcounter{page}{1}
\setcounter{section}{0}

\renewcommand{\theequation}{S\arabic{equation}}
\renewcommand{\thefigure}{S\arabic{figure}}
\renewcommand{\thetable}{S\arabic{table}}
\renewcommand{\thesection}{S\arabic{section}}
\renewcommand{\thetheorem}{S.\arabic{theorem}}
\renewcommand{\bibnumfmt}[1]{[S#1]}
\renewcommand{\citenumfont}[1]{S#1}

\tableofcontents

\newpage

\section{Proof of Proposition~\ref{prop_prior}}
\label{S-supp:prop1}
\begin{proof}
To prove (\ref{eqn:prior1}), simply notice that 
\[
    \begin{array}{lll}
         \mathbb{E}_{ \theta_1,..,\theta_K}(\beta_j)  &=&  \mathbb{E}_{ \theta_1,..,\theta_K}(  \mathbb{E}_{ \theta_1,..,\theta_K}(\beta_j \,|\, \sigma^2,\lambda_j^2, \tau^2   ) )  \\
         & =  &\mathbb{E}_{ \theta_1,..,\theta_K}( \sum_{k=1}^{K} \eta_k \text{\small$\frac{\norm{\widehat{\bbeta}_0}}{\norm{\widehat{\bomega}^{(k)}}}$} \widehat{\omega}^{(k)}_j )\\
            &=& \sum_{k=1}^{K} \text{\small$\frac{\norm{\widehat{\bbeta}_0}}{\norm{\widehat{\bomega}^{(k)}}}$} \widehat{\omega}^{(k)}_j\mathbb{E}_{ \theta_1,..,\theta_K}(\eta_k   )
    \end{array}
\]
and the claim follows by the properties of the Dirichlet distribution, which proves (\ref{eqn:prior1}).
    
\end{proof}
\newpage
\section{MCMC algorithm for TRADER}
\label{S-supp:mcmc}

We developed a specialized MCMC algorithm to enhance computational efficiency, particularly for large $p$. This design ensures that TRADER remains scalable, effectively accommodating scenarios where $p$ reaches several thousand or more. We drop the superscript of $\Xb^{(0)}$ and $\yb^{(0)}$ and denote it as $\Xb$ and $\yb$ for simplicity.

To clarify notation, let $\blambda = (\lambda_1, \ldots, \lambda_p)$, $\bmeta = (\eta_1, \ldots, \eta_K)$, and $\bOmega = (\bomega_1, \ldots, \bomega_K)$. The joint posterior distribution of TRADER can then be expressed as:
\begin{equation*}
\begin{aligned}
\pi(\bbeta, \bmeta, \blambda, \tau, \sigma^2 \mid \yb, \Xb, \bOmega) 
&\propto (\sigma^2)^{-\frac{n}{2}} \exp\left[ -\frac{1}{2\sigma^2} (\yb - \Xb\bbeta)^{\top} (\yb - \Xb\bbeta) \right] \\
&\quad \times \prod_{j=1}^p \left( \sigma^2 \lambda_j^2 \tau^2 \right)^{-\frac{1}{2}}
\exp\left[ -\frac{ \left( \beta_j - \sum_{k=1}^K \eta_k \omega_{kj} \right)^2 }{ 2\sigma^2 \lambda_j^2 \tau^2 } \right] \\
&\quad \times \prod_{k=1}^K \eta_k^{\theta_k-1} \left( 1 - \sum_{k=1}^K \eta_k \right)^{\zeta-1} \\
&\quad \times \prod_{j=1}^p \frac{1}{1 + \lambda_j^2} \\
&\quad \times (\sigma^2)^{-1}.
\end{aligned}
\end{equation*}

Our MCMC sampling algorithm proceeds as follows. For all parameter updates except $\beta_j$, we employ the slice sampling method of \citet{neal2003slice} (see also \cite{polson2011thebayesian} for applications with HS priors). We consider two strategies for selecting $\tau$: a fixed approach, in which $\tau$ is specified a priori by the user, and a hyperprior approach, in which $\tau$ is assigned a truncated $\mathcal{C}^+(0,1)$ prior on $(1/p, 1)$. Both approaches are based on \cite{vanderpas2017uncertainty}.

 \begin{enumerate}
    \item \textbf{Initialization:} Set initial values for $\blambda$, $\bbeta$, $\tau$, $\bmeta$, and $\sigma^2$.

    \item  \textbf{Update each} $\beta_j$ $(j=1,\ldots,p)$: The marginal posterior is $\beta_j$
    \begin{equation*}
        \beta_j \mid  \bbeta_{-j},\bmeta,\lambda_j, \sigma^2,\Xb, \yb, \bOmega
        \sim \mathcal{N}\left( \widetilde{\sigma}_j^2 \widetilde{\mu}_j,\; \widetilde{\sigma}_j^2 \right)
    \end{equation*}
    where
    $$
    \begin{aligned}
        \widetilde{\sigma}_j^2 &= \sigma^2\left( (\Xb^\top\Xb)_{jj} + \frac{1}{\lambda_j^2\tau^2} \right),\\
        \widetilde{\mu}_j &= \frac{1}{\sigma^2} \left[ (\Xb^\top\yb)_j - \sum_{l \neq j} (\Xb^\top\Xb)_{lj} \beta_l + \sum_{k=1}^K \frac{ \eta_k \omega_{kj} }{ \lambda_j^2 \tau^2 } \right].
    \end{aligned}
    $$

    \item \textbf{Update each} $\lambda_j$ $(j = 1,\ldots,p)$: The conditional posterior for $\lambda_j$ is
    \begin{equation*}
        \begin{aligned}
            \pi\bigl(\lambda_j \mid \blambda_{-j}, \bbeta_{-j}, \bmeta, \tau, \sigma^2,\Xb, \yb, \bOmega \bigr) 
            \propto\; & \exp \left( -\frac{\left(\beta_j - \sum_{k=1}^K \eta_k \omega_{kj}\right)^2}{2\sigma^2 \lambda_j^2 \tau^2} \right) \\
            & \times \frac{1}{\lambda_j (1 + \lambda_j^2)} \times \mathbb{I}\bigl[\lambda_j > 0\bigr]
        \end{aligned}
    \end{equation*}

    Let $v_j=\lambda_j^{-2}$. To sample $v_j$, perform the following two steps:
    \begin{enumerate}
        \item Sample $u \mid v_j \sim \mathcal{U}\left(0, \dfrac{1}{1+v_j}\right)$
        \item Sample $v_j \mid u \sim \mathrm{Exp}\left( \dfrac{\left(\beta_j - \sum_{k=1}^K \eta_k \omega_{kj}\right)^2}{2\sigma^2\tau^2} \right) \cdot \mathbb{I}\left[ 0 < v_j < \dfrac{1-u}{u} \right]$
    \end{enumerate}

    \item \textbf{Update each} $\eta_k$ $(k=1,\ldots,K)$: The conditional posterior for $\eta_k$ can be written as
    \begin{equation*}
        \begin{aligned}
            \rho(\eta_{k}\mid\bmeta_{-k},\blambda, \bbeta, \tau, \sigma^2,\Xb, \yb, \bOmega) 
            & \propto \exp\left\{ -\sum_{j=1}^{p}\frac{\left(\beta_{j} - \sum_{i=1}^{K}\eta_{i}\omega_{ij}\right)^{2}}{2\sigma^{2}\lambda_{j}^{2}\tau^{2}} \right\} \\
            & \quad \times \eta_{k}^{\theta_{k}-1} \left(1-\sum_{i=1}^{K}\eta_{i}\right)^{\zeta-1}\\
            & \propto \exp\left\{ -\frac{1}{2\sigma^{2}\tau^{2}}\left[ a_k \eta_{k}^{2} + 2 b_k \eta_{k} \right] \right\} \\
            & \quad \times \eta_{k}^{\theta_{k}-1} \left(1-\sum_{i=1}^{K}\eta_{i}\right)^{\zeta-1}
        \end{aligned}
    \end{equation*}
    where
    $$
    \begin{aligned}
        a_{k} &= \sum_{j=1}^{p}\frac{\omega_{kj}^{2}}{\lambda_{j}^{2}},\\
        b_{k} &= \sum_{j=1}^{p}\frac{\omega_{kj} \left(\sum_{l \neq k}\eta_l\omega_{lj} - \beta_{j}\right)}{\lambda_{j}^{2}}.
    \end{aligned}
    $$
    The conditional posterior for $\eta_k$ is then
    \begin{equation*}
        \pi(\eta_{k}\mid \bmeta_{-k},\blambda, \bbeta, \tau, \sigma^2,\Xb, \yb, \bOmega) \propto \exp\left\{ -\frac{a_{k}(\eta_{k} + b_{k})^{2}}{2\sigma^{2}\tau^{2}} \right\} \times \eta_{k}^{\theta_{k}-1} \left(1-\sum_{i=1}^{K}\eta_{i}\right)^{\zeta-1}.
    \end{equation*}
    We again take the slice sampling procedure to sample $\eta_k$:
    \begin{enumerate}
        \item Sample $u \sim \mathcal{U}\left(0,\, \exp\left\{ -\frac{a_{k}(\eta_{k} + b_{k})^{2}}{2\sigma^{2}\tau^{2}} \right\}\right)$
        \item Sample $\delta_{k} \sim \mathrm{Beta}(\theta_{k}, \zeta)\, \mathbb{I}\left[\kappa_L, \kappa_U\right]$, where $\delta_k$ is an auxiliary variable with $ \delta_{k} =\eta_{k}/(1 - \sum_{l \neq k} \eta_l)$, and 
        \begin{equation*}
            \begin{aligned}
                \kappa_L &= \max\left\{
                    \frac{ -\sqrt{ -\dfrac{2\sigma^{2}\tau^{2} \ln u}{a_{k}} } - b_{k} }{ 1 - \sum_{l \neq k} \eta_l },\; 0
                \right\},\\
                \kappa_U &= \min\left\{
                    \frac{ \sqrt{ -\dfrac{2\sigma^{2}\tau^{2} \ln u}{a_{k}} } - b_{k} }{ 1 - \sum_{l \neq k} \eta_l },\; 1
                \right\}.\\
            \end{aligned}
        \end{equation*}

    \end{enumerate}

    \item \textbf{Update} $\tau$: The conditional posterior for $\tau$ is
    $$
        \pi(\tau \mid \bmeta,\blambda, \bbeta, \sigma^2,\Xb, \yb, \bOmega) \propto \tau^{-p} \exp\left\{ -\sum_{j=1}^{p} \frac{(\beta_{j} - \sum_{k=1}^K \eta_k \omega_{kj})^{2}}{2\sigma^{2}\lambda_j^{2}\tau^{2}} \right\}
        \cdot \frac{1}{1 + \tau^{2}}\, \mathbb{I}\left[a, b\right].
    $$
    
    Letting $\phi = 1/\tau^2$, this becomes
    $$
    \begin{aligned}
        \pi(\phi \mid  \bmeta,\blambda, \bbeta, \sigma^2,\Xb, \yb, \bOmega) \propto& \phi^{\frac{p-1}{2}} \exp\left\{ -\sum_{j=1}^{p} \frac{(\beta_{j} - \sum_{k=1}^K \eta_k \omega_{kj})^{2}}{2\sigma^{2}\lambda_{j}^{2}} \phi \right\}\\
        &\times \frac{1}{1 + \phi}\, \mathbb{I}\left[ \phi \in \left( \frac{1}{b^{2}},\, \frac{1}{a^{2}} \right) \right].
    \end{aligned}
    $$
    
    The following slice sampling scheme was proposed:
    \begin{enumerate}
        \item Sample $u\mid\phi \sim \mathcal{U}\left(0, \frac{1}{1+\phi}\right)$,
        \item Sample $\phi\mid u \sim \Gamma\left(\frac{p+1}{2}, \sum_{j=1}^p\frac{\left(\beta_j-\sum_{k=1}^K \eta_k \omega_{kj}\right)^2}{2\sigma^2\lambda_j^2}\right) \cdot \mathbb{I}\left[\frac{1}{b^2}, \min\left\{\frac{1}{a^2}, \frac{1-u}{u}\right\}\right]$.
    \end{enumerate}
\end{enumerate}

\newpage
\section{Proof of Theorem~\ref{thm1}}
\begin{proof}
For simplicity, we drop the superscript of $\Xb^{(0)}$ and $\yb^{(0)}$ and denote it as $\Xb$ and $\yb$. The proof closely follows Theorem A.5 from \cite{song2023nearly}. We  notice that the true model is equivalent to 
\[
\widetilde{\yb} := \yb - \Xb\widetilde{\bomega} \,\sim \, \mathcal{N}(\Xb \bb^*, \sigma^{*2} \Ib),
\]
where $\bb^* =  \bbeta^* - \widetilde{\bomega} $.    Also, writing $\bb =  \bbeta - \widetilde{\bomega}$, the Bayesian model is equivalent to

\[
\begin{array}{lll}
	\widetilde{\yb} &\sim& \mathcal{N}(\Xb \bb, \sigma^2 \Ib)        \\
	b_j   &\sim  & \mathcal{N}( 0,  \sigma^2 \lambda_j^2  \tau^2  )\\
	\lambda_j &\sim & \text{Cauchy}^{+}(0,1)\\
	\sigma^2 &\sim & \text{Inverse-Gamma}(\nu,\nu).
\end{array}
\]
Next, let $\widetilde{p}$ such that $\widetilde{p}\asymp \widetilde{s} = \widetilde{s}(r_{\sup})$,  $a_n = M \sqrt{\widetilde{s}\log p /n }/p$, 
and 
\[
\begin{array}{lll}
	A_n& =& \{ \text{at least} \,\,\widetilde{p}\, \text{entries of} \, \,\vert \bb/\sigma\vert  \,\,\text{are larger than } \,\,a_n \} \cup  \\
	& &  \{  \|\bb- \bb^*\|    \geq  (7/2 + \sqrt{\lambda_0}) \sigma^*\epsilon_n\} \cup  \{     \frac{\sigma^2}{ \sigma^{*2} } > \frac{1+\epsilon_n}{1- \epsilon_n }  \,\,\text{or} \frac{\sigma^2}{ \sigma^{*2} } < \frac{1-\epsilon_n}{1+ \epsilon_n } 
	\,\,\} \\
	B_n& =& \{ \text{at least} \,\,\widetilde{p}\, \text{entries of} \, \,\vert \bb/\sigma\vert  \,\,\text{are larger than } \,\,a_n \}\\
	
\end{array}
\]
and $C_n   =   A_n \backslash B_n $. We also write
\[
\widetilde{\xi}  = \{ j\,:\, \vert   \beta_j^* - \widetilde{\omega}_j \vert > r_\text{sup} \},\,\,\,\, \Gamma \,:=\,\{ j\,:\, \vert \beta_j^* - \widetilde{\omega}_j \vert >0 \},
\]
and notice that $\widetilde{s} = \vert \widetilde{\xi}\vert$.

Next, we proceed to check the conditions in Lemma A.4 in \cite{song2023nearly}. We use the notation $\mathcal{I}\{C\}$ to denote the indicator function of a set $C$.\\
\textbf{Step 1}.  Let  
\[
	\phi_n^{\prime}=  \underset{ \widetilde{\xi}\subset \xi,\,\vert \xi \vert \leq \widetilde{p} +\widetilde{s} }{\max}\,\, \mathcal{I}\left\{  \left\vert  \frac{  \widetilde{\yb}^{\top}(\Ib-\Hb_{\xi}) \widetilde{\yb}  }{ \sigma^{*2}(n -\vert \xi \vert)  } -1 \right\vert \geq   3 \epsilon_n    \right\}   \\
\]
and 
\[
	\widetilde{\phi}_n =  \underset{ \widetilde{\xi}\subset \xi,\,\vert \xi \vert \leq \widetilde{p} +\widetilde{s} }{\max}\,\, \mathcal{I}\left\{ \|  (\Xb_{\xi}^{\top }  \Xb_{\xi}  )^{-1}\Xb_{\xi}^{\top}\widetilde{\yb}  - \bb^*_{\xi}  \|  \geq   2 \sigma^* \epsilon_n    \right\}   \\
\]
where $\Hb_{\xi} = \Xb_{\xi} (\Xb_{\xi}^{\top }  \Xb_{\xi}  )^{-1}\Xb_{\xi}^{\top}    $. Then let $\phi_n = \max\{ \phi_n^{\prime}, \widetilde{\phi}_n \}$.

Next, we observe that for any $\xi $ such that $\widetilde{\xi} \subset \xi$ and $\vert \xi \vert \leq \widetilde{p} + \widetilde{s}$, letting $\bepsilon = \widetilde{\yb}- \Xb \bb^*$, it holds that 
\[
\begin{array}{lll}
	\widetilde{\yb}^{\top}(\Ib-\Hb_{\xi}) \widetilde{\yb}  &  =  &   [  \Xb_{\widetilde{\xi}} \bb^*_{\widetilde{\xi}}\,+\,  \Xb_{\Gamma\backslash \widetilde{\xi}} \bb^*_{\Gamma\backslash\widetilde{\xi}} \,+\, \bepsilon  ]^{\top}(\Ib-\Hb_{\xi}) \widetilde{\yb} [  \Xb_{\widetilde{\xi}} \bb^*_{\widetilde{\xi}}\,+\,  \Xb_{\Gamma\backslash \widetilde{\xi}} \bb^*_{\Gamma\backslash\widetilde{\xi}} \,+\,\bepsilon  ] \\
	&  =  &  \bepsilon^{\top} (\Ib -\Xb_{\xi}) \bepsilon \,+\,   2 \bepsilon^{\top} (\Ib-\Hb_{\xi})\Xb_{ \Gamma\backslash \widetilde{\xi} } \bb^*_{  \Gamma \backslash\widetilde{\xi} }    \,+\,  \| \Xb_{\Gamma\backslash \widetilde{\xi}} \bb^*_{\Gamma\backslash\widetilde{\xi}}\|^2  \\
	& = : & \mathcal{A}_1(\bepsilon)  \,+\, \mathcal{A}_2(\bepsilon) \,+\,  \| \Xb_{\Gamma\backslash \widetilde{\xi}} \bb^*_{\Gamma\backslash\widetilde{\xi}}\|^2.
\end{array}  
\]
Now, 
\begin{equation}
	\label{eqn:e1}
	\|(\Ib-\Hb_{\xi})\Xb_{ \Gamma\backslash \widetilde{\xi} } \bb^*_{  \Gamma \backslash \widetilde{\xi} }  \| \,\leq \| \Xb_{ \Gamma\backslash \widetilde{\xi} } \bb^*_{  \Gamma \backslash \widetilde{\xi} }  \|\,\leq \, \sqrt{n s} r_{\sup}
\end{equation}
where the second inequality follows from Assumption \ref{as1}. Hence,
\[
\frac{\| \Xb_{ \Gamma\backslash \widetilde{\xi} } \bb^*_{  \Gamma \backslash \widetilde{\xi} }  \|^2}{ \sigma^{*2}(n -\vert \xi \vert)}\,\leq \,  \frac{n s r_{\sup}^{2} }{\sigma^{*2}(n -\vert \xi \vert)} \,\leq\, \epsilon_n  
\]
by our condition on $r$. Moreover,  
\begin{equation}
	\label{eqn:e5}
	\begin{array}{lll}
		\displaystyle \mathbb{E}_{ (\bb^*,\sigma^{*2}) } \left( \mathcal{I}\left\{  \left\vert  \frac{  \widetilde{\yb}^{\top}(\Ib-\Hb_{\xi}) \widetilde{\yb}  }{ \sigma^{*2}(n -\vert \xi \vert)  } -1 \right\vert \geq   3 \epsilon_n    \right\}  \right)  & =  &  \displaystyle  \mathbb{P}\left(  \left\vert  \frac{  \widetilde{\yb}^{\top}(\Ib-\Hb_{\xi}) \widetilde{\yb}  }{ \sigma^{*2}(n -\vert \xi \vert)  } -1 \right\vert \geq   3 \epsilon_n \right) \\
		& = &\displaystyle \mathbb{P}\left(   \left\vert  \mathcal{A}_1(\xi) /[ \sigma^{*2}(n- \vert \xi\vert)  ]  \,-\,1\right\vert      \geq \epsilon_n  \right) \,+\,\\
		& &\displaystyle \mathbb{P}\left(   \left\vert  \mathcal{A}_2(\xi) /[ \sigma^{*2}(n- \vert \xi\vert)  ] \right\vert      \geq \epsilon_n  \right)   \\
		& =:& \mathcal{B}_1 \,+\,\mathcal{B}_2.
	\end{array}
\end{equation}
To bound $\mathcal{B}_1$, notice that 
\begin{equation}
	\label{eqn:e3}
	\begin{array}{lll}
		\displaystyle     \mathcal{B}_1  & =  & \displaystyle  \mathbb{P}(  \vert   \chi_{n -\vert \xi \vert }^2  - (n -\vert \xi \vert)   \vert  \geq n -\vert \xi \vert)   \epsilon_n   ) \\
		&\leq  &\displaystyle   \exp( -c_1 n \epsilon_n^2 )
	\end{array}
\end{equation}
for a positive constant $c_1>0$, 
where $\xi_k^2 $ denotes the chi-squared distribution with $k$ degrees of freedom, and the
last inequality follows from the Bernstein inequality.

To bound $\mathcal{B}_2$, observe that 
\begin{equation}
	\label{eqn:e4}
	\begin{array}{lll}
		\displaystyle  \mathcal{B}_2  & =  &\displaystyle  \mathbb{P}\left(   \left\vert  \mathcal{A}_2(\xi) /[ \sigma^{*2}(n- \vert \xi\vert)  ] \right\vert      \geq \epsilon_n  \right)\\
		& =&\displaystyle \mathbb{P}\left(  \left\vert \frac{ \bepsilon^{\top} (\Ib-\Hb_{\xi}) \Xb_{ \Gamma \backslash \widetilde{\xi}   }\bb^*_{  \Gamma \backslash \widetilde{\xi}  } }{\sigma^*\| (\Ib-\Hb_{\xi}) \Xb_{ \Gamma \backslash \widetilde{\xi}   }\bb^*_{  \Gamma \backslash \widetilde{\xi}  } \|  }    \right\vert   \geq  \frac{ \epsilon_n\sigma^* (n- \vert \xi\vert  ) }{2\| (\Ib-\Hb_{\xi}) \Xb_{ \Gamma \backslash \widetilde{\xi}   }\bb^*_{  \Gamma \backslash \widetilde{\xi}  } \| }  \right)   \\
		& \leq&  \displaystyle  2\exp\left(  -\frac{  \sigma^{*2} \epsilon_n^2   (n- \vert \xi\vert)^2  }{ 8 \| (\Ib-\Hb_{\xi}) \Xb_{ \Gamma \backslash \widetilde{\xi}   }\bb^*_{  \Gamma \backslash \widetilde{\xi}  } \|^2 }  \right)\\
		& \leq&    \displaystyle  2\exp\left( - \frac{  \sigma^{*2} \epsilon_n^2   (n- \vert \xi\vert)^2  }{ 8 ( \sqrt{s} r_{\sup} )^2 }  \right)\\
		& \leq& \displaystyle \exp( -c_2  n \epsilon_n^2)
	\end{array}
\end{equation}
where the first inequality is followed by the Gaussian tail inequality, the third by (\ref{eqn:e1}), and the last holds for some constant $c_2>0$ since  $r_{\sup} = O(1/ \sqrt{s})$.

Therefore, combining (\ref{eqn:e5}), (\ref{eqn:e3}) and (\ref{eqn:e4}), we obtain that there exists a constant $c_3>0$ such that
\begin{equation}
	\label{eqn:e6}
	\displaystyle \mathbb{E}_{ (\bb^*,\sigma^{*2}) } \left( \mathcal{I}\left\{  \left\vert  \frac{  \widetilde{\yb}^{\top}(\Ib-\Hb_{\xi}) \widetilde{\yb}  }{ \sigma^{*2}(n -\vert \xi \vert)  } -1 \right\vert \geq   3 \epsilon_n    \right\}  \right)\,\leq\,  \exp(-c_3 n \epsilon_n^2 ).
\end{equation}

Furthermore,  for $\xi $ satisfying $\widetilde{\xi} \subset \xi$  and $\vert \xi\vert \leq \widetilde{p} + \widetilde{s} \lesssim n \epsilon^2  $, we have that  
\begin{equation}
	\label{eqn:e7}
	\begin{array}{lll}
		\displaystyle    \|  (\Xb_{\xi}^{\top} \Xb_{\xi})^{-1}\Xb_{\xi}^{\top} \widetilde{\yb}  -  \bb_{\xi}^*  \|  &  =  &      \displaystyle   \|  (\Xb_{\xi}^{\top} \Xb_{\xi})^{-1}\Xb_{\xi}^{\top}[  \Xb_{\xi} \bb_{\xi}^* + \Xb_{ \Gamma\backslash \xi }\bb_{ \Gamma\backslash \xi }^*  + \bepsilon     ] -  \bb_{\xi}^*  \|\\
		& \leq &    \displaystyle \|  (\Xb_{\xi}^{\top} \Xb_{\xi})^{-1}\Xb_{\xi}^{\top} \bepsilon\| \,+\,   \|  (\Xb_{\xi}^{\top} \Xb_{\xi})^{-1}\Xb_{\xi}^{\top}  \Xb_{ \Gamma\backslash \xi }\bb_{ \Gamma\backslash \xi}^* \|\\
		& \leq &    \displaystyle \|  (\Xb_{\xi}^{\top} \Xb_{\xi})^{-1}\Xb_{\xi}^{\top} \bepsilon\| \,+\,  \sqrt{ \lambda_{\max}( (\Xb_{\xi}^{\top} \Xb_{\xi} )^{-1} ) }  \|  \Xb_{ \Gamma\backslash \xi }\bb_{ \Gamma\backslash \xi }^* \|\\
		& \leq &    \displaystyle \|  (\Xb_{\xi}^{\top} \Xb_{\xi})^{-1}\Xb_{\xi}^{\top} \bepsilon\| \,+\,  \sqrt{ \lambda_{\min}^{-1}( \Xb_{\xi}^{\top} \Xb_{\xi}  ) }  \|  \Xb_{ \Gamma\backslash \xi}\bb_{ \Gamma\backslash \xi }^* \|\\
		&\leq &   \displaystyle \|  (\Xb_{\xi}^{\top} \Xb_{\xi})^{-1}\Xb_{\xi}^{\top} \bepsilon\| \,+\,   \|  \Xb_{ \Gamma\backslash \xi } \bb_{ \Gamma\backslash \xi }^* \|/ \sqrt{n \lambda_0} \\  
		&\leq &   \displaystyle \|  (\Xb_{\xi}^{\top} \Xb_{\xi})^{-1}\Xb_{\xi}^{\top} \bepsilon\| \,+\, \| \bb_{ \Gamma\backslash \widetilde{\xi}}^* \|_1 \sqrt{n}/ \sqrt{n \lambda_0}   \\  
		&=&   \displaystyle \|  (\Xb_{\xi}^{\top} \Xb_{\xi})^{-1}\Xb_{\xi}^{\top} \bepsilon\| \,+\, \frac{\sum_{j\,:\, \vert b_j^*\vert \leq r  } \vert b_j^*\vert }{\lambda_0}   \\  
		&\leq &   \displaystyle \|  (\Xb_{\xi}^{\top} \Xb_{\xi})^{-1}\Xb_{\xi}^{\top} \bepsilon\| \,+\, \sigma^*\epsilon_n  \\  
	\end{array}
\end{equation}
where the fourth inequality follows from Assumption \ref{as3}, if $M$ is chosen large enough. Hence, 
\begin{equation}
	\label{eqn:e8}
	\begin{array}{lll}
		\displaystyle  \mathbb{E}_{ (\bb^*,\sigma^{*2}) } \mathcal{I}\left\{ \|  (\Xb_{\xi}^{\top }  \Xb_{\xi}  )^{-1}\Xb_{\xi}^{\top}\widetilde{\yb}  - \bb^*_{\xi}  \|  \geq   2 \sigma^* \epsilon_n    \right\}      & = & \displaystyle \mathbb{P}\left( \|  (\Xb_{\xi}^{\top }  \Xb_{\xi}  )^{-1}\Xb_{\xi}^{\top}\widetilde{\yb}  - \bb^*_{\xi}  \|  \geq   2 \sigma^* \epsilon_n  \right)\\
		& \displaystyle \leq &\mathbb{P}\left( \|  (\Xb_{\xi}^{\top }  \Xb_{\xi}  )^{-1}\Xb_{\xi}^{\top}\bepsilon \|  \geq    \sigma^* \epsilon_n  \right)\\
		& \displaystyle \leq &\mathbb{P}\left(\chi_{ \vert\xi \vert}^2  \geq    n \lambda_0 \epsilon_n^2 \right)\\
		& \displaystyle \leq& \exp(- c_4 n \epsilon_n^2)
	\end{array}
\end{equation}
for a constant $c_4>0$, and where the last two inequalities follow as in Lemma 1 of \cite{armagan2013posterior}.

Therefore, combining (\ref{eqn:e6}) and (\ref{eqn:e8}), we obtain,
\begin{equation}
	\label{eqn:e9}
	\begin{array}{lll}
		\displaystyle    \mathbb{E}_{ (\bb^*,\sigma^{*2}) }  \phi_n  & \leq &\displaystyle   \mathbb{E}\bigg(  \underset{ \widetilde{\xi}\subset \xi,\,\vert \xi \vert \leq \widetilde{p} +\widetilde{s}  }{\sum} \bigg( \mathcal{I}\left\{  \left\vert  \frac{  \widetilde{\yb}^{\top}(\Ib-\Hb_{\xi}) \widetilde{\yb}  }{ \sigma^{*2}(n -\vert \xi \vert)  } -1 \right\vert \geq   3 \epsilon_n    \right\}  \,+\,     \\
		&&   \mathcal{I}\left\{ \|  (\Xb_{\xi}^{\top }  \Xb_{\xi}  )^{-1}\Xb_{\xi}^{\top}\widetilde{\yb}  - \bb^*_{\xi}  \|  \geq   2 \sigma^* \epsilon_n    \right\}  \bigg)   \bigg)  \\
			& \leq&    (\widetilde{p} + \widetilde{s}) \binom{p}{\widetilde{p}+\widetilde{s}}[  \exp(-c_3 n \epsilon_n^2)  \,+\, \exp(-c_4 n \epsilon_n^2) ].
	\end{array}
\end{equation}
Hence, we can choose $\widetilde{p}\asymp \widetilde{s} $ such that \[
\log(  \widetilde{p}  +\widetilde{s} ) + (\widetilde{p} + \widetilde{s}) \log p < c_5 n \epsilon_n^2 
\]
for a small enough constant $c_5$, such that for some $c_6>0$, we have that 
\begin{equation}
	\label{eqn:e10}
		\mathbb{E}_{ (\bb^*,\sigma^{*2}) }  \phi_n \,\leq\, \exp(-c_6  n \epsilon_n^2).
	\end{equation}
\textbf{Step 2.} We start by noticing that $C_n \subset C_{n,1}   \cup C_{n,2}$, where
	\[
	C_{n,1}\,=\,\left\{  \frac{\sigma^2}{ \sigma^{*2} } \geq \frac{1+\epsilon_n}{1-\epsilon_n} \,\,\text{or}\,\,\frac{\sigma^2}{ \sigma^{*2} } \leq \frac{1-\epsilon_n}{1+\epsilon_n} \right\}\cap \{ \text{at most}\,\, \widetilde{p}\,\,\text{entries of } \,\,\vert \bb/\sigma\vert \,\,\text{are larger than or equal to } \,\,a_n \}
	\]
	and 
	\[
	\begin{array}{lll}
		C_{n,2}&=&\left\{  \frac{\sigma^2}{ \sigma^{*2} } \leq \frac{1+\epsilon_n}{1-\epsilon_n}, \,\,\| \bb^*-\bb\| \geq (  7/2 + \sqrt{\lambda_0} )\sigma^*\epsilon_n   \right\}\cap \\
		&&\{ \text{at most}\,\, \widetilde{p}\,\,\text{entries of } \,\,\vert \bb/\sigma\vert \,\,\text{are larger than or equal to } \,\,a_n \}.
	\end{array}
	\]
	Then
	\[
	\begin{array}{lll}
		\underset{(\bb,\sigma^2 )  \in C_n }{\sup }\mathbb{E}_{(\bb, \sigma^2) }(1-\phi_n)       &= &    \underset{(\bb,\sigma^2 )  \in C_n }{\sup }\mathbb{E}_{(\bb, \sigma^2) }\min\{1-\phi_n^{\prime},1-\widetilde{\phi}_n \}      \\
		& \leq & \max\left\{   \underset{(\bb,\sigma^2 )  \in C_{n,1} }{\sup }\mathbb{E}_{(\bb, \sigma^2) }(1-\phi_n^{\prime}),  \underset{(\bb,\sigma^2 )  \in C_{n,2} }{\sup }\mathbb{E}_{(\bb, \sigma^2) }(1-\widetilde{\phi}_n) \right\}. 
	\end{array}
	\]
	Moreover,   let  $\xi^{\prime} =   \xi^{\prime}(\bb) \,:=\,  \{j\,:\, \vert b_j/\sigma\vert > a_n     \} \cup \widetilde{\xi}$ and $\xi^{\prime c} \,=\,   \{1,\ldots,p\} \backslash \xi^{\prime} $.  Then, for any $(\bb,\sigma^2) \in  C_{n,1} \cup C_{n,2}$, it holds that 
	$\vert \xi^{\prime}(\bb) \vert \leq  \widetilde{p} +  \widetilde{s}$, and $\| \Xb_{ \xi^{\prime c} } \bb_{ \xi^{\prime c}   }\| \leq \sqrt{n p} \|  \bb_{ \xi^{\prime c}   }\| \,\leq\, \sqrt{n}p\sigma a_n \,=\, \sigma M \sqrt{\widetilde{s} \log p}\,\leq\,  \sigma \sqrt{n} \epsilon_n$. Also, proceeding exactly as on Page 433 of \cite{song2023nearly}, we obtain that 
	\begin{equation}
		\label{eqn:e11}
		\underset{(\bb,\sigma^2 )  \in C_{n,1} }{\sup }\mathbb{E}_{(\bb, \sigma^2) }(1-\phi_n^{\prime}) \,\leq\, \exp(-c_7 n\epsilon_n^2)
	\end{equation}
	for some constant $c_7>0$.
	
	Furthermore, 
	\begin{equation}
		\label{eqn:e12}
		\begin{array}{l}
			\underset{(\bb,\sigma^2 )  \in C_{n,2} }{\sup }\mathbb{E}_{(\bb, \sigma^2) }(1-\widetilde{\phi}_n)  \\
			\leq  \underset{(\bb,\sigma^2 )  \in C_{n,2} }{\sup }\mathbb{E}_{(\bb, \sigma^2) } \mathcal{I}\{ \| ( \Xb_{\xi^{\prime} }^{\top}\Xb_{\xi^{\prime} }  )^{-1}\Xb_{\xi^{\prime} }^{\top} \widetilde{\yb }  - \bb_{\xi^{\prime}}^* \| \leq \sigma^*\epsilon_n \}  \\
			=   \underset{(\bb,\sigma^2 )  \in C_{n,2} }{\sup }\mathbb{P}_{(\bb, \sigma^2) } (\|( \Xb_{\xi^{\prime} }^{\top}\Xb_{\xi^{\prime} }  )^{-1}\Xb_{\xi^{\prime} }^{\top} \widetilde{\yb }  - \bb_{\xi^{\prime}}^* \| \leq \sigma^*\epsilon_n )  \\
			=   \underset{(\bb,\sigma^2 )  \in C_{n,2} }{\sup }\mathbb{P}_{(\bb, \sigma^2) } (\|( \Xb_{\xi^{\prime} }^{\top}\Xb_{\xi^{\prime} }  )^{-1}\Xb_{\xi^{\prime} }^{\top} \bepsilon + \bb_{\xi^{\prime} } +  ( \Xb_{\xi^{\prime} }^{\top}\Xb_{\xi^{\prime} }  )^{-1}\Xb_{\xi^{\prime} }^{\top} \Xb_{\xi^{\prime c} } \bb_{\xi^{\prime c} }   - \bb_{\xi^{\prime}}^* \| \leq \sigma^*\epsilon_n )  \\  
			\leq        \underset{(\bb,\sigma^2 )  \in C_{n,2} }{\sup }\mathbb{P}_{(\bb, \sigma^2) } ( (\|( \Xb_{\xi^{\prime} }^{\top}\Xb_{\xi^{\prime} }  )^{-1}\Xb_{\xi^{\prime} }^{\top} \bepsilon \|\geq  
			\|\bb_{\xi^{\prime}} - \bb^{*}_{\xi^{\prime } } \|  - \|  ( \Xb_{\xi^{\prime} }^{\top}\Xb_{\xi^{\prime} }  )^{-1}\Xb_{\xi^{\prime} }^{\top} \Xb_{\xi^{\prime c} } \bb_{\xi^{\prime c} }\| - \sigma^* \epsilon_n       )  \\    
			=       \underset{(\bb,\sigma^2 )  \in C_{n,2} }{\sup }\mathbb{P}_{(\bb, \sigma^2) } ( (\|( \Xb_{\xi^{\prime} }^{\top}\Xb_{\xi^{\prime} }  )^{-1}\Xb_{\xi^{\prime} }^{\top}(\bepsilon/\sigma)  \|\geq  
			[\|\bb_{\xi^{\prime}} - \bb^{*}_{\xi^{\prime } } \|  - \|  ( \Xb_{\xi^{\prime} }^{\top}\Xb_{\xi^{\prime} }  )^{-1}\Xb_{\xi^{\prime} }^{\top} \Xb_{\xi^{\prime c} } \bb_{\xi^{\prime c} }\| - \sigma^* \epsilon_n ]/\sigma      ).  \\    
		\end{array}
	\end{equation}
	However, if  $(\bb,\sigma^2) \in C_{n,2}$, then 
	\[
	\begin{array}{lll}
		\displaystyle \|  \bb- \bb^*\|    &   \leq   &  \displaystyle \| \bb_{ \xi^{\prime} } -   \bb^*_{ \xi^{\prime} } \| \,+\,    \| \bb_{ \xi^{\prime c} } -   \bb^*_{ \xi^{\prime c } } \| \\
		&\leq  &   \displaystyle \| \bb_{ \xi^{\prime} } -   \bb^*_{ \xi^{\prime} } \| \,+\,    \| \bb_{ \xi^{\prime c} }\|  + \|   \bb^*_{ \xi^{\prime c } } \| \\
		& \leq & \displaystyle \| \bb_{ \xi^{\prime} } -   \bb^*_{ \xi^{\prime} } \| \,+\,    \sqrt{p} a_n  \sigma \,+\, \sqrt{   \sum_{  j\,:\, \vert b^*_j\vert \leq r_{\sup}  }  \vert b_j^*\vert^2      }\\
		& \leq &  \displaystyle\| \bb_{ \xi^{\prime} } -   \bb^*_{ \xi^{\prime} } \| \,+\,    \sqrt{p} a_n \sigma \,+\, \sum_{  j\,:\, \vert b^*_j\vert \leq r_{\sup}  }  \vert b_j^*\vert\\
		& \leq &   \displaystyle\| \bb_{ \xi^{\prime} } -   \bb^*_{ \xi^{\prime} } \| \,+\,    \frac{\sigma^* \epsilon_n}{4}\,+\,\frac{ \sigma^*\epsilon_n}{4},
	\end{array}
	\]
	provided that $M$ is chosen large enough and since $\sigma \leq \sigma^*  \sqrt{  (1+\epsilon_n)/(1-\epsilon_n)  }$ . Hence, if  $(\bb,\sigma^2) \in C_{n,2}$, then 
	\[
	\| \bb_{ \xi^{\prime} } -   \bb^*_{ \xi^{\prime} } \|   \,\geq \, (3+ \sqrt{\lambda_0} )\sigma^* \epsilon_n. 
	\]
	Moreover,
	\begin{equation}
	\begin{array}{lll}
		\|  ( \Xb_{\xi^{\prime} }^{\top}\Xb_{\xi^{\prime} }  )^{-1}\Xb_{\xi^{\prime} }^{\top} \Xb_{\xi^{\prime c} } \bb_{\xi^{\prime c} }\|    & \leq  &  \sqrt{  \lambda_{\max}( (\Xb_{\xi^{\prime } }^{\top} \Xb_{\xi^{\prime} }    )^{-1}  )  }   \| \Xb_{\xi^{\prime c} } \bb_{\xi^{\prime c} }\|    \\
		& \leq & \frac{\| \Xb_{\xi^{\prime c} } \bb_{\xi^{\prime c} }\|   }{ \sqrt{n \lambda_0}  }\\
		& \leq &\frac{\sqrt{n p}  \sigma a_n  }{ \sqrt{n \lambda_0}  }\\
		& \leq & \sigma\epsilon_n.
	\end{array}    
	\end{equation}
	As a result,
	\begin{equation}
		\label{eqn:e14}
		\begin{array}{l}
			\underset{(\bb,\sigma^2 )  \in C_{n,2} }{\sup }\mathbb{E}_{(\bb, \sigma^2) }(1-\widetilde{\phi}_n)  \\
			\leq    \underset{(\bb,\sigma^2 )  \in C_{n,2} }{\sup }\mathbb{P}_{(\bb, \sigma^2) }(  |( \Xb_{\xi^{\prime} }^{\top}\Xb_{\xi^{\prime} }  )^{-1}\Xb_{\xi^{\prime} }^{\top}(\bepsilon/\sigma)  \| \geq \epsilon_n  )\\
			\leq  \exp(- c_8 n \epsilon_n^2).
		\end{array}
	\end{equation}
	
	Combining (\ref{eqn:e11}) with (\ref{eqn:e14})
	\begin{equation}
		\underset{(\bb,\sigma^2 )  \in C_n }{\sup }\mathbb{E}_{(\bb, \sigma^2) }(1-\phi_n) \,\leq\, \exp( -c_9 n \epsilon_n^2 )
	\end{equation}
	for some constant $c_9>0$.\\
	\textbf{Step 3.}  We observe that
	\begin{equation*}
		\pi(B_n )\,=\,  \mathbb{P}(V \geq  \widetilde{p}  ) 
	\end{equation*}
	where 
	\[
	V \,=\,   \sum_{j=1}^{ p } \mathcal{I}\{ \vert b_j/\sigma\vert \geq  a_n  \}. 
	\]
	Then
	\begin{equation}
		\label{eqn:20}
		\begin{array}{lll}
			\mathbb{P}(\vert b_j/\sigma\vert\geq   a_n  \,|\,   \sigma, \eta  )    & = &    \displaystyle \int_{a_n }^{\infty}  \int_0^{\infty}  \frac{1}{  \sqrt{ 2\pi  }  \lambda  \tau } \exp\left(   -\frac{ ( \sigma x  )^2  }{2 \lambda^2 \sigma^2\tau^2  }  \right)\frac{2}{\pi(1+\lambda^2)} \dif\lambda\dif x   \,+\, \\
			& &     \displaystyle    \int_{-\infty}^{-a_n}  \int_0^{\infty}  \frac{1}{  \sqrt{ 2\pi  }  \lambda   \tau } \exp\left(   -\frac{ (\sigma x  )^2  }{2 \lambda^2 \sigma^2 \tau^2 } \right)\frac{2}{\pi(1+\lambda^2)} \dif\lambda\dif x\\
			& = &    \displaystyle  \frac{1}{\tau  }\int_{a_n }^{\infty}  \int_0^{\infty}  \frac{1}{  \sqrt{ 2\pi  }  \lambda   } \exp\left(   -\frac{ x^2  }{2 \lambda^2 \tau^2 }  \right)\frac{2}{\pi(1+\lambda^2)} \dif \lambda \dif x \,+\, \\
			& &     \displaystyle  \frac{1}{\tau   }\int_{-\infty}^{-a_n}  \int_0^{\infty}  \frac{1}{  \sqrt{ 2\pi  }  \lambda } \exp\left(   -\frac{ x^2 }{2 \lambda^2 \tau^2 } \right)\frac{2}{\pi(1+\lambda^2)} \dif \lambda\dif x \\
			& \leq  &    \displaystyle  \frac{C}{\tau }\int_{a_n }^{\infty} \log\left( 1+ \frac{2\tau^2}{x^2 } \right)   \dif x   \,+\, \\
			& &     \displaystyle  \frac{C}{\tau    }\int_{-\infty}^{-a_n}  \log\left( 1+ \frac{2\tau^2}{x^2} \right)  \dif x \\
			& \leq  &    \displaystyle  \frac{2C}{\tau  }\int_{a_n }^{\infty}   \frac{\tau^2}{ x^2 }  \dif x \,+\, \frac{2C}{\tau  }\int_{-\infty}^{-a_n} \frac{\tau^2}{ x^2 }   \dif x\\
			& \leq& \displaystyle 4C \tau/a_n \\
			&\leq & c_{10} \tau/a_n
		\end{array}
	\end{equation}
	where the first inequality follows by the proof of Theorem 1 in \cite{carvalho2010horseshoe}, and $C>0$ is a constant, and $c_{10} =  4C$. Hence, if $ \tau = p^{-u_1}$
	for $u_1 >0$ large enough, then  $1/(c_{10} p\tau/a_n    )\geq O(p^u)$  for some $u>0$.  Then as in Page 433 of \cite{song2023nearly}, we obtain that
	\[
	\pi(B_n) \,\leq\,  \exp(  -c_{11}  n\epsilon_n^2 ) 
	\]
	for some $c_{11} >0$.\\
	\textbf{Step 4.} Let $f^*$ be the density of the true model and $m(\yb)$ the marginal density of $\yb$ after integrating out the prior in the Bayesian model. Then, as in  Page 435 in \cite{song2023nearly}, in order to obtain 
	\[
	P^*(  \pi( m( \widetilde{\yb} )/f^*( \widetilde{\yb})   \geq \exp(-c_{12} n \epsilon_n^2)  ) \,\geq \, \exp(-c_{13} n \epsilon_n^2 )
	\]
	for some positive constants $c_{12}$ and $c_{13}$, it is enough to show that 
	\[
	P^*(  \pi(\{  \| \widetilde{\yb} -\Xb \bb\|^2/2\sigma^2 + n \log(\sigma /\sigma^*) \,< \,  \| \widetilde{\yb} -\Xb \bb^*\|^2/2\sigma^{*2} \,+\,c_{12} n \epsilon_n^2/2   \}) \geq  e^{  -c_{12} n\epsilon_n^2/2   }  )\,\geq\, 1 \,-\, \exp\left( -c_{13} n \epsilon_n^2  \right).
	\]
	Then proceeding as in Page 435 in \cite{song2023nearly}, it is enough to bound 
	\[
	-\log(\pi( \{ 0\leq  \sigma^2 -\sigma^{*2} \leq \eta_1 \epsilon_n\}  )   - \log \pi( \|(\bb^*-\bb)/\sigma \|_1 < 2\eta_2\epsilon_n  \,|\,   \sigma \in [\sigma^*,\sigma^{*} + \eta_1 \epsilon_n^2] )
	\]
	for small positive constants $\eta_1$ and $\eta_2$. 
	
	Then 
 \[
 \begin{array}{l}
	\{   \sigma \in [\sigma^*,\sigma^{*} + \eta_1 \epsilon_n^2]  \,\,\text{and}\,\, \|\bb -\bb^*\|_1 \leq  2\eta_2 \epsilon_n   \}\\
	\supseteq   \{   \sigma \in [\sigma^*,\sigma^{*} + \eta_1 \epsilon_n^2]  \,\,\text{and}\,\, \|(\bb_{ \widetilde{\xi}   } -\bb^*_{\widetilde{\xi}})/\sigma\|_1 \leq  \eta_2 \epsilon_n  \,\,\text{and}\,\,    \|\bb_{ \widetilde{\xi}^c   }/\sigma \|_1\leq  \eta_2 \epsilon_n/2 \,\,\text{and}\,\, \| \bb_{\widetilde{\xi}^c }^*\|_1/\sigma^* \,\leq \,   \eta_2 \epsilon_n/2    \}\\
	\supseteq   \{   \sigma \in [\sigma^*,\sigma^{*} + \eta_1 \epsilon_n^2]  \,\,\text{and}\,\, \|(\bb_{ \widetilde{\xi}   } -\bb^*_{\widetilde{\xi}})/\sigma\|_1 \leq  \eta_2 \epsilon_n  \,\,\text{and}\,\,    \|\bb_{ \widetilde{\xi}^c   }/\sigma \|_1\leq  \eta_2 \epsilon_n/2    \}\\
	\supseteq  \{  \sigma \in [\sigma^*,\sigma^{*} + \eta_1 \epsilon_n^2]  \,\,\text{and}\,\,  b_j/\sigma \in [b_j^*/\sigma -  \eta_2 \epsilon_n/\widetilde{s} ,  b_j^*/\sigma +  \eta_2 \epsilon_n/\widetilde{s}   ]   \,\forall j \in \widetilde{\xi}   \,\,\text{and}\,\,    \|\bb_{ \widetilde{\xi}^c   }/\sigma \|_1\leq  \eta_2 \epsilon_n/2   \}
 \end{array}
 \]
	where the second contention follows since
	\[
	\| \bb_{\widetilde{\xi}^c }^*\|_1/\sigma^* \,\leq \,   \eta_2 \epsilon_n/2
	\]
	if $M$ is chosen large enough. 
	
	Furthermore, same argument from \cite{song2023nearly} shows that 
	\[
	-\log(\pi( \{ 0\leq  \sigma^2 -\sigma^{*2} \leq \eta_1 \epsilon_n\}  )    \,\leq \, \delta_1 n \epsilon_n^2
	\]
	for an arbitrary constant $\delta_1$ if $M$ is chosen large enough.  
	
	Moreover, it holds that 
	\[
	\begin{array}{lll}
		\pi(  \{  \|\bb_{ \widetilde{\xi}^c   }/\sigma \|_1\leq  \eta_2 \epsilon_n/2  \,|\,  \sigma \in [\sigma^*,\sigma^{*} + \eta_1 \epsilon_n^2]   \})   &\geq   &    \pi(  \{ \vert b_j /\sigma \vert \leq  \eta_2 \epsilon_n/(2p)   \,\, \forall j \notin \widetilde{\xi}   \}  \,|\,  \sigma \in [\sigma^*,\sigma^{*} + \eta_1 \epsilon_n^2]   )    \\
		& \geq & (1-  c_{10} \tau/a_n )^p \,\rightarrow 1.  
	\end{array}
	\]
	where the second inequality follows from the calculation in (\ref{eqn:20}).
	
	Finally, for  $E \asymp \|\bb^*\|_{\infty}/\sigma^*$ and a positive constant $c_{14}$, we have that 
	\[
	\begin{array}{l}
		\pi(   b_j/\sigma \in [b_j^*/\sigma -  \eta_2 \epsilon_n/\widetilde{s} ,  b_j^*/\sigma +  \eta_2 \epsilon_n/\widetilde{s}   ],   \,  \,\forall j \in \widetilde{\xi} \,\,  | \,\,  \sigma \in [\sigma^*,\sigma^{*} + \eta_1 \epsilon_n^2]) \\
		\displaystyle        \geq  \left[  \frac{2 \eta_2 \epsilon_n }{\widetilde{s}}  \underset{x \in [-E,E] }{\inf}   \int_0^{\infty}  \frac{1}{  \sqrt{ 2\pi  }  \lambda  \tau } \exp\left(   -\frac{ x^2  }{2 \lambda^2 \tau^2  }  \right)\frac{2}{\pi(1+\lambda^2)} \dif\lambda    \right]^{\widetilde{s}}\\
		\geq \displaystyle \left[  \frac{c_{14} \eta_2 \epsilon_n }{\widetilde{s} \tau }  \underset{x \in [-E,E] }{\inf}  \log\left(1   +    \frac{4 \tau^2}{x^2}  \right)    \right]^{\widetilde{s}}\\
		=  \displaystyle \left[  \frac{c_{14} \eta_2 \epsilon_n }{\widetilde{s}  \tau  }   \log\left(1   +    \frac{4 \tau^2}{E^2}  \right)    \right]^{\widetilde{s}}\\
		\geq   \displaystyle \left[  \frac{c_{14} \eta_2 \epsilon_n }{\widetilde{s}}  \frac{4 \tau}{ E^2+4\tau^2  }   \right]^{\widetilde{s}}\\
		\geq   \displaystyle \left[  \frac{c_{14} \eta_2 \epsilon_n }{\widetilde{s}}  \frac{2 \tau}{ E^2 }   \right]^{\widetilde{s}}
	\end{array}
	\]
	where the second inequality follows from Theorem 1 in \cite{carvalho2010horseshoe}. Therefore, 
	\[
	-\log  \pi(   b_j/\sigma \in [b_j^*/\sigma -  \eta_2 \epsilon_n/\widetilde{s} ,  b_j^*/\sigma +  \eta_2 \epsilon_n/\widetilde{s}   ],   \,  \,\forall j \in \widetilde{\xi}   |   \sigma \in [\sigma^*,\sigma^{*} + \eta_1 \epsilon_n^2]) \,\leq\,c_{16} n \epsilon_n^2
	\]
	for some positive constant $c_{16}$. The claim then follows by an application of Lemma A.4 in \cite{song2023nearly} or \cite{bernardo1998information}.
\end{proof}
\newpage

\section{Proof of Corollary \ref{cor:n0assumption}}
\label{S-cor:n0assumption-proof}
\begin{proof}
When the target sample size $n$ grows and source data remain fixed, TRADER remains consistent. In some cases, it even achieves the optimal convergence rate, albeit with potentially larger constants. Specifically, Theorem \ref{thm1} shows that TRADER's contraction rate depends on the alignment between source and target through a parameter $r$, which must satisfy: 
    \[
     r \leq C\min\left\{\frac{1}{\sqrt{s} }, \sigma^*\left(\frac{ \widetilde{s}(r) \log p }{s^2 n }\right)^{1/4}    \right\}. 
    \]
    As $n$ increases, this simplifies to: 
    \[
     r \leq C\, \sigma^*\left(\frac{ \widetilde{s}(r) \log p }{s^2 n }\right)^{1/4}.  
    \]
    Here, $\widetilde{s}(r)$ counts the number of coordinates where the deviation between $\widetilde{\bomega}$ and $\bbeta^*$ exceeds a threshold $r$, reflecting source-target approximation quality. The quantity $\widetilde{s}(r)$ is bounded by:
    \[
    \widetilde{s}(r)\,\leq\,  s + \vert \{j\,:\,  \widetilde{\omega}_j \neq 0 \}\vert.
    \]
    When the source is entirely uninformative, with $\widetilde{s}(r) = s + \left| { j : \widetilde{\omega}_j \neq 0 } \right|$, and $n$ increases, Theorem \ref{thm1} bounds TRADER's contraction rate by:
    \[
    \sqrt{ \frac{(s + \vert \{j\,:\,  \widetilde{\omega}_j \neq 0 \}\vert)\log p }{n} },
    \]
    which can be worse than the rate without source information (i.e., target-only HS):
    \[\sqrt{ \frac{s\log p }{n} }.
    \]
    However, when $\left| \{ j : \widetilde{\omega}_j \neq 0 \} \right| = O(s)$, both rates are of the same order. Thus, Theorem \ref{thm1} confirms TRADER's consistency as $n$ grows. When $\left| { j : \widetilde{\omega}_j \neq 0 } \right| = O(s)$, TRADER's convergence rate aligns with the target-only setting, potentially with a larger constant, but improves with favorable source-target alignment when $\widetilde{s}(r)$ is below its upper bound. 
\end{proof}

\newpage
\section{Proof of Theorem~\ref{thm2}}

\begin{proof}
    The proof is similar to that of Theorem \ref{thm1}.   We let 
\[
    \bmu \,=\, \sum_{k=1}^K  (\eta_k-\eta^*_K )\widetilde{\bomega}^{(k)}.
\]

Then notice that the true model is equivalent to 
\[
      \widetilde{\yb} := \yb - \Xb \bomega^* \,\sim \, \mathcal{N}( \Xb  \bb^*, \sigma^{*2} \Ib),
\]
 where $\bb^* =  \bbeta^* - \widetilde{\bomega} $.    Also, writing $\bb =  \bbeta - \widetilde{\bomega}$, the Bayesian model is equivalent to

 \[
     \begin{array}{lll}
       \widetilde{\yb} &\sim& \mathcal{N}( \Xb  \bb, \sigma^2 \Ib)        \\
         b_j   &\sim  & \mathcal{N}( \mu_j,  \sigma^2 \lambda_j^2  \tau^2  )\\
         (\eta_1,\eta_2,\ldots,\eta_{K+1}) & \sim &  \text{Dirichlet}(\theta_1,\ldots,\theta_K,\zeta)   ,\\
         \lambda_j &\sim & \text{Cauchy}^{+}(0,1)\\
         \sigma^2 &\sim & \text{Inverse-Gamma}(\nu,\nu).
     \end{array}
\]
Next, let $\widetilde{p}$ be determined later such that $\widetilde{p}\asymp \widetilde{s} := \widetilde{s}(r)$,  $a_n = M \sqrt{\widetilde{s}\log p /n }/p$, 
and let 
\[
 \begin{array}{lll}
     A_n& =& \{ \text{at least} \,\,\widetilde{p}\, \text{entries of} \, \,\vert \bb/\sigma\vert  \,\,\text{are larger than } \,\,a_n \} \cup  \\
      & &    \{  \|\bbeta- \bbeta^*\|    \geq  (3 + \sqrt{\lambda_0}) \sigma^*\epsilon_n\} \cup  \{     \frac{\sigma^2}{ \sigma^{*2} } > \frac{1+\epsilon_n}{1- \epsilon_n }   \,\,\text{or} \frac{\sigma^2}{ \sigma^{*2} } < \frac{1-\epsilon_n}{1+ \epsilon_n } 
      \,\,\}\\
      B_n& =& \{ \text{at least} \,\,\widetilde{p}\, \text{entries of} \, \,\vert \bb/\sigma\vert  \,\,\text{are larger than } \,\,a_n \}\\
      
 \end{array}
\]
and $C_n   =   A_n \backslash B_n $. We also write
\[
 \widetilde{\xi}  = \{ j\,:\, \vert   \beta_j^* - \widetilde{\omega}_j \vert > r^* \},\,\,\,\, \Gamma \,:=\,\{ j\,:\, \vert \beta_j^* - \widetilde{\omega}_j \vert >0 \},
\]
\[
Z_{\text{pre}} := \{ j \in \{1,\ldots,p\} \,:\,   \widetilde{\omega}_j^{(k)} = 0 \, \,\forall k \in \{1,\ldots,K\}  \},
\]
and notice that $\widetilde{s} = \vert \widetilde{\xi}\vert$.\\
\textbf{Step 1}.  Let  
\[
       \phi_n^{\prime}=  \underset{ \widetilde{\xi}\subset \xi,\,\vert \xi \vert \leq \widetilde{p} +\widetilde{s} }{\max}\,\, \mathcal{I}\left\{  \left\vert  \frac{  \widetilde{\yb}^{\top}(\Ib-\Hb_{\xi}) \widetilde{\yb}  }{ \sigma^{*2}(n -\vert \xi \vert)  } -1 \right\vert \geq   3 \epsilon_n    \right\}   \\
\]
and 
\[
       \widetilde{\phi}_n =  \underset{ \widetilde{\xi}\subset \xi,\,\vert \xi \vert \leq \widetilde{p} +\widetilde{s} }{\max}\,\, \mathcal{I}\left\{ \|  (\Xb_{\xi}^{\top }  \Xb_{\xi}  )^{-1}\Xb_{\xi}^{\top}\widetilde{\yb}  - \bb^*_{\xi}  \|  \geq   2 \sigma^* \epsilon_n    \right\}   \\
\]
and $\phi_n = \max\{\phi_n,\phi_n^{\prime} \}$. Then Step 1 in the proof of Theorem \ref{thm1} shows that
for some $c_6>0$, we have that 
\begin{equation}
    \label{eqn:e50}
          \mathbb{E}_{ (\bb^*,\sigma^{*2}) }  \phi_n \,\leq\, \exp(-c_6  n \epsilon_n^2).
\end{equation}
for some constant $c_6>0$.\\
\textbf{Step 2}. We notice that $C_n \subset C_{n,1}   \cup C_{n,2}$, where
\[
 C_{n,1}\,=\,\left\{  \frac{\sigma^2}{ \sigma^{*2} } \geq \frac{1+\epsilon_n}{1-\epsilon_n} \,\,\text{or}\,\,\frac{\sigma^2}{ \sigma^{*2} } \leq \frac{1-\epsilon_n}{1+\epsilon_n} \right\}\cap \{ \text{at most}\,\, \widetilde{p}\,\,\text{entries of } \,\,\vert \bb/\sigma\vert \,\,\text{are larger than or equal to } \,\,a_n \}
\]
and 
\[
 \begin{array}{lll}
    C_{n,2}&=&\left\{  \frac{\sigma^2}{ \sigma^{*2} } \leq \frac{1+\epsilon_n}{1-\epsilon_n}, \,\,\| \bb^*-\bb\| \geq (  7/2 + \sqrt{\lambda_0} )\sigma^*\epsilon_n   \right\}\cap \\
    &&\{ \text{at most}\,\, \widetilde{p}\,\,\text{entries of } \,\,\vert \bb/\sigma\vert \,\,\text{are larger than or equal to } \,\,a_n \}.
 \end{array}
\]
and as in the proof of Step 2 of Theorem \ref{thm1},
\begin{equation}
    \label{eqn:e13}
     \underset{(\bb,\sigma^2 )  \in C_n }{\sup }\mathbb{E}_{(\bb, \sigma^2) }(1-\phi_n) \,\leq\, \exp( -c_9 n \epsilon_n^2 )
\end{equation}
for some constant $c_9>0$.\\
\textbf{Step 3}. We observe that
\[
 \pi(B_n  ) \,=\, \mathbb{P}(V \geq  \widetilde{p}   )
\]
where 
\[
V \,=\,   \sum_{j=1}^{ p } \mathcal{I}\{ \vert b_j\vert \geq  \sigma  a_n  \}. 
\]
Now, define let us define the set
\[
Z_{\text{pre}} := \{ j \in \{1,\ldots,p\} \,:\,   \widetilde{\omega}_j^{(k)} = 0 \, \,\forall k \in \{1,\ldots,K\}  \}.
\]
Clear, if $j \in Z_{\text{pre}}$ then $\mu_j =0$. Hence, as in (\ref{eqn:20}),
\[
	\mathbb{P}(\vert b_j/\sigma\vert\geq   a_n  \,|\,   \sigma, \eta  )   \,\leq\,  c_{10} \tau/a_n
\]
for some constant $c_{10}>0$.

Therefore, if $ \tau = p^{-u_1}$
 for $u_1 >0$ large enough, then  $1/(c_{10} p\tau/a_n    )\geq O(p^u)$  for some $u>0$.  Then
 \begin{equation}
     \label{eqn:e24}
     \begin{array}{lll}
      \pi(B_n)    &\leq    &  \displaystyle \mathcal{I}\{ \vert Z_{\text{pre}}\vert \geq \widetilde{p}/2  \}\cdot \mathbb{P}\left(  \sum_{j\in  Z_{\text{pre}} } \mathcal{I}\{ \vert b_j\vert \geq  \sigma  a_n  \} \geq \widetilde{p}/2\right )\,+\,\\
          & & \displaystyle\mathcal{I}\{ \vert \{1,\ldots, p,\}\backslash Z_{\text{pre}}\vert \geq \widetilde{p}/2  \}\cdot \mathbb{P}\left(  \sum_{j\in  \{1,\ldots, p\}\backslash Z_{\text{pre}}}\mathcal{I}\{ \vert b_j\vert \geq  \sigma  a_n  \} \geq \widetilde{p}/2\right)\\
     &=&    \displaystyle \mathcal{I}\{ \vert Z_{\text{pre}}\vert \geq \widetilde{p}/2  \}\cdot \mathbb{P}\left(  \sum_{j\in  Z_{\text{pre}} } \mathcal{I}\{ \vert b_j\vert \geq  \sigma  a_n  \} \geq \widetilde{p}/2\right )\,+\,\\
     \end{array}
 \end{equation}
  since $\vert \{1,\ldots, p,\}\backslash Z_{\text{pre}}\vert  < \widetilde{p}/2$ by (\ref{eqn:extra-cond}). 

However, 
 \[
 \sum_{j\in  Z_{\text{pre}} } \mathcal{I}\{ \vert b_j\vert \geq  \sigma  a_n  \} \,\,\sim \,\,\text{Binom}(  \vert  Z_{\text{pre}}  \vert , q_1 )
 \]
 with $q_1 \leq c_{10}\tau/a_n $. Hence,  if $ \tau = p^{-u_1}$
 for $u_1 >0$ large enough, then  $1/(c_{10} p\tau/a_n    )\geq O(p^u)$  for some $u>0$.  Then
 \begin{equation}
     \begin{array}{lll}
      \pi(B_n)   
           & \leq& \displaystyle \exp(-c_{11} n \epsilon_n^2 )
     \end{array}
 \end{equation}
 where the last inequality follows as in Page 434 of \cite{song2023nearly}, thus using Lemma A.3 from \cite{song2023nearly}.\\
\textbf{Step 4.} This proceeds as Step 4 in the proof of Theorem \ref{thm1}, except for  
$$E \geq 2 p^\gamma  \geq 2\max\{\|\bb^*\|_{\infty},K\max_{k=1,\ldots,K}\| \widetilde{\bomega}\|_{\infty} \}/\sigma^*,$$ 
and a positive constant $c_{14}$, we have that 
\[
  \begin{array}{l}
      \pi(   b_j/\sigma \in [b_j^*/\sigma -  \eta_2 \epsilon_n/\widetilde{s} ,  b_j^*/\sigma +  \eta_2 \epsilon_n/\widetilde{s}   ],   \,  \,\forall j \in \widetilde{\xi} \,  \,|\,\,   \sigma \in [\sigma^*,\sigma^{*} + \eta_1 \epsilon_n^2]) \\
\displaystyle        \geq  \left[  \frac{2 \eta_2 \epsilon_n }{\widetilde{s}} \int   \underset{x \in [-E,E] }{\inf}   \int_0^{\infty}  \frac{1}{  \sqrt{ 2\pi  }  \lambda  \tau } \exp\left(   -\frac{ (x- \mu_j/\sigma )^2  }{2 \lambda^2 \tau^2  }  \right)\frac{2}{\pi(1+\lambda^2)}  \pi(\mu_j) \,\dif\lambda  \dif \mu_j \right]^{\widetilde{s}}  \\
 \geq \displaystyle \left[  \frac{c_{14} \eta_2 \epsilon_n }{\widetilde{s} \tau }  \underset{x \in [-E,E] }{\inf}  \log\left(1   +    \frac{4 \tau^2}{x^2}  \right)    \right]^{\widetilde{s}}\\
 =  \displaystyle \left[  \frac{c_{14} \eta_2 \epsilon_n }{\widetilde{s}  \tau  }   \log\left(1   +    \frac{4 \tau^2}{E^2}  \right)    \right]^{\widetilde{s}}\\
  \geq   \displaystyle \left[  \frac{c_{14} \eta_2 \epsilon_n }{\widetilde{s}}  \frac{4 \tau}{ E^2+4\tau^2  }   \right]^{\widetilde{s}}\\
  \geq   \displaystyle \left[  \frac{c_{14} \eta_2 \epsilon_n }{\widetilde{s}}  \frac{2 \tau}{ E^2 }   \right]^{\widetilde{s}}
  \end{array}
\]
where the second inequality follows from Theorem 1 in \cite{carvalho2010horseshoe} and the fact that $\vert \mu_j \vert \leq K \max_{k=1,\ldots,K}\| \widetilde{\bomega}\|_{\infty} $. Therefore, 
\[
-\log  \pi(   b_j/\sigma \in [b_j^*/\sigma -  \eta_2 \epsilon_n/\widetilde{s} ,  b_j^*/\sigma +  \eta_2 \epsilon_n/\widetilde{s}   ],   \,  \,\forall j \in \widetilde{\xi} \,  |\,   \sigma \in [\sigma^*,\sigma^{*} + \eta_1 \epsilon_n^2]) \,\leq\,c_{16} n \epsilon_n^2
\]
for some positive constant $c_{16}$. The claim then follows by an application of Lemma A.4 in \cite{song2023nearly} or Lemma 6 in  \cite{bernardo1998information}.
\end{proof}

\newpage
\section{Proof of Theorem~\ref{thm3}}
Theorem \ref{thm3} is a direct consequence of the following theorem.

\begin{theorem}
    \label{S-thm4}
  Suppose that $\sigma^{*2} =\sigma^2$ is known and consider the setting of Theorem \ref{thm1} with no prior on $\sigma^2$.  Let $j \in \{1\ldots,p\}$ and 
 \begin{equation}
     \label{eqn:e78_2}
       g_j(\bbeta_{-j} )\,=\,  \beta_j^* \,+\,   ( \xb_j^{(0)\top}  \xb_j^{(0)}   )^{-1}  \xb_j^{(0)\top} \bepsilon^{(0)} \,+\, ( \xb_j^{(0)\top}  \xb_j^{(0)}   )^{-1}  \xb_j^{(0)\top} \Xb_{-j}^{(0)}(\bbeta_{-j}^* -\bbeta_{-j} ).
 \end{equation}
  In addition, assume that $\| \xb_j^{(0)} \|\asymp \sqrt{n}$, and  there exists $\overline{p}> \max\{\tilde{s}(r_{\sup}),s\}$ such that 
  \[
    \lambda_{\max}\,:=\,\Lambda_{\max}\left( \frac{ \Xb_{\xi}^{(0)\top} \Xb_{\xi}^{(0)}}{n }\right) \,=\,  O(1),
  \]
  for all $\xi \subset\{1,\ldots, p\}$ with $\vert \xi\vert < \overline{p}$. 
  
  \textbf{(i)} Suppose that $\vert \tilde{\omega}_j -  \beta_j^*\vert  > C \sigma \epsilon_n$ for some constant  $C >1$  and let us assume that $\vert \beta_j^* -\tilde{\omega}_j\vert^2 = o( \sigma p^2  )$.
  Also, with $a_n \asymp \sqrt{ \tilde{s}(r_{\sup}) \log p/n}/p$  for a constant  $\tilde{C}>0$ let 
  \[
R_{-j}\,:=\,\{ \bbeta_{-j}  \in \mathbb{R}^p\,:\,    \vert \{\ell \neq j\,:\, \vert \beta_{\ell} -\tilde{\omega}_{\ell} \vert >a_n    \} \vert\leq \tilde{C} \tilde{s}(r_{\sup}),\,\,\|\bbeta_{-j} -\bbeta_{-j}^*\|\leq c_1\sigma\epsilon_n    \}
  \]
  and 
    \[
R(\bbeta_{-j})\,:=\,\{ \beta_{j} \in \mathbb{R}\,:\,   \vert\beta_j  - g_j(\bbeta_{-j})\vert\leq \tilde{C} \sigma\epsilon_n, \, \}.
  \]
  Consider the marginal pdf $f_j$  of a random variable $\beta_j$ generated as 
\begin{equation}
    \label{eqn:post_approx_2}
      \begin{array}{lll}
    \beta_j  \,|\,   \bbeta_{-j} &\sim   & \mathcal{N}\left( g_{j}(\bbeta_{-j}),\sigma_j^2\right),\\
    \bbeta_{-j}  & \sim &  \frac{ \pi(\bbeta_{-j}|\,\yb^{(0)},\Xb^{(0)},\sigma^2 )}{ \int_{R_{-j}}  \pi( \bbeta_{-j}|\,\yb^{(0)},\Xb^{(0)},\sigma^2 ) \dif \bbeta_{-j}},\\
  \end{array}
\end{equation}
  where $\sigma_j^2 =  \sigma^2  ( \xb_j^{(0)\top}  \xb_j^{(0)})^{-1} $. Then, with probability approaching one, we have 
   \begin{equation}
       \label{eqn:e77_2}
       \| f_j -   \pi(\beta_j \,|\, \Xb^{(0)},\yb^{(0)}, \sigma^2  )\|_{ \mathrm{TV} }  \,=\,  O\left( \frac{ \vert \beta_j^* -\tilde{\omega}_j\vert^2  }{\sigma p^2 }    \,+\,  \exp(- c_2n \epsilon_n^2)     \,+\, \frac{ \sigma\epsilon_n}{\vert \beta_j^* - \tilde{\omega}_j \vert  }  \right).
   \end{equation}
\textbf{(ii)} If instead $\vert \tilde{\omega}_j -  \beta_j^*\vert  \leq C \sigma \epsilon_n$, then in the construction of $\beta_j  \,|\,   \bbeta_{-j} $ in (\ref{eqn:post_approx_2}) we  have 
\[
p(\beta_j  \,|\,   \bbeta_{-j}  \,) \,\propto   \, \mathcal{I}(\beta_j \in R(\bbeta_{-j}) ) \,\mathcal{N}(\beta_j; g_j(\bbeta_{-j} ),\sigma_j^2   ) \,\cdot \,h(\beta_j )
 \]
where $\mathcal{I}(\cdot)$ is the indicator function of its input, $\tilde{C}$ is a different constant than in the previous case, $\tilde{\omega}_j \in R(\bbeta_{-j})$, the function $h$ satisfies $\underset{\beta_j \rightarrow \tilde{\bomega}_j}{\lim} h(\beta_j)  \,=\,\infty$, and the corresponding $f_j$ satisfies 
  \begin{equation}
       \label{eqn:e955_@}
       \| f_j -   \pi(\beta_j\,|\,\Xb^{(0)},\yb^{(0)}, \sigma^2  )\|_{ \mathrm{TV} }  \,=\,  O\left( \frac{ \epsilon_n^2 }{p^2 }  \,+\,  \exp\left(- c_2n \epsilon_n^2 \right)     \right).
   \end{equation}
 
\end{theorem}
\begin{proof}
We  drop the superscript of $\Xb^{(0)}$ and $\yb^{(0)}$ and denote them as $\Xb$ and $\yb$. We also
 use the notation in the proof of Theorem \ref{thm1} and focus first on the case $\vert \tilde{\omega}_j -  \beta_j^*\vert  > C \sigma \epsilon_n$. Then
let
 \[
R_{-j}\,:=\,\{ \bb_{-j} \in \mathbb{R}^{p-1}\,:\,    \vert \{\ell \neq j\,:\, \vert b_{\ell} \vert >a_n    \} \vert\leq \tilde{C} \tilde{s}(r_{\sup}),\,\,\|\bb_{-j} -\bb_{-j}^*\|\leq c_1\sigma\epsilon_n    \}
  \]
  and 
    \[
R(\bb_{-j})\,:=\,\{ b_{j} \in \mathbb{R}\,:\,   \vert b_j - g_j(\bb_{-j})\vert\leq \tilde{C}  \sigma \epsilon_n, \, \, \bb_{-j} \in R_{-j}\},
  \]
  where we write
    \[
      g_j(\bb_{-j} )\,=\,  b_j^* \,+\,   ( \xb_j^{\top}\xb_j)^{-1}  \xb_j^{\top} \varepsilon \,+\, ( \xb_j^{\top}  \xb_j   )^{-1}  \xb_j^{\top} \Xb_{-j}(\bb_{-j}^* -\bb_{-j}).
    \]
\textbf{Step 1}. The following events hold \\
\[
\mathcal{E}\,=\,\left\{ \vert ( \xb_j^{\top}  \xb_j)^{-1}  \xb_j^{\top} \bepsilon \vert \,\leq\, \frac{ c \sigma}{\sqrt{n}}   
  \right\},
\]
for some constant $c>0$, with probability approaching one, by the sub-Gaussian tail inequality.  Hence, from now, we assume that $\mathcal{E}$ holds.
Also, for $b_j \in R_{-j}$,
\[
\begin{array}{lll}
\vert  ( \xb_j^{\top}\xb_j)^{-1}  \xb_j^{\top} \Xb_{-j}(\bb_{-j}^* -\bb_{-j} ) \vert &\leq     &   \frac{\| \Xb_{-j}(\bb_{-j}^* -\bb_{-j} )\|}{ \|\xb_j\|}\\
&\leq     & C_2 \frac{   \| \Xb_{-j}(\bb_{-j}^* -\bb_{-j} )\|}{ \sqrt{n}}\\
 &\leq &C_2 \sqrt{\lambda_{\max}}   \| \bb_{-j}^* -\bb_{-j} \| \,+\, \frac{ \sqrt{n}pa_n }{\sqrt{n}}\\
  & \leq& C_2 \sqrt{\lambda_{\max}}   \| \bb_{-j}^* -\bb_{-j} \| \,+\, C^{\prime}\epsilon_n\\
\end{array}
\]
for some constants $C_2, C^{\prime}>0$.\\
\textbf{Step 2}. If $\vert b_j - g_j(\bb_{-j}) \vert > \tilde{C}\sigma \epsilon_n$ and  $\bb_{-j} \in R_{-j}$, then, by Step 1,\\ 
\[
   \begin{array}{lll}
   \displaystyle \tilde{C}\sigma \epsilon_n  & \leq  &  \displaystyle  \vert b_j - b_j^* \vert \,+\, \vert b_j^* - g_j(\bb_{-j}) \vert \\
   & \leq  &  \displaystyle  \vert b_j - b_j^* \vert\,+\, C_2 \sqrt{\lambda_{\max}}   \| \bb_{-j}^* -\bb_{-j} \|  \,+\, C^{\prime}\epsilon_n \,+\,\frac{ c \sigma}{ \sqrt{n}}\\
   \end{array}
\]
which implies 
\[
\vert b_j - b_j^* \vert  >c_1\sigma \epsilon_n
\]
if $\tilde{C}$ is large enough.\\
\textbf{Step 3}. Moreover, if $\vert b_j - g_j(\bb_{-j}) \vert \leq  \tilde{C}\sigma \epsilon_n$ and  $\bb_{-j} \in R_{-j}$, then, by Step 1, 
\[
   \begin{array}{lll}
   \displaystyle  \vert b_j - b_j^* \vert  & \leq&  \displaystyle \vert b_j - g_j(\bb_{-j})  \vert  \,+\,\vert g_j(\bb_{-j})  - b_j^* \vert \\
   & \leq&  \displaystyle   \tilde{C}\sigma \epsilon_n    \,+\, C_2 \sqrt{\lambda_{\max}}   \| \bb_{-j}^* -\bb_{-j} \|  \,+\, C^{\prime}\epsilon_n \,+\,\frac{ c \sigma}{ \sqrt{n}}\\
   &\leq & \displaystyle C_3 \sigma\epsilon_n
   \end{array}%
\]
for some constant $C_3>0$.\\
\textbf{Step 4}. Next, notice that \\
    \[
    \pi(b_j\,|\,\bb_{-j}, \Xb,\tilde{\yb} ,\sigma^2 )  =   \frac{ \phi \left(  \frac{b_j - g_j( \bb_{-j})  }{\sigma_j} \right)   \pi_j(b_j| \sigma^2)  }{  \int  \phi \left(  \frac{a - g_j( \bb_{-j})  }{\sigma_j} \right)\pi_j(a |\sigma^2) \dif a  }    
    \]
    where $\sigma_j^2 \,=\,  \sigma^2 ( \xb_j^{\top}\xb_j)^{-1} $, $\phi$ is the pdf of the standard normal distribution,  $\pi_j(\cdot|\sigma^2)$ is the prior on $b_j$.
    
 Then for any measurable  $A \subset \mathbb{R}$,  by Fubini's theorem,
 \[
\begin{array}{lll}
    \displaystyle    \int_{  A} \pi(b_j\,|\,\Xb, \tilde{\yb} ,\sigma^2 ) \dif b_j &=& \displaystyle \int_{R_{-j} }  \int_{  A}  \pi(b_j\,|\,\bb_{-j},   \Xb, \tilde{\yb} ,\sigma^2 )  \pi(\bb_{-j} |\Xb, \tilde{\yb},\sigma^2)  \dif b_j  \dif \bb_{-j}\,+\,   \\
  & &\displaystyle   \int_{R_{-j}^c }  \int_{  A}  \pi(b_j\,|\,\bb_{-j},   \Xb, \tilde{\yb} ,\sigma^2 )   \pi(\bb_{-j} |\Xb,\tilde{\yb},\sigma^2)   \dif b_j  \dif \bb_{-j}
\end{array}
 \]
 and
 
  \begin{equation}
      \label{eqn:e61}
          \begin{array}{l}
          \displaystyle      \int_{R_{-j}^c }  \int_{  A}  \pi(b_j\,|\,\bb_{-j},   \Xb, \tilde{\yb} ,\sigma^2 )  \pi(\bb_{-j} |\yb,\Xb,\sigma^2) \dif b_j  \dif \bb_{-j} \\
          \leq         \displaystyle  \int_{R_{-j}^c }  \int  \pi(b_j\,|\,\bb_{-j},   \Xb, \tilde{\yb} ,\sigma^2 ) \pi(\bb_{-j} |\yb,\Xb,\sigma^2)  \dif b_j  \dif \bb_{-j} \\
            <      \displaystyle \exp(- c_2n \epsilon_n^2 )   
      \end{array}
  \end{equation}
where the second inequality does not depend on $A$ and follows by the proof of  Theorem \ref{thm1} and it holds with probability approaching.

Next, notice
  \begin{equation}
      \label{eqn:e62}
          \begin{array}{l}
          \displaystyle    
          \int_{R_{-j} }  \int_{  A}  \pi(b_j\,|\,\bb_{-j},   \Xb, \tilde{\yb} ,\sigma^2 )  \pi(\bb_{-j} |\tilde{\yb},\Xb,\sigma^2)  \dif b_j  \dif \bb_{-j} \\
        \displaystyle =    \int_{R_{-j} }  \int_{  A \cap \{ b_j \in R(\bb_{-j}) \}  }  \pi(b_j\,|\,\bb_{-j},   \Xb, \tilde{\yb} ,\sigma^2 )  \pi(\bb_{-j} |\tilde{\yb},\Xb,\sigma^2)  \dif b_j  \dif \bb_{-j}  \,+\,\\
                \displaystyle \,\,\,\,\,  \int_{R_{-j} }  \int_{  A \cap \{ b_j \in R(\bb_{-j})^c \}  }  \pi(b_j\,|\,\bb_{-j},   \Xb, \tilde{\yb} ,\sigma^2 )  \pi(\bb_{-j} |\yb,\Xb,\sigma^2)  \dif b_j  \dif \bb_{-j}. \\
          \end{array}
  \end{equation}

  However,

  \begin{equation}
      \label{eqn:e63}
      \begin{array}{l}
    \displaystyle      \int_{R_{-j} }  \int_{  A \cap \{ b_j \in R(\bb_{-j})^c \}  }  \pi(b_j\,|\,\bb_{-j},   \Xb, \tilde{\yb} ,\sigma^2 )  \pi(\bb_{-j} |\tilde{\yb},\Xb,\sigma^2)  \dif b_j  \dif \bb_{-j}\\
    \displaystyle \leq \int_{R_{-j} }  \int_{  A \cap \{ b_j \,:\, \vert b_j - b_j^* \vert > c_{1}\sigma\epsilon_n  \}  }  \pi(b_j\,|\,\bb_{-j},   \Xb, \tilde{\yb} ,\sigma^2 )  \pi(\bb_{-j} |\tilde{\yb},\Xb,\sigma^2)  \dif b_j  \dif \bb_{-j}\\
       \displaystyle \leq \int  \int_{  A \cap \{ b_j \,:\, \vert b_j - b_j^* \vert > c_{1}\sigma\epsilon_n  \}  }  \pi(b_j\,|\,\bb_{-j},   \Xb, \tilde{\yb} ,\sigma^2 )  \pi(\bb_{-j} |\,\tilde{\yb},\Xb,\sigma^2)  \dif b_j  \dif \bb_{-j}\\
 \displaystyle \leq    \pi( A \cap \{ b_j \,:\, \vert b_j - b_j^* \vert > c_{1}\sigma\epsilon_n  \}  |\,  \Xb, \tilde{\yb} ,\sigma^2 )\\
  \displaystyle \leq    \pi( \vert b_j - b_j^* \vert > c_{1}\sigma\epsilon_n   |\,  \Xb, \tilde{\yb} ,\sigma^2 )\\
   \displaystyle     \leq   \exp(-c_2 n \epsilon_n^2)
       \end{array}
  \end{equation}
  where the second inequality follows from Step 2, and the last inequality holds, independently of $A$, with high probability by Theorem \ref{thm1}.\\
 \textbf{Step 5}. Moreover, for any $x \in   R(\bb_{-j})$ with $\bb_{-j} \in R_{-j}$, Step 3 implies that $x \in [ b_j^*- C_3\sigma\epsilon_n,  b_j^*  +C_3\sigma\epsilon_n]$. Hence, 

   \[
      \begin{array}{lll}
\displaystyle  \frac{\pi_j(x |\sigma^2) }{\pi_j(b^*_j|\sigma^2)} & =& \displaystyle\frac{   \int_0^{\infty}  \frac{1}{\sqrt{2\pi}\lambda  }   \exp\left( -\frac{x^2}{ 2 \sigma^2 \tau^2 \lambda^2  }   \right)\frac{2   }{\pi(1+\lambda^2)}\dif\lambda     } { \int_0^{\infty}  \frac{1}{\sqrt{2\pi}\lambda  }   \exp\left( -\frac{ b_j^{*2} }{ 2 \sigma^2 \tau^2 \lambda^2  }   \right)\frac{2   }{\pi(1+\lambda^2)}\dif\lambda }.
      \end{array}
   \]
 
Then, by the mean value theorem,  there exists $x^{\prime}$ between $x$ and $b_j^*$ such that
    \begin{equation}
  \label{eqn:e66}
      \begin{array}{lll}
\displaystyle  \left\vert  1 \,-\,  \frac{\pi_j(x |\sigma^2) }{\pi_j(b^*_j|\sigma^2)}\right \vert  & \leq & \displaystyle\frac{ 2 \int_0^{\infty}  \frac{1}{\sqrt{2\pi}\lambda  }   \frac{\vert x^{\prime} \vert^2}{  \sigma^2 \tau^2 \lambda^2 \vert x^{\prime}\vert  }   \exp\left( -\frac{x^{\prime 2}}{ 2 \sigma^2 \tau^2 \lambda^2  }   \right)\frac{2   }{\pi(1+\lambda^2)}\dif\lambda     } { \int_0^{\infty}  \frac{1}{\sqrt{2\pi}\lambda  }   \exp\left( -\frac{ b_j^{*2} }{ 2 \sigma^2 \tau^2 \lambda^2  }   \right)\frac{2   }{\pi(1+\lambda^2)}\dif\lambda } \cdot \vert  x -b_j^*\vert\\
& \leq & \displaystyle\frac{ 2 \int_0^{\infty}  \frac{1}{\sqrt{2\pi}\lambda  }   \exp\left( -\frac{c_0 x^{\prime 2}}{ 2 \sigma^2 \tau^2 \lambda^2  }   \right)\frac{2   }{\pi(1+\lambda^2)}\dif\lambda     } { 
  \vert x^{\prime} \vert\int_0^{\infty}  \frac{1}{\sqrt{2\pi}\lambda  }   \exp\left( -\frac{ b_j^{*2} }{ 2 \sigma^2 \tau^2 \lambda^2  }   \right)\frac{2   }{\pi(1+\lambda^2)}\dif\lambda } \cdot \vert  x -b_j^*\vert\\
  &\leq& \displaystyle \frac{  4\log\left(1 +  \frac{\tau^2\sigma^2}{c_0 \vert x^{\prime}\vert^2 } \right)  }{    \vert x^{\prime} \vert  \log\left(1 +  \frac{2\tau^2 \sigma^2}{ b_j^{*2} }\right) }\cdot \vert  x -b_j^*\vert\\
   & \leq& \frac{4  \frac{\tau^2\sigma^2}{c_0 \vert x^{\prime}\vert^2 }  }{  \frac{2\frac{\tau^2 \sigma^2}{ b_j^{*2} } }{1+  2\frac{\tau^2 \sigma^2}{ b_j^{*2} }   }   }\frac{\vert  x -b_j^*\vert}{ \vert x^{\prime}\vert }\\ 
    & \leq& \displaystyle  \frac{6    b^{*2}_j    \,\vert x^{\prime} -b_j^*\vert   }{  \vert x^{\prime}\vert^3 c_0   }\\
        & \leq& \displaystyle  \frac{ c_0^{\prime} \sigma\epsilon_n  }{ \vert b_j^*\vert },\\
   \end{array}
  \end{equation}
  where the second inequality follows from the inequality $t^2 \leq e^{t^2/2.1}$ for all $t\in \mathbb{R}$, $c_0$ is positive constant,  the third inequality follows by Theorem 1 in \cite{carvalho2010horseshoe},  the fourth inequality follows by the inequality $ \frac{t}{1+t} \leq \log(1+t) \leq t$ for all $t>0$,  the fifth inequality follows by our Assumption on $b^*_j$ (with large enough $C$)and our choice of $\tau$ in the proof of Theorem \ref{thm1}, and last inequality by our assumption on $b_j^*$ and with $c_0^{\prime}$ a positive constant.

Therefore, by (\ref{eqn:e66}), for $\bb_{-j} \in R_{-j}$,
 \begin{equation}
     \label{eqn:e67}
     \begin{array}{l}
 \displaystyle      \bigg\vert    \int_{ x \in   R(\bb_{-j}) }   \frac{\phi( (x - g_j(\bb_{-j}))/\sigma_j  )}{\sigma_j} \frac{\pi_j(x |\sigma^2)}{\pi_j(b_j^*|\sigma^2)  } \dif x - \int_{ x \in   R(\bb_{-j}) }   \frac{\phi( (x - g_j(\bb_{-j}))/\sigma_j  )}{\sigma_j} \dif x \bigg\vert\\
 \displaystyle   \leq   \underset{x \in [ b_j^* -C_3\sigma\epsilon_n  , b_j^* +C_3\sigma\epsilon_n  ]  }{\max}   \left\vert  1 \,-\,  \frac{\pi_j(x |\sigma^2) }{\pi_j(b^*_j|\sigma^2)}\right \vert  \\
  \displaystyle  = O\left(  \frac{ \sigma\epsilon_n}{ \vert b_j^{*}\vert  } \right)\\
     \end{array}
 \end{equation}
 Also, for $b_j \in R(\bb_{-j}) $ with $\bb_{-j} \in R_{-j}$, by (\ref{eqn:e66}),
  \begin{equation}
     \label{eqn:e68}
     \begin{array}{l}
 \displaystyle      \bigg\vert    \frac{\phi( (\beta_j - g_j(\bb_{-j}))/\sigma_j  )}{\sigma_j} \frac{\pi_j(x |\sigma^2)}{\pi_j(b_j^*|\sigma^2)  } -    \frac{\phi( (b_j - g_j(\bb_{-j}))/\sigma_j  )}{\sigma_j} \bigg\vert\\
 \displaystyle   \leq    \frac{\phi( (\beta_j - g_j(\bb_{-j}))/\sigma_j  )}{\sigma_j}  \frac{ c_0^{\prime} \sigma\epsilon_n  }{ \vert b_j^*\vert }.\\
     \end{array}
 \end{equation}\\
\textbf{Step 6}. Furthermore, letting $\Omega = \{\bb_{-j} \in R_{-j}   \}$ and $\Omega(\bb_{-j})  \,=\,\{b_j \in R(\bb_{-j}) \}$, we obtain that \\
\begin{equation*}
    \begin{array}{l}
    \displaystyle  \,\,\\
    T(A)
  \displaystyle \,=\, \bigg\vert   \int_{R_{-j} }  \int_{  A \cap \{ b_j \in R(\bb_{-j}) \}  }   \pi(b_j\,|\,\bb_{-j},\Xb,\tilde{\yb} ,\sigma^2 ) \pi(\bb_{-j}| \Xb,\yb,\sigma^2)   \dif b_j  \dif \bb_{-j}    \,-\,\\
\displaystyle \,\,\,\,\,\,\int \int_{  A}   \frac{\mathcal{I}(\Omega) \,\mathcal{I}(\Omega(\bb_{-j})  ) \cdot \left[ \phi( (b_j - g_j(\bb_{-j}) /\sigma_j)/\sigma_j\right]\cdot \left[  \pi_j(b_j |\sigma^2)  \right] \cdot \pi(\bb_{-j}| \Xb,\tilde{\yb},\sigma^2)  }{      \int_{ x\in  R(\bb_{-j})  }   \frac{\phi( (x - g_j(\bb_{-j}))/\sigma_j  )}{\sigma_j} \pi_j(x |\sigma^2) \dif x } \bigg\vert \\
    \end{array}
\end{equation*}

satisfies

\begin{equation}
  \label{eqn:e65.4}
        \begin{array}{l}
    \displaystyle T(A) \,\,\\
    \displaystyle=\,
  \bigg\vert\int \int_{  A}  \frac{\mathcal{I}(\Omega) \,\mathcal{I}(\Omega(\bb_{-j})  ) \cdot \left[ \phi( (b_j - g_j(\bb_{-j}) /\sigma_j)/\sigma_j\right]\cdot   \pi_j(b_j |\sigma^2)\cdot \pi(\bb_{-j}| \Xb,\tilde{\yb},\sigma^2)  }{    \int   \frac{\phi( (x - g_j(\bb_{-j}))/\sigma_j  )}{\sigma_j} \pi_j(x |\sigma^2) \dif x    }  \dif b_j\dif \bb_{-j}  \,-\,\\
\displaystyle \,\,\,\,\,\,\int \int_{  A}   \frac{\mathcal{I}(\Omega) \,\mathcal{I}(\Omega(\bb_{-j})  ) \cdot \left[ \phi( (b_j - g_j(\bb_{-j}) /\sigma_j)/\sigma_j\right]\cdot  \pi_j(b_j |\sigma^2) \cdot \pi(\bb_{-j}| \Xb,\tilde{\yb},\sigma^2)  }{     \int_{ x\in  R(\bb_{-j})  }   \frac{\phi( (x - g_j(\bb_{-j}))/\sigma_j  )}{\sigma_j} \pi_j(x |\sigma^2) \dif x   }  \dif b_j \dif \bb_{-j} \bigg\vert. \\
    \end{array}
\end{equation}

However,
\begin{equation*}
    \begin{array}{l}
      \displaystyle \bigg\vert  \int_{ x\in  R(\bb_{-j})  }   \frac{\phi( (x - g_j(\bb_{-j}))/\sigma_j  )}{\sigma_j} \pi_j(x |\sigma^2) \dif x \,-\,  \int  \frac{\phi( (x - g_j(\bb_{-j}))/\sigma_j  )}{\sigma_j} \pi_j(x |\sigma^2) \dif x \bigg\vert \\\
    \displaystyle \,=\, \int_{ g(\bb_{-j})+ \tilde{C}\sigma \epsilon_n    }^{\infty}   \frac{\phi( (x - g_j(\bb_{-j}))/\sigma_j  )}{\sigma_j} \pi_j(x |\sigma^2) \dif x\,+\,\int_{-\infty}^{ g(\bb_{-j})- \tilde{C}\sigma \epsilon_n    }   \frac{\phi( (x - g_j(\bb_{-j}))/\sigma_j  )}{\sigma_j} \pi_j(x |\sigma^2) \dif x.\\
    \end{array}
\end{equation*}
And we see that for some constants $C_4, C_5>0$, we have that
\begin{equation*}
    \begin{array}{l}
      \displaystyle  \int_{ g(\bb_{-j})+ \tilde{C}\sigma \epsilon_n    }^{\infty}   \frac{\phi( (x - g_j(\bb_{-j}))/\sigma_j  )}{\sigma_j} \pi_j(x |\sigma^2) \dif x\,\\
       \displaystyle\,\leq \,   \frac{\phi( \tilde{C} \sigma \epsilon_n /\sigma_j  )}{\sigma_j}   \int_{ g(\bb_{-j})+ \tilde{C}\sigma \epsilon_n    }^{\infty}   \pi_j(x |\sigma^2) \dif x\,\\
        \displaystyle \,\leq \, \frac{C_4 \sqrt{n} \exp(- \tilde{C}^2C_5^2 \tilde{s}(r_{\sup})\log p       ) }{\sigma }  \int_{ g(\bb_{-j})+ \tilde{C}\sigma \epsilon_n    }^{\infty}   \pi_j(x |\sigma^2) \dif x\,\\
             \displaystyle \,\leq \, \frac{C_4 \sqrt{n} \exp(- \tilde{C}^2C_5^2 \tilde{s}(r_{\sup})\log p       ) }{\sigma } \\
\displaystyle \,\leq\, \frac{\tau}{2p^2}
    \end{array}
\end{equation*}
if $\tilde{C}$ is chosen large enough.

Therefore, we obtain that  
\begin{equation}
\label{eqn:e82}
    \begin{array}{l}
      \displaystyle \bigg\vert  \int_{ x\in  R(\bb_{-j})  }   \frac{\phi( (x - g_j(\bb_{-j}))/\sigma_j  )}{\sigma_j} \pi_j(x |\sigma^2) \dif x \,-\,  \int  \frac{\phi( (x - g_j(\bb_{-j}))/\sigma_j  )}{\sigma_j} \pi_j(x |\sigma^2) \dif x \bigg\vert \\
    \displaystyle \,\leq \,  \frac{\tau}{p^2},
    \end{array}
\end{equation}
for some constant $\tilde{C}_2>0$.

Since $\sigma_j \,\asymp \, \frac{\sigma}{\sqrt{n}} \leq \sigma\epsilon_n$, then for $\tilde{C}$ large enough, it holds that 
  \begin{equation*}
  \int_{ x \in  R(\bb_{-j})  }   \frac{\phi( (x - g_j(\bb_{-j}))/\sigma_j  )}{\sigma_j} \dif x \,=\, \int_{ g_j(\bb_{-j}) - \tilde{C}\sigma\epsilon_n  }^{ g_j(\bb_{-j}) + \tilde{C}\sigma\epsilon_n }   \frac{\phi( (x - g_j(\bb_{-j}))/\sigma_j  )}{\sigma_j} \dif x \,\geq \, \frac{3}{4}.
 \end{equation*}
And so by (\ref{eqn:e66}),
\begin{equation}
  \label{eqn:e85}
  \int_{ x \in  R(\bb_{-j})  }   \frac{\phi( (x - g_j(\bb_{-j}))/\sigma_j  )}{\sigma_j}  \frac{\pi_j(x|\sigma) }{ \pi_j(b_j^*|\sigma) }    \dif x \,\geq \, \frac{1}{2}.
 \end{equation}
Hence, 
  \begin{equation}
  \label{eqn:e65}
  \int_{ x \in  R(\bb_{-j})  }   \frac{\phi( (x - g_j(\bb_{-j}))/\sigma_j  )}{\sigma_j} \pi_j(x|\sigma)    \dif x \,\geq \, \frac{\sigma\tau }{4b_j^{*2} } >   \frac{8 \tau }{ p^2 }
 \end{equation}
since $b_j^{*2} = o( \sigma p^2 )$ and by Theorem 1 in \cite{carvalho2010horseshoe}, 
\[
  \begin{array}{lll}
    \pi_j(b_j^*|\sigma)   &\geq   &  \displaystyle  \frac{1}{2\tau \sigma}  \log\left(  1+      \frac{2\tau^2 \sigma^2}{  b_j^{*2} }    \right)  \\
       & \geq &\displaystyle  \frac{1}{2\tau \sigma}  \frac{  2\sigma^2\tau^2/b_j^{*2}   }{1+  2\sigma^2\tau^2/b_j^{*2}} \\
        & \geq& \displaystyle \frac{\sigma\tau }{2b_j^{*2} }.
  \end{array}
\]
As a result,
\begin{equation}
    \label{eqn:e80}
    \begin{array}{lll}
       \displaystyle T(A)&\leq &\ \displaystyle   \int \int_A \mathcal{I}(\Omega) \,\mathcal{I}(\Omega(\bb_{-j})  ) \cdot \left[ \phi( (b_j - g_j(\bb_{-j}) /\sigma_j)/\sigma_j\right]\cdot   \pi_j(b_j |\sigma^2)\cdot \pi(\bb_{-j}| \Xb,\tilde{\yb},\sigma^2)\\
      && \displaystyle \left[ \int_{ x\in  R(\bb_{-j})  }   \frac{\phi( (x - g_j(\bb_{-j}))/\sigma_j  )}{\sigma_j} \pi_j(x |\sigma^2) \dif x \right]^{-1} \cdot\left[ \int \frac{\phi( (x - g_j(\bb_{-j}))/\sigma_j  )}{\sigma_j} \pi_j(x |\sigma^2) \dif x \right]^{-1}\cdot \\
   && \displaystyle  \,\,\,\,\,\,\,\frac{ \tau }{ p^2 }  \dif b_{j}\,\dif \bb_{-j} \\
    &\leq &\displaystyle  2 \int \int_A \mathcal{I}(\Omega) \,\mathcal{I}(\Omega(\bb_{-j})  ) \cdot \left[ \phi( (b_j - g_j(\bb_{-j}) /\sigma_j)/\sigma_j\right]\cdot   \pi_j(b_j |\sigma^2)\cdot \pi(\bb_{-j}| \Xb,\tilde{\yb},\sigma^2) \cdot \\
  &&\displaystyle  \left[ \int_{ x\in  R(\bb_{-j})  }   \frac{\phi( (x - g_j(\bb_{-j}))/\sigma_j  )}{\sigma_j} \pi_j(x |\sigma^2) \dif x \right]^{-2} \cdot\\
   && \displaystyle \,\,\,\,\,\,\, \frac{ \tau }{ p^2}  \dif b_{j}\,\dif \bb_{-j} \\
     &\leq &\displaystyle  2  \int \mathcal{I}(\Omega)\left[ \int_{ x\in  R(\bb_{-j})  }   \frac{\phi( (x - g_j(\bb_{-j}))/\sigma_j  )}{\sigma_j} \pi_j(x |\sigma^2) \dif x \right]^{-1} \cdot  \pi(\bb_{-j} |, \Xb,\tilde{\yb},\sigma^2) \dif \bb_{-j} \\
   && \displaystyle\,\,\,\,\,\,\, \frac{ \tau }{ p^2   }\\
    & \leq& \displaystyle  O\left(  \frac{ b_j^{*2} }{ \sigma p^2 }  \right)
    \end{array}
\end{equation}
where the first inequality follows from (\ref{eqn:e82}), the second inequality follows by (\ref{eqn:e82}) and (\ref{eqn:e65}), and the third by (\ref{eqn:e65}).

 Therefore, we  focus on bounding $\Delta(A)$ given as 
    \begin{equation}
      \label{eqn:e64}
      \begin{array}{l}
    \displaystyle   \Delta(A) \,=\,  \\
        \displaystyle\bigg\vert    \int \int_{  A}   \frac{\mathcal{I}(\Omega) \,\mathcal{I}(\Omega(\bb_{-j})  ) \cdot \left[ \phi( (b_j - g_j(\bb_{-j}) /\sigma_j)/\sigma_j\right]\cdot \left[  \pi_j(b_j |\sigma^2) /\pi_j(b^*_j |\sigma^2) \right] \cdot \pi(\bb_{-j}| \Xb,\tilde{\yb},\sigma^2)  }{    \int_{ x\in  R(\bb_{-j})  }   \frac{\phi( (x - g_j(\bb_{-j}))/\sigma_j  )}{\sigma_j} \frac{\pi_j(x |\sigma^2)}{\pi_j(b_j^*|\sigma^2)  } \dif x    }  \dif b_j\dif \bb_{-j}   \\
    \displaystyle   \,-\,  \int    \int_{  A }    \frac{ \mathcal{I}(\Omega) \,\mathcal{I}(\Omega(\bb_{-j})  ) \cdot\phi( (b_j - g_j(\bb_{-j}) /\sigma_j)/\sigma_j \,\pi(\bb_{-j}| \Xb,\tilde{\yb},\sigma^2)}{     \int_{ x \in  R(\bb_{-j}) }   \frac{\phi( (x - g_j(\bb_{-j}))/\sigma_j  )}{\sigma_j}  \dif x     }    \dif b_j \dif \bb_{-j}     \bigg\vert\\
      \end{array}
  \end{equation}\\
\textbf{Step 7}. From (\ref{eqn:e64}),  we obtain\\
 \begin{equation}
     \label{eqn:e69}
     \begin{array}{l}
         \Delta(A) \leq \\
 \displaystyle \bigg\vert    \int \int_A  \frac{\mathcal{I}(\Omega) \,\mathcal{I}(\Omega(\bb_{-j}) ) \cdot \left[ \phi( (b_j - g_j(\bb_{-j}) /\sigma_j)/\sigma_j\right]\cdot \left[  \pi_j(b_j |\sigma^2) /\pi_j(b^*_j |\sigma^2) \right] \cdot \pi(\bb_{-j}| \Xb,\tilde{\yb},\sigma^2)  }{     \int_{ x \in R(\bb_{-j})  }   \frac{\phi( (x - g_j(\bb_{-j}))/\sigma_j  )}{\sigma_j} \frac{\pi_j(x |\sigma^2)}{\pi_j(b_j^*|\sigma^2)  } \dif x    } \dif b_j  \dif \bb_{-j}    \\
  \displaystyle   \,-\,  \int \int_A  \frac{\mathcal{I}(\Omega) \,\mathcal{I}( \Omega(\bb_{-j}) ) \cdot \left[ \phi( (b_j - g_j(\bb_{-j}) /\sigma_j)/\sigma_j\right]\cdot \left[  \pi_j(b_j |\sigma^2) /\pi_j(b^*_j |\sigma^2) \right] \cdot \pi(\bb_{-j}| \Xb,\tilde{\yb},\sigma^2)  }{      \int_{ x \in R(\bb_{-j})  }   \frac{\phi( (x - g_j(\bb_{-j}))/\sigma_j  )}{\sigma_j} \dif x    }  \dif b_j \dif \bb_{-j}     \bigg\vert \\
\,+\,           \displaystyle\bigg\vert    \int \int_A  \frac{\mathcal{I}(\Omega) \,\mathcal{I}(\Omega(\bb_{-j})) \cdot \left[ \phi( (b_j - g_j(\bb_{-j}) /\sigma_j)/\sigma_j\right]\cdot \left[  \pi_j(b_j |\sigma^2) /\pi_j(b^*_j |\sigma^2) \right] \cdot \pi(\bb_{-j}| \Xb,\tilde{\yb},\sigma^2)  }{    \int_{x \in R(\bb_{-j}) }   \frac{\phi( (x - g_j(\bb_{-j}))/\sigma_j  )}{\sigma_j} \dif x   } \dif b_j \dif \bb_{-j}  \\
     \displaystyle    \,-\,     \int \int_A    \frac{ \mathcal{I}(\Omega) \,\mathcal{I}( \Omega(\bb_{-j}) ) \cdot\phi( (b_j - g_j(\bb_{-j}) /\sigma_j)/\sigma_j \,\pi(\bb_{-j}| \Xb,\tilde{\yb},\sigma^2)}{  \int_{ x\in R(\bb_{-j}) }   \frac{\phi( (x - g_j(\bb_{-j}))/\sigma_j  )}{\sigma_j} \dif x      }  \dif b_j \dif \bb_{-j}       \bigg\vert\\ 
    \displaystyle      \,=:\,      \Delta_1(A) \,+\,     \Delta_2(A), 
     \end{array}
 \end{equation}
 and we proceed to bound $\Delta_1(A)$ and $\Delta_2(A)$. Then,

\begin{equation*}
     \label{eqn:e70}
     \begin{array}{lll}
         \Delta_1(A) &\leq &  \displaystyle 
  \int \int_A  \mathcal{I}(\Omega) \,\mathcal{I}(\Omega(\bb_{-j}) ) \cdot \left[ \phi( (b_j - g_j(\bb_{-j}) /\sigma_j)/\sigma_j\right]\cdot \left[  \pi_j(b_j |\sigma^2) /\pi_j(b^*_j |\sigma^2) \right] \cdot  \\
 & &\displaystyle \,\,\,\,\, \cdot \pi(\bb_{-j}| \Xb,\tilde{\yb},\sigma^2) \cdot\\
 &&\displaystyle  \frac{1}{     \int_{ x \in R(\bb_{-j}) }   \frac{\phi( (x - g_j(\bb_{-j}))/\sigma_j  )}{\sigma_j} \frac{\pi_j(x |\sigma^2)}{\pi_j(b_j^*|\sigma^2)  } \dif x   }\cdot
         \frac{1}{   \int_{ x \in R(\bb_{-j}) }   \frac{\phi( (x - g_j(\bb_{-j}))/\sigma_j  )}{\sigma_j} \dif x}\cdot\\ 
 &&\displaystyle \left\vert   \int_{ x \in R(\bb_{-j}) }   \frac{\phi( (x - g_j(\bb_{-j}))/\sigma_j  )}{\sigma_j} \frac{\pi_j(x |\sigma^2)}{\pi_j(b_j^*|\sigma^2)  }  \dif x  \,-\, \int_{ x \in R(\bb_{-j}) }   \frac{\phi( (x - g_j(\bb_{-j}))/\sigma_j  )}{\sigma_j}  \dif x \right\vert   \, db_{j} \dif \bb_{-j}  \\
  & \leq& \displaystyle 
  \int \int_A  \mathcal{I}(\Omega) \,\mathcal{I}(\Omega(\bb_{-j}) ) \cdot \left[ \phi( (b_j - g_j(\bb_{-j}) /\sigma_j)/\sigma_j\right]\cdot \left[  \pi_j(b_j |\sigma^2) /\pi_j(b^*_j |\sigma^2) \right] \cdot  \\
 & &\displaystyle \,\,\,\,\, \cdot \pi(\bb_{-j}| \Xb,\tilde{\yb},\sigma^2) \cdot\\
 &&\displaystyle  \frac{1}{     \int_{ x \in R(\bb_{-j}) }   \frac{\phi( (x - g_j(\bb_{-j}))/\sigma_j  )}{\sigma_j} \frac{\pi_j(x |\sigma^2)}{\pi_j(b_j^*|\sigma^2)  } \dif x   }\cdot
         \frac{1}{   \int_{ x \in R(\bb_{-j}) }   \frac{\phi( (x - g_j(\bb_{-j}))/\sigma_j  )}{\sigma_j} \dif x} db_{j} \dif \bb_{-j}  \cdot\\ 
          & &\displaystyle O\left(  \frac{ \sigma\epsilon_n}{ \vert b_j^{*} \vert   } \right)
       \end{array}
 \end{equation*}
where the second inequality follows from (\ref{eqn:e67}).  Hence, from (\ref{eqn:e85}) and (\ref{eqn:e67}),

\begin{equation}
     \begin{array}{lll}
         \Delta_1(A) &\leq &  \displaystyle 
  \int \int_A  \mathcal{I}(\Omega) \,\mathcal{I}(\Omega(\bb_{-j}) ) \cdot \left[ \phi( (b_j - g_j(\bb_{-j}) /\sigma_j)/\sigma_j\right]\cdot \left[  \pi_j(b_j |\sigma^2) /\pi_j(b^*_j |\sigma^2) \right] \cdot  \\
 & &\displaystyle \,\,\,\,\, \cdot \pi(\bb_{-j}| \Xb,\tilde{\yb},\sigma^2) \cdot\\
 &&\displaystyle  \frac{4}{  \bigg[   \int_{ x \in R(\bb_{-j}) }   \frac{\phi( (x - g_j(\bb_{-j}))/\sigma_j  )}{\sigma_j} \frac{\pi_j(x |\sigma^2)}{\pi_j(b_j^*|\sigma^2)  } \dif x \bigg]^2  } \dif b_{j} \dif \bb_{-j} \cdot \\
          & &\displaystyle O\left(  \frac{ \sigma\epsilon_n}{ \vert b_j^{*} \vert  } \right)\\
           & \leq&\displaystyle  \int  \frac{4 \mathcal{I}(\Omega) \pi(\bb_{-j} | \Xb,\tilde{\yb},\sigma^2 )  }{  \bigg[   \int_{ x \in R(\bb_{-j}) }   \frac{\phi( (x - g_j(\bb_{-j}))/\sigma_j  )}{\sigma_j} \frac{\pi_j(x |\sigma^2)}{\pi_j(b_j^*|\sigma^2)  } \dif x \bigg]  }   \dif \bb_{-j}    \cdot O\left(  \frac{ \sigma\epsilon_n}{\vert b_j^{*} \vert  } \right)\\
            & = & \displaystyle O\left(  \frac{ \sigma\epsilon_n}{\vert b_j^{*} \vert  } \right).\\
       \end{array}
 \end{equation}\\
\textbf{Step 8}. Next, we bound $\Delta_2(A)$. We observe that by (\ref{eqn:e68}), \\
\begin{equation}
    \label{eqn:e71}
    \begin{array}{lll}
         \Delta_2(A)& \leq  & \displaystyle    \int\int_A    \frac{ \mathcal{I}(\Omega) \,\mathcal{I}(\Omega(b_j)  ) \cdot\phi( (b_j - g_j(\bb_{-j}) /\sigma_j)/\sigma_j \,\pi(\bb_{-j}| \Xb,\yb,\sigma^2)}{  \   \int_{x \in R(\bb_{-j}) }   \frac{\phi( (x - g_j(\bb_{-j}))/\sigma_j  )}{\sigma_j} \dif x      }  \dif b_j  \dif \bb_{-j}    \cdot\\
         & & \displaystyle\frac{ c_0^{\prime} \sigma\epsilon_n  }{ \vert b_j^*\vert } \\
          & \leq& \displaystyle  O\left(  \frac{ \sigma\epsilon_n}{ b_j^{*}  } \right). \\
    \end{array}
\end{equation}\\
\textbf{Step 9}. Next we see that\\
\begin{equation*}
    \begin{array}{l}
   \displaystyle     \bigg\vert \int \int_A    \frac{ \mathcal{I}(\Omega) \,\mathcal{I}( \Omega(\bb_{-j}) ) \cdot\phi( (b_j - g_j(\bb_{-j}) /\sigma_j)/\sigma_j \,\pi(\bb_{-j}| \Xb,\tilde{\yb},\sigma^2)}{  \int_{ x\in R(\bb_{-j}) }   \frac{\phi( (x - g_j(\bb_{-j}))/\sigma_j  )}{\sigma_j} \dif x      }  \dif b_j \dif \bb_{-j}  \,-\, \\
         \displaystyle  \int \int_A    \frac{ \mathcal{I}(\Omega) \,\mathcal{I}( \Omega(\bb_{-j}) ) \cdot\phi( (b_j - g_j(\bb_{-j}) /\sigma_j)/\sigma_j \,\pi(\bb_{-j}| \Xb,\tilde{\yb},\sigma^2)}{  \left[\int_{ x\in R(\bb_{-j}) }   \frac{\phi( (x - g_j(\bb_{-j}))/\sigma_j  )}{\sigma_j} \dif x   \right]\cdot \int_{R_{-j}} \pi(\bb_{-j} | \,\Xb,\tilde{\yb},\sigma^2 )  \dif \bb_{-j}    }    \dif b_j \dif \bb_{-j} 
        \bigg\vert\\
     \displaystyle \leq    \int \int_A    \frac{ \mathcal{I}(\Omega) \,\mathcal{I}( \Omega(\bb_{-j}) ) \cdot\phi( (b_j - g_j(\bb_{-j}) /\sigma_j)/\sigma_j \,\pi(\bb_{-j}| \Xb,\tilde{\yb},\sigma^2)}{  \int_{ x\in R(\bb_{-j}) }   \frac{\phi( (x - g_j(\bb_{-j}))/\sigma_j  )}{\sigma_j} \dif x      } db_{j}\dif \bb_{-j} \cdot \\
     \displaystyle\,\,\,\,\,\,\,\,\,\,\, \frac{ \int_{R_{-j}^c } \pi(\bb_{-j} | \,\Xb,\tilde{\yb},\sigma^2 )  \dif \bb_{-j}   }{\int_{R_{-j}} \pi(\bb_{-j} | \,\Xb,\tilde{\yb},\sigma^2 )  \dif \bb_{-j}  }  \cdot\\
     \displaystyle  \leq   2 \int \int_A    \frac{ \mathcal{I}(\Omega) \,\mathcal{I}( \Omega(\bb_{-j}) ) \cdot\phi( (b_j - g_j(\bb_{-j}) /\sigma_j)/\sigma_j \,\pi(\bb_{-j}| \Xb,\tilde{\yb},\sigma^2)}{  \int_{ x\in R(\bb_{-j}) }   \frac{\phi( (x - g_j(\bb_{-j}))/\sigma_j  )}{\sigma_j} \dif x      }  \cdot \exp(-c_2 n \epsilon_n^2 )\dif b_j \dif \bb_{-j},  \\
     
    \end{array}
\end{equation*}
where the second inequality holds with high probability by Theorem \ref{thm1}, and so 
\begin{equation}
    \label{eqn:e72}
    \begin{array}{l}
   \displaystyle     \bigg\vert \int \int_A    \frac{ \mathcal{I}(\Omega) \,\mathcal{I}( \Omega(\bb_{-j}) ) \cdot\phi( (b_j - g_j(\bb_{-j}) /\sigma_j)/\sigma_j \,\pi(\bb_{-j}| \Xb,\tilde{\yb},\sigma^2)}{  \int_{ x\in R(\bb_{-j}) }   \frac{\phi( (x - g_j(\bb_{-j}))/\sigma_j  )}{\sigma_j} \dif x      }  \dif b_j \dif \bb_{-j}  \,-\, \\
   \displaystyle  \int \int_A    \frac{ \mathcal{I}(\Omega) \,\mathcal{I}( \Omega(\bb_{-j}) ) \cdot\phi( (b_j - g_j(\bb_{-j}) /\sigma_j)/\sigma_j \,\pi(\bb_{-j}| \Xb,\tilde{\yb},\sigma^2)}{  \left[\int_{ x\in R(\bb_{-j}) }   \frac{\phi( (x - g_j(\bb_{-j}))/\sigma_j  )}{\sigma_j} \dif x   \right]\cdot \int_{R_{-j}} \pi(\bb_{-j} | \,\Xb,\tilde{\yb},\sigma^2 )  \dif \bb_{-j}    }    \dif b_j \dif \bb_{-j} 
        \bigg\vert\\
         \displaystyle \,\leq\, 2 \exp(-c_2 n \epsilon_n^2 ). \\
     
    \end{array}
\end{equation}\\
\textbf{Step 10}. Notice that\\
\begin{equation}
    \label{eqn:100}
      \begin{array}{l}
            \displaystyle \bigg\vert  \int \int_A    \frac{ \mathcal{I}(\Omega) \,\mathcal{I}( \Omega(\bb_{-j}) ) \cdot\phi( (b_j - g_j(\bb_{-j}) /\sigma_j)/\sigma_j \,\pi(\bb_{-j}| \Xb,\tilde{\yb},\sigma^2)}{  \left[\int_{ x\in R(\bb_{-j}) }   \frac{\phi( (x - g_j(\bb_{-j}))/\sigma_j  )}{\sigma_j} \dif x   \right]\cdot \int_{R_{-j}} \pi(\bb_{-j} | \,\Xb,\tilde{\yb},\sigma^2 )  \dif \bb_{-j}    }  \dif b_j \dif \bb_{-j}  \,-\,\\
        \displaystyle   \int \int_A    \frac{ \mathcal{I}(\Omega) \,\mathcal{I}( \Omega(\bb_{-j}) ) \cdot\phi( (b_j - g_j(\bb_{-j}) /\sigma_j)/\sigma_j \,\pi(\bb_{-j}| \Xb,\tilde{\yb},\sigma^2)}{  \left[\int   \frac{\phi( (x - g_j(\bb_{-j}))/\sigma_j  )}{\sigma_j} \dif x   \right]\cdot \int_{R_{-j}} \pi(\bb_{-j} | \,\Xb,\tilde{\yb},\sigma^2 )  \dif \bb_{-j}    }  \dif b_j \dif \bb_{-j} \bigg\vert  \\
        \displaystyle   \leq  \int \int_A    \frac{ \mathcal{I}(\Omega) \,\mathcal{I}( \Omega(\bb_{-j}) ) \cdot\phi( (b_j - g_j(\bb_{-j}) /\sigma_j)/\sigma_j \,\pi(\bb_{-j}| \Xb,\tilde{\yb},\sigma^2)}{  \left[\int_{ x\in R(\bb_{-j}) }   \frac{\phi( (x - g_j(\bb_{-j}))/\sigma_j  )}{\sigma_j} \dif x   \right]\cdot \int_{R_{-j}} \pi(\bb_{-j} | \,\Xb,\tilde{\yb},\sigma^2 )  \dif \bb_{-j}    }\cdot
        \\ 
        \displaystyle\,\,\,\,\,\,\,\,\left\vert   \int_{ x\in R(\bb_{-j}) }   \frac{\phi( (x - g_j(\bb_{-j}))/\sigma_j  )}{\sigma_j} \dif x \,-\, \int   \frac{\phi( (x - g_j(\bb_{-j}))/\sigma_j  )}{\sigma_j} \dif x \right\vert \dif b_j \dif \bb_{-j} \\  
              \displaystyle   \leq  \int \int_A    \frac{ \mathcal{I}(\Omega) \,\mathcal{I}( \Omega(\bb_{-j}) ) \cdot\phi( (b_j - g_j(\bb_{-j}) /\sigma_j)/\sigma_j \,\pi(\bb_{-j}| \Xb,\tilde{\yb},\sigma^2)}{  \left[\int_{ x\in R(\bb_{-j}) }   \frac{\phi( (x - g_j(\bb_{-j}))/\sigma_j  )}{\sigma_j} \dif x   \right]\cdot \int_{R_{-j}} \pi(\bb_{-j} | \,\Xb,\tilde{\yb},\sigma^2 )  \dif \bb_{-j}    }\cdot O( \exp(-c_3 n \epsilon_n^2 ) )\\
              \displaystyle\leq  O( \exp(-c_3 n \epsilon_n^2 ) )
      \end{array}
\end{equation}
where the second inequality does not depend on $A$ and holds by the Gaussian tail inequality for some constant $c_3>0$.\\
\textbf{Step 11}. The remaining term to bound is
\begin{equation}
    \label{S-eqn:101}
   \begin{array}{l}
   \displaystyle \Delta_3(A)\,=:\,\\
        \displaystyle  \bigg\vert \int \int_A    \frac{ \mathcal{I}(\Omega) \,\mathcal{I}( \Omega(\bb_{-j}) ) \cdot\phi( (b_j - g_j(\bb_{-j}) /\sigma_j)/\sigma_j \,\pi(\bb_{-j}| \Xb,\tilde{\yb},\sigma^2)}{  \left[\int   \frac{\phi( (x - g_j(\bb_{-j}))/\sigma_j  )}{\sigma_j} \dif x   \right]\cdot \int_{R_{-j}} \pi(\bb_{-j} | \,\Xb,\tilde{\yb},\sigma^2 )  \dif \bb_{-j}    }  \dif b_j \dif \bb_{-j}  \,-\, \\
   \displaystyle \int \int_A    \frac{\phi( (b_j - g_j(\bb_{-j}) /\sigma_j)/\sigma_j \,\pi(\bb_{-j}| \Xb,\tilde{\yb},\sigma^2)}{  \left[\int   \frac{\phi( (x - g_j(\bb_{-j}))/\sigma_j  )}{\sigma_j} \dif x   \right]\cdot \int_{R_{-j}} \pi(\bb_{-j} | \,\Xb,\tilde{\yb},\sigma^2 )  \dif \bb_{-j}    }  \dif b_j \dif \bb_{-j} \bigg\vert\\
    \displaystyle  \leq \int \int_A    \frac{ \mathcal{I}(\Omega^c) \,\cdot\phi( (b_j - g_j(\bb_{-j}) /\sigma_j)/\sigma_j \,\pi(\bb_{-j}| \Xb,\tilde{\yb},\sigma^2)}{  \left[\int   \frac{\phi( (x - g_j(\bb_{-j}))/\sigma_j  )}{\sigma_j} \dif x   \right]\cdot \int_{R_{-j}} \pi(\bb_{-j} | \,\Xb,\tilde{\yb},\sigma^2 )  \dif \bb_{-j}    }  \dif b_j \dif \bb_{-j}  \,+\,\\
    \displaystyle\,\,\,\,\,\, \int \int_A    \frac{ \,\mathcal{I}( \Omega(\bb_{-j})^c ) \cdot\phi( (b_j - g_j(\bb_{-j}) /\sigma_j)/\sigma_j \,\pi(\bb_{-j}| \Xb,\tilde{\yb},\sigma^2)}{  \left[\int   \frac{\phi( (x - g_j(\bb_{-j}))/\sigma_j  )}{\sigma_j} \dif x   \right]\cdot \int_{R_{-j}} \pi(\bb_{-j} | \,\Xb,\tilde{\yb},\sigma^2 )  \dif \bb_{-j}    }  \dif b_j \dif \bb_{-j}.\\
   \end{array}
\end{equation}
Hence, 
\begin{equation}
    \label{eqn:102}
    \begin{array}{lll}
     \displaystyle \Delta_3(A)&\leq\,&\displaystyle \frac{ \int_{R_{-j}^c} \pi(\bb_{-j} | \,\Xb,\tilde{\yb},\sigma^2 )  \dif \bb_{-j}  }{ \int_{R_{-j}} \pi(\bb_{-j} | \,\Xb,\tilde{\yb},\sigma^2 )  \dif \bb_{-j} }\,+\,\\
     &&\displaystyle\int \int   \frac{ \,\mathcal{I}( \Omega(\bb_{-j})^c ) \cdot\phi( (b_j - g_j(\bb_{-j}) /\sigma_j)/\sigma_j \,\pi(\bb_{-j}| \Xb,\tilde{\yb},\sigma^2)}{  \left[\int   \frac{\phi( (x - g_j(\bb_{-j}))/\sigma_j  )}{\sigma_j} \dif x   \right]\cdot \int_{R_{-j}} \pi(\bb_{-j} | \,\Xb,\tilde{\yb},\sigma^2 )  \dif \bb_{-j}    }  \dif b_j \dif \bb_{-j}.\\
       &\leq& \displaystyle \frac{ \int_{R_{-j}^c} \pi(\bb_{-j} | \,\Xb,\tilde{\yb},\sigma^2 )  \dif \bb_{-j}  }{ \int_{R_{-j}} \pi(\bb_{-j} | \,\Xb,\tilde{\yb},\sigma^2 )  \dif \bb_{-j} } \,+\,   \int    \frac{ \,O( \exp(-c_3 n \epsilon_n^2 ) )\cdot\,\pi(\bb_{-j}| \Xb,\tilde{\yb},\sigma^2)}{  \int_{R_{-j}} \pi(\bb_{-j} | \,\Xb,\tilde{\yb},\sigma^2 )  \dif \bb_{-j}    }   \dif \bb_{-j}  \\
       & = & O(\exp(-c_2 n \epsilon_n^2 )  \,+\,\exp(-c_3 n \epsilon_n^2 )   )
    \end{array}
\end{equation}
where the last equality holds uniformly for all $A$, with probability approaching one, by Theorem \ref{thm1}.\\
\textbf{Step 12}. Therefore, combining  (\ref{eqn:e61}), (\ref{eqn:e62}), (\ref{eqn:e63}), (\ref{eqn:e64}), (\ref{eqn:e69}), (\ref{eqn:e70}), (\ref{eqn:e71}), (\ref{eqn:e72}), (\ref{eqn:100}) and (\ref{eqn:102}), we obtain\\
\[
  \begin{array}{l}
       \displaystyle   \underset{A \subset \mathbb{R}\,\,\text{measurable}  }{\sup}\,\left\vert \int_A f_j(b_j)\dif b_j   -  \int_A  \pi(b_j |  \tilde{\yb},\Xb,\sigma^2) \dif b_j  \right \vert \\
    \displaystyle      \,=\, O\left(\frac{b_j^{*2}}{  \sigma  p^2  }     \,+\,  \exp(- c_4n \epsilon_n^2/2 )     \,+\, \frac{ \sigma\epsilon_n}{\vert \beta_j^* - \tilde{\omega}_j \vert  }  \right),
  \end{array}
\]
with probability approaching one, and with $c_4= \min\{c_2,c_3\}$.  Thus, the claim in (\ref{eqn:e77}) follows.\\
\textbf{Small $\vert \beta_j^*-\tilde{\omega}_j\vert$ case.}
Next we consider the case $\vert \tilde{\omega}_j -  \beta_j^*\vert \leq  C \sigma \epsilon_n$. Towards this end, notice that with the notation in the original statement, we have that 
   \[
    \pi(\beta_j\,|\,\beta_{-j}, \Xb,\yb ,\sigma^2 )  =   \frac{ \phi \left(  \frac{\beta_j - g_j( \beta_{-j})  }{\sigma_j} \right)   \pi_j((\beta_j - \tilde{\omega}_j)|\sigma)  }{  \int  \phi \left(  \frac{a - g_j( \beta_{-j})  }{\sigma_j} \right)\pi_j((a- \tilde{\omega}_j)|\sigma ) \dif a  }.    
    \]
  Then, with high probability, for any measurable set $A$, as in Equation (\ref{eqn:e61}),
    \[
      \int_{R_{-j}^c }  \int_{  A}  \pi(b_j\,|\,\bb_{-j},   \Xb, \tilde{\yb} ,\sigma^2 )  \pi(\bb_{-j} |\yb,\Xb,\sigma^2) \dif b_j  \dif \bb_{-j}  \,\leq\, \exp(- c_2n \epsilon_n^2 ). 
    \]
    Also, as in (\ref{eqn:e63}), 

      \begin{equation}
      \begin{array}{l}
    \displaystyle      \int_{R_{-j} }  \int_{  A \cap \{ b_j \in R(\bb_{-j})^c \}  }  \pi(b_j\,|\,\bb_{-j},   \Xb, \tilde{\yb} ,\sigma^2 )  \pi(\bb_{-j} |\yb,\Xb,\sigma^2)  \dif b_j  \dif \bb_{-j}\\
   \displaystyle     \leq   \exp(-c_2 n \epsilon_n^2).
       \end{array}
  \end{equation}
Furthermore (\ref{eqn:e82}) still holds. Thus, 
\begin{equation}
\label{eqn:e92}
    \begin{array}{l}
      \displaystyle \bigg\vert  \int_{ x\in  R(\bb_{-j})  }   \frac{\phi( (x - g_j(\bb_{-j}))/\sigma_j  )}{\sigma_j} \pi_j(x |\sigma^2) \dif x \,-\,  \int  \frac{\phi( (x - g_j(\bb_{-j}))/\sigma_j  )}{\sigma_j} \pi_j(x |\sigma^2) \dif x \bigg\vert \\
    \displaystyle \,\leq \,  \frac{\tau}{p^2  }.
    \end{array}
\end{equation}

Moreover, for $\bb_{-j} \in R_{-j}$,  for some constants $C_5,K>0$
\begin{equation}
\label{eqn:e93}
    \begin{array}{l}
    \displaystyle   \int  \frac{\phi( (x - g_j(\bb_{-j}))/\sigma_j  )}{\sigma_j} \pi_j(x |\sigma^2) \dif x  \\
     \displaystyle\,\geq \,  \frac{K}{2\sigma \tau}  \int  \frac{\phi( (x - g_j(\bb_{-j}))/\sigma_j  )}{\sigma_j}  \log\left( 1+ \frac{2 \tau^2\sigma^2}{x^2}  \right)\dif x\\
      \displaystyle \geq  \frac{K}{2\sigma \tau}  \int  \frac{\phi( (x - g_j(\bb_{-j}))/\sigma_j  )}{\sigma_j} \frac{2 \tau^2\sigma^2 }{2 \tau^2\sigma^2 +x^2}\dif x\\
           \displaystyle \geq  \frac{K}{2\sigma \tau}  \int_{g_j(\bb_{-j}) - \sigma_j}^{g_j(\bb_{-j}) + \sigma_j}  \frac{\phi( (x - g_j(\bb_{-j}))/\sigma_j  )}{\sigma_j} \frac{2 \tau^2\sigma^2 }{2 \tau^2\sigma^2 +x^2}\dif x\\
             \displaystyle \geq  \frac{K\,\phi(1)}{2\sigma \tau \sigma_j}  \int_{g_j(\bb_{-j}) - \sigma_j}^{g_j(\bb_{-j}) + \sigma_j}   \frac{2 \tau^2\sigma^2  }{2 \tau^2\sigma^2 +x^2}\dif x\\
                 \displaystyle \geq  \frac{K\,\phi(1)}{2\sigma \tau \sigma_j}  \int_{g_j(\bb_{-j}) - \sigma_j}^{g_j(\bb_{-j}) + \sigma_j}   \frac{2 \tau^2\sigma^2  }{2 \tau^2\sigma^2 + C_5\sigma^2 \epsilon_n^2}\dif x\\
                        \displaystyle \geq  \frac{K\,\phi(1)}{\sigma \tau }   \frac{2 \tau^2\sigma^2  }{2 \tau^2\sigma^2 + C_5\sigma^2 \epsilon_n^2}\\
                                  \displaystyle \geq  \frac{K\,\phi(1)}{\sigma \tau }   \frac{ \tau^2 \sigma^2 }{C_5\epsilon_n^2}\\
                                        \displaystyle \geq  \frac{K\,\phi(1)\sigma  \tau}{ C_5\epsilon_n^2}  \\
                                        \displaystyle \gg \frac{\tau}{p^2}
    \end{array}
\end{equation}
where second inequality follows by Theorem 1 in \cite{carvalho2010horseshoe}.

Therefore, 
\begin{equation*}
    \begin{array}{l}
    \displaystyle  \,\,\\
    U(A)
  \displaystyle \,:=\, \bigg\vert   \int_{R_{-j} }  \int_{  A \cap \{ b_j \in R(\bb_{-j}) \}  }   \pi(b_j\,|\,\bb_{-j},\Xb,\tilde{\yb} ,\sigma^2 ) \pi(\bb_{-j}| \Xb,\tilde{\yb},\sigma^2)   \dif b_j  \dif \bb_{-j}    \,-\,\\
\displaystyle \,\,\,\,\,\,\int \int_{  A}   \frac{\mathcal{I}(\Omega) \,\mathcal{I}(\Omega(\bb_{-j})  ) \cdot \left[ \phi( (b_j - g_j(\bb_{-j}) /\sigma_j)/\sigma_j\right]\cdot \left[  \pi_j(b_j |\sigma^2)  \right] \cdot \pi(\bb_{-j}| \Xb,\tilde{\yb},\sigma^2)  }{      \int_{ x\in  R(\bb_{-j})  }   \frac{\phi( (x - g_j(\bb_{-j}))/\sigma_j  )}{\sigma_j} \pi_j(x |\sigma^2) \dif x }  \dif b_j  \dif \bb_{-j}\bigg\vert \\
    \displaystyle=\,
\bigg\vert\int \int_{  A}    \frac{\mathcal{I}(\Omega) \,\mathcal{I}(\Omega(\bb_{-j})  ) \cdot \left[ \phi( (b_j - g_j(\bb_{-j}) /\sigma_j)/\sigma_j\right]\cdot   \pi_j(b_j |\sigma^2)\cdot \pi(\bb_{-j}| \Xb,\tilde{\yb},\sigma^2)  }{    \int   \frac{\phi( (x - g_j(\bb_{-j}))/\sigma_j  )}{\sigma_j} \pi_j(x |\sigma^2) \dif x    }  \dif b_j\dif \bb_{-j}  \,-\,\\
\displaystyle \,\,\,\,\,\,\int \int_{  A}   \frac{\mathcal{I}(\Omega) \,\mathcal{I}(\Omega(\bb_{-j})  ) \cdot \left[ \phi( (b_j - g_j(\bb_{-j}) /\sigma_j)/\sigma_j\right]\cdot  \pi_j(b_j |\sigma^2) \cdot \pi(\bb_{-j}| \Xb,\tilde{\yb},\sigma^2)  }{     \int_{ x\in  R(\bb_{-j})  }   \frac{\phi( (x - g_j(\bb_{-j}))/\sigma_j  )}{\sigma_j} \pi_j(x |\sigma^2) \dif x   }  \dif b_j \dif \bb_{-j} \bigg\vert \\
\displaystyle \,\leq\,\tilde{C}_3\left[ \frac{\epsilon_n^2}{\sigma \tau}  \right]\cdot\left[ \frac{ \tau }{ p^2  }.\right] \\
\displaystyle \,=\,\tilde{C}_3\frac{\epsilon_n^2}{p^2}
    \end{array}
\end{equation*}
  where the inequality follows from (\ref{eqn:e92}) and (\ref{eqn:e93}) and with $\tilde{C}_3$ a positive constant.

  Finally,
  \begin{equation*}
    \begin{array}{l}
\displaystyle  \bigg\vert   \int \int_{  A}   \frac{\mathcal{I}(\Omega) \,\mathcal{I}(\Omega(\bb_{-j})  ) \cdot \left[ \phi( (b_j - g_j(\bb_{-j}) /\sigma_j)/\sigma_j\right]\cdot \left[  \pi_j(b_j |\sigma^2)  \right] \cdot \pi(\bb_{-j}| \Xb,\tilde{\yb},\sigma^2)  }{      \int_{ x\in  R(\bb_{-j})  }   \frac{\phi( (x - g_j(\bb_{-j}))/\sigma_j  )}{\sigma_j} \pi_j(x |\sigma^2) \dif x }   \dif b_j  \dif \bb_{-j} \,-\,\\
\displaystyle \,\,\,\,\,\,\int \int_{  A}   \frac{\mathcal{I}(\Omega) \,\mathcal{I}(\Omega(\bb_{-j})  ) \cdot \left[ \phi( (b_j - g_j(\bb_{-j}) /\sigma_j)/\sigma_j\right]\cdot \left[  \pi_j(b_j |\sigma^2)  \right] \cdot \pi(\bb_{-j}| \Xb,\tilde{\yb},\sigma^2)  }{   \left[\int_{ x\in  R(\bb_{-j})  }   \frac{\phi( (x - g_j(\bb_{-j}))/\sigma_j  )}{\sigma_j} \pi_j(x |\sigma^2) \dif x\right] \int_{R_{-j}} \pi(b_j| \, \Xb,\tilde{\yb},\sigma^2)    }  \dif b_j  \dif \bb_{-j}\bigg\vert \\
 \leq 2 \exp(-c_2 n \epsilon_n^2 ). \\\\
    \end{array}
\end{equation*}

The claim then follows combining all the pieces together with $h(\beta_j) = \pi_j(b_j)$. 

\end{proof}

\newpage
\section{Additional simulation result}
\subsection{Simulation metrics}
\label{S-supp:simu-metric}
\begin{itemize}
\item Average estimation error: 

\begin{itemize}
\item \textbf{Bayesian}:
$$
\text{Estimation error} =\frac{1}{p}\sum_{j=1}^p(\widehat{\beta}_j^\text{mean}-\beta_j^*)^2,
$$ where $\widehat{\beta}_j^\text{mean}$ is the posterior mean of $\beta_j$.

\item \textbf{Frequentist}:
$$
\text{Estimation error} =\frac{1}{p}\sum_{j=1}^p(\widehat{\beta}_j-\beta_j^*)^2,
$$ where $\widehat{\beta}_j$ is the point estimate of $\beta_j$.
\end{itemize}

\item Power: 

\begin{itemize}
    \item \textbf{Bayesian}:   
    \[
    \text{Power}_{\text{Bayes}} = \frac{|\{j: \beta_j^* \neq 0 \text{ and } \text{CL}_j \geq 0.95\}|}{|\{j: \beta_j^* \neq 0\}|}.
    \]

    \item \textbf{Frequentist}:  
    \[
    \text{Power}_{\text{Freq}} = \frac{|\{j: \beta_j^* \neq 0 \text{ and } p_j \leq 0.05\}|}{|\{j: \beta_j^* \neq 0\}|},
    \]  
    where \(p_j\) is the \(p\)-value for $\beta_j$.
\end{itemize}

\item False discovery rate: 
\begin{itemize}
    \item \textbf{Bayesian}:
    \[
    \text{FDR}_\text{Bayes} = 1 - \frac{|\{j: \beta_j^* \neq 0 \text{ and } \text{CL}_j \geq 0.95\}|}{|\{j: \text{CL}_j \geq 0.95\}|}.
    \]  

    \item \textbf{Frequentist}:
    \[
    \text{FDR}_\text{Freq}= 1 - \frac{|\{j: \beta_j^* \neq 0 \text{ and } p_j \leq 0.05\}|}{|\{j: p_j \leq 0.05\}|}.
    \]  
\end{itemize}

\item Computation time: the time difference (in seconds) between start of calculating point estimate and the end of calculating the confidence/credible interval.

\end{itemize}

\newpage
\subsection{Simulation Setting I: Increased heterogeneity}
We gradually increase the level of heterogeneity between the source datasets and the target dataset. Specifically, $h=\{10,15\}$. 

\begin{figure}[ht]
	\centering
	\includegraphics[width=\linewidth]{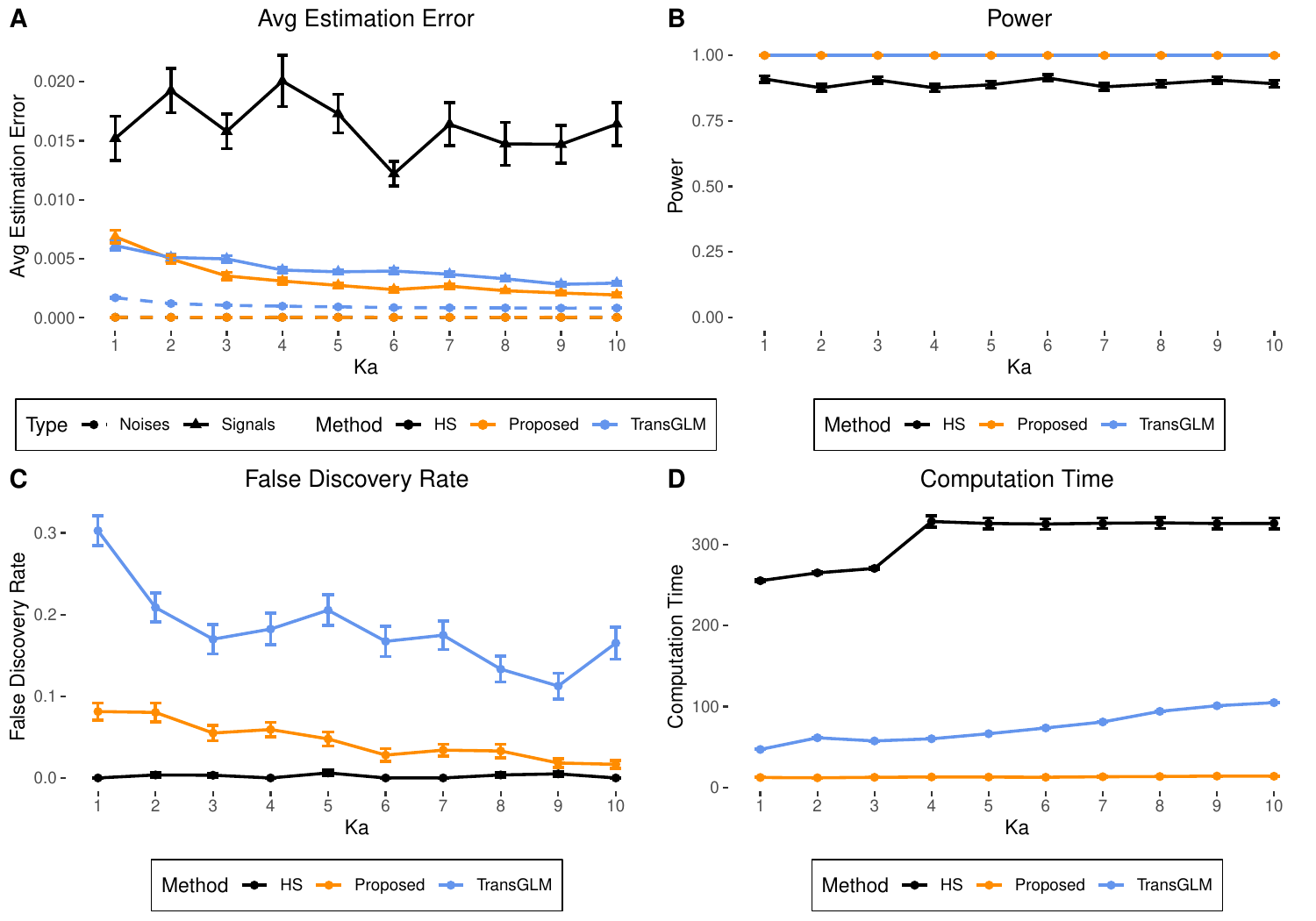}
        \caption{\textbf{Simulation Setting I} ($h = 10$ and other parameters are the same with Figure \ref{fig:stratified-covtype2}). Average estimation error, power, FDR, and computation time of proposed TRADER (orange), TransGLM (blue; implemented with 80 CPU cores per simulation), and target-only HS (black) over 100 simulations. Results are separated by true noise (dots) and signal (triangles), with error bars denoting the standard deviations.
        }
	\label{S-fig:simu-setting-I-h-10}
\end{figure}

\begin{figure}
	\centering
	\includegraphics[width=\linewidth]{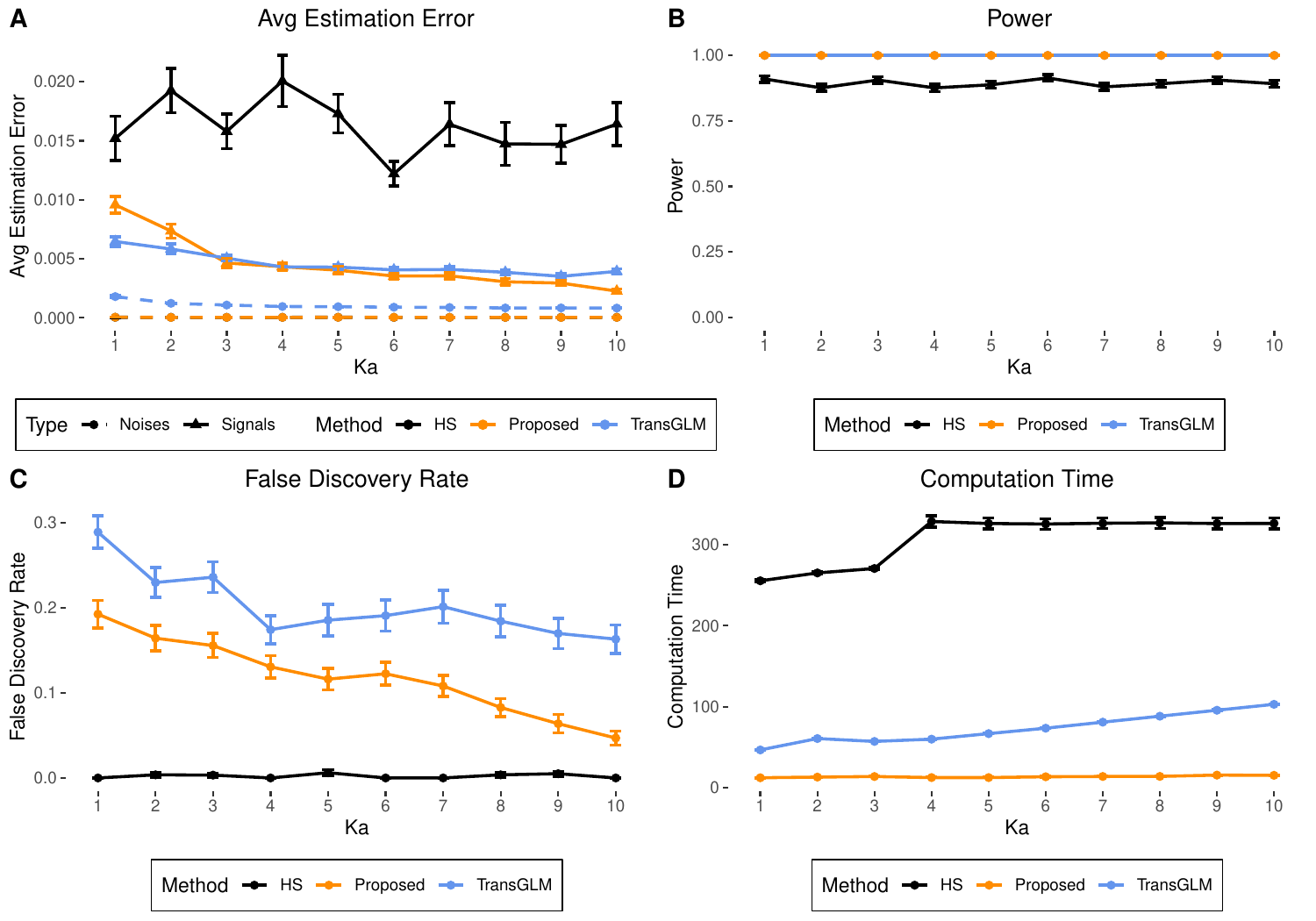}
        \caption{\textbf{Simulation Setting I} ($h = 15$ and other parameters are the same with Figure \ref{fig:stratified-covtype2}). Average estimation error, power, FDR, and computation time of proposed TRADER (orange), TransGLM (blue; implemented with 80 CPU cores per simulation), and target-only HS (black) over 100 simulations. Results are separated by true noise (dots) and signal (triangles), with error bars denoting the standard deviations.
        }
    \label{S-fig:simu-setting-I-h-15}
\end{figure}
\newpage
\subsection{Simulation Setting II: Sources with different correlations and signal strength}
\label{S-supp:source-with-diff-corr}
We evaluated TRADER's performance on Simulation Setting II with different parameter settings.
\begin{figure}
	\centering
	\includegraphics[width=\textwidth]{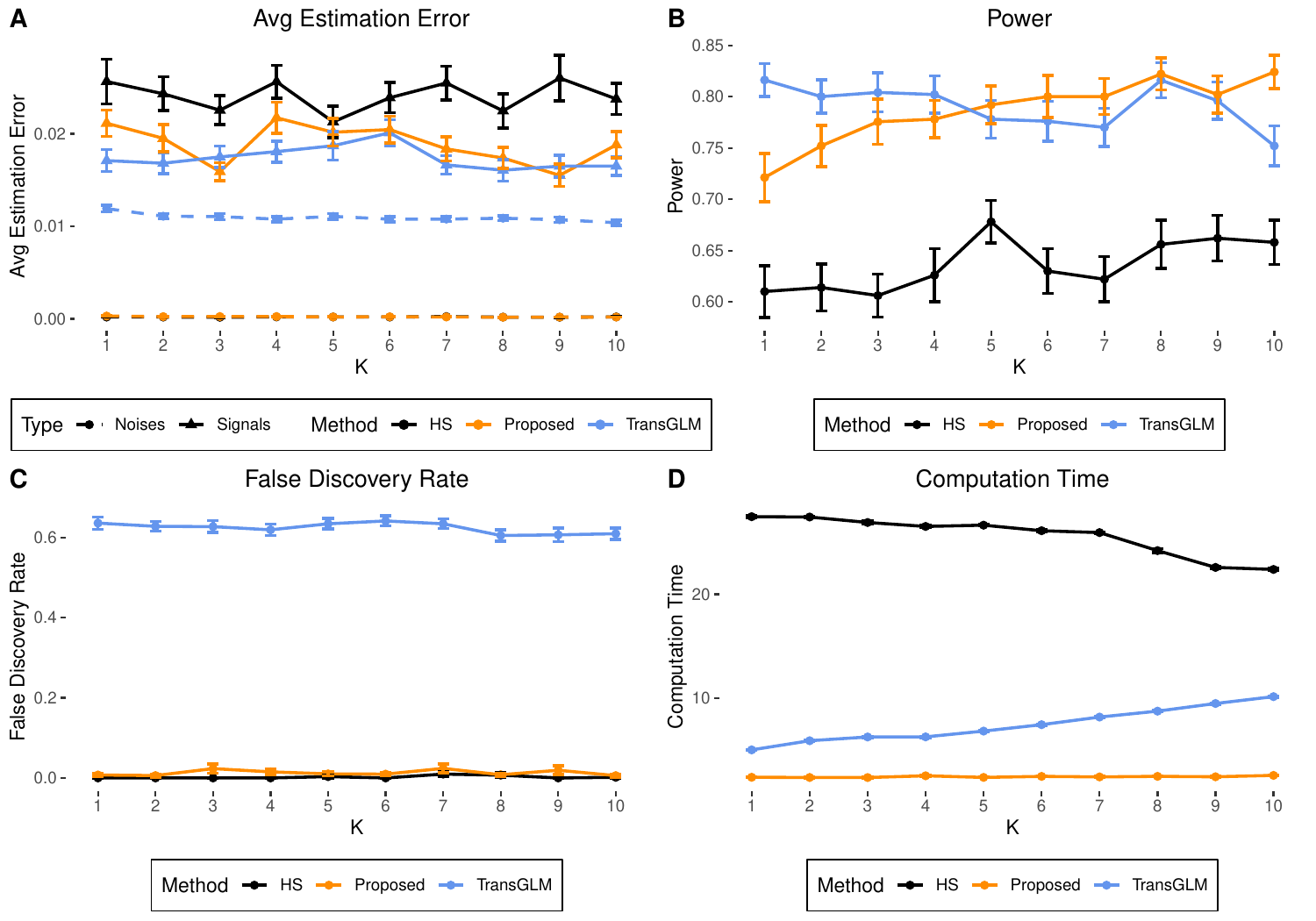}
    \caption{Average estimation error, power, false discovery rate (FDR), and computational time of proposed TRADER (orange), TransGLM (blue; implemented with 20 additional CPU cores per simulation), and target-only HS (black) over 100 simulations under Simulation Setting II. $K = 1,\cdots,10$, $p = 200$, with fixed scale ratios $\alpha_\text{tgt}/\alpha_{\text{src},k} = 1$ with fixed correlation at $\rho_k = 0.3$ for $k = 1, \ldots, 10$, with error bars denoting the standard deviations.}
	\label{S-fig:additional-setting-II-rho-0.3-alpha-1}
\end{figure}
\newpage
\begin{figure}[t]
	\centering
	\includegraphics[width=\textwidth]{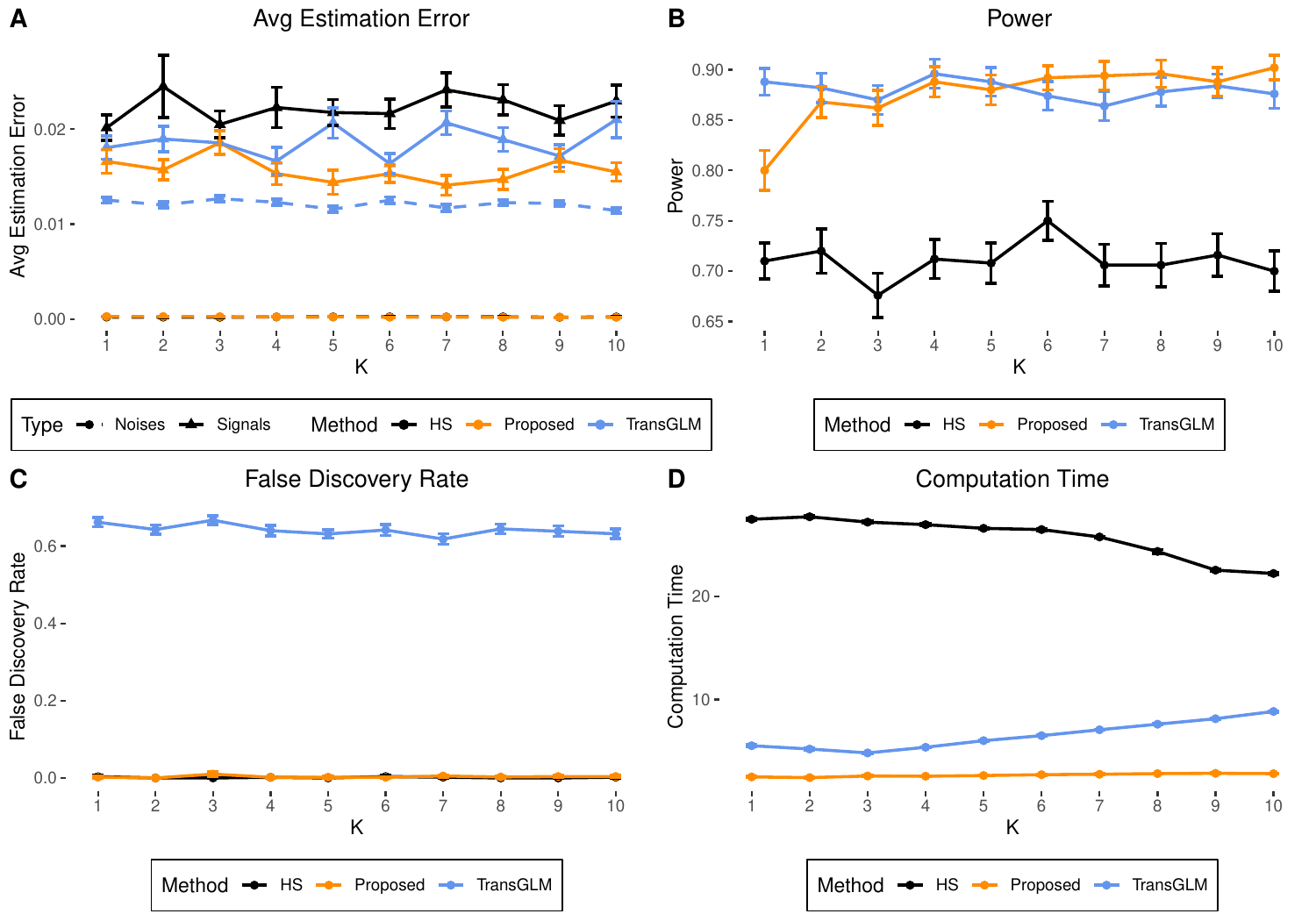}
	\caption{Average estimation error, power, false discovery rate (FDR), and computational time of proposed TRADER (orange), TransGLM (blue; implemented with 20 additional CPU cores per simulation), and target-only HS (black) over 100 simulations under Simulation Setting II. Each panel shows results for $k=1,\ldots,K$ assuming $\alpha_\text{tgt}/\alpha_{\text{src},k}=1/2$, $\rho_k=0.8\times\bm{1}_K$. Error bars indicate standard deviations.}
	\label{S-fig:additional-setting-II-rho-0.8-alpha-2}
\end{figure}

\newpage

\subsection{Simulation Setting III: Comparison with a hierarchical model}
\label{S-supp:ModelH}
We compare TRADER with a hierarchical model $\mathcal{H}$ as follows. We denote 
$$
\widetilde{\mathbf{X}}=\left(\begin{array}{c}
\mathbf{X}^{(1)} \\
\mathbf{X}^{(2)} \\
\vdots \\
\mathbf{X}^{(K)}
\end{array}\right), \quad \widetilde{\mathbf{y}}=\left(\begin{array}{c}
\mathbf{y}^{(1)} \\
\mathbf{y}^{(2)} \\
\vdots \\
\mathbf{y}^{(K)}
\end{array}\right),
$$ 
where $\widetilde{\Xb} \in \mathbb{R}^{(\sum_{k=1}^K n_k) \times p}$ and $\widetilde{\yb} \in \mathbb{R}^{\sum_{k=1}^K n_k}$.

Model $\mathcal{H}$ is used for the likelihood of the data:
$$
\begin{aligned}
\yb^{(0)} \mid \Xb^{(0)}, \bomega, \bz, \sigma^2 &\sim \mathcal{N}\left(\Xb^{(0)}\left(\bomega+\bz\right),\sigma^2\Ib\right), \\
\widetilde{\yb} \mid \widetilde{\Xb}, \bomega, \widetilde{\sigma}^2 &\sim \mathcal{N}\left(\widetilde{\Xb}\bomega,\widetilde{\sigma}^2\Ib\right),
\end{aligned}
$$
where $\bomega+\bz$ represents the target parameter $\bbeta$.

We place independent horseshoe priors on both $\bomega$ and $\bz$. Specifically,
$$
\begin{aligned}
z_j \mid \sigma,\lambda_j,\tau & \sim \mathcal{N}\left(0, \sigma^2\lambda_j^2\tau^2\right),\\
\omega_j \mid \widetilde{\sigma}, \widetilde{\lambda}_j, \widetilde{\tau} & \sim \mathcal{N}\left(0, \widetilde{\sigma}^2\widetilde{\lambda}_j^2\widetilde{\tau}^2\right),
\end{aligned}
$$
with
$$
\begin{aligned}
    \lambda_j &\sim \text{Cauchy}^+(0,1), \tau \sim \text{Cauchy}^+(0,1),\\
    \widetilde{\lambda_j} &\sim \text{Cauchy}^+(0,1), \widetilde{\tau} \sim \text{Cauchy}^+(0,1).
\end{aligned}
$$
For $\sigma^2$ and $\widetilde{\sigma}^2$, we use weakly informative inverse-Gamma priors consistent with the main manuscript. Specifically,
$$
\sigma^2 \sim \text{Inv-Gamma}(\nu,\nu), \widetilde{\sigma}^2 \sim \text{Inv-Gamma}(\psi,\psi),
\text{ where } \nu,\psi>0.$$
We evaluated the performance of TRADER relative to Model \(\mathcal{H}\) in two settings, one with uninformative sources (Simulation Setting III in the main manuscript) and another ideal case with all informative sources.
\begin{figure}[ht]
    \centering
        \includegraphics[width=0.8\linewidth]{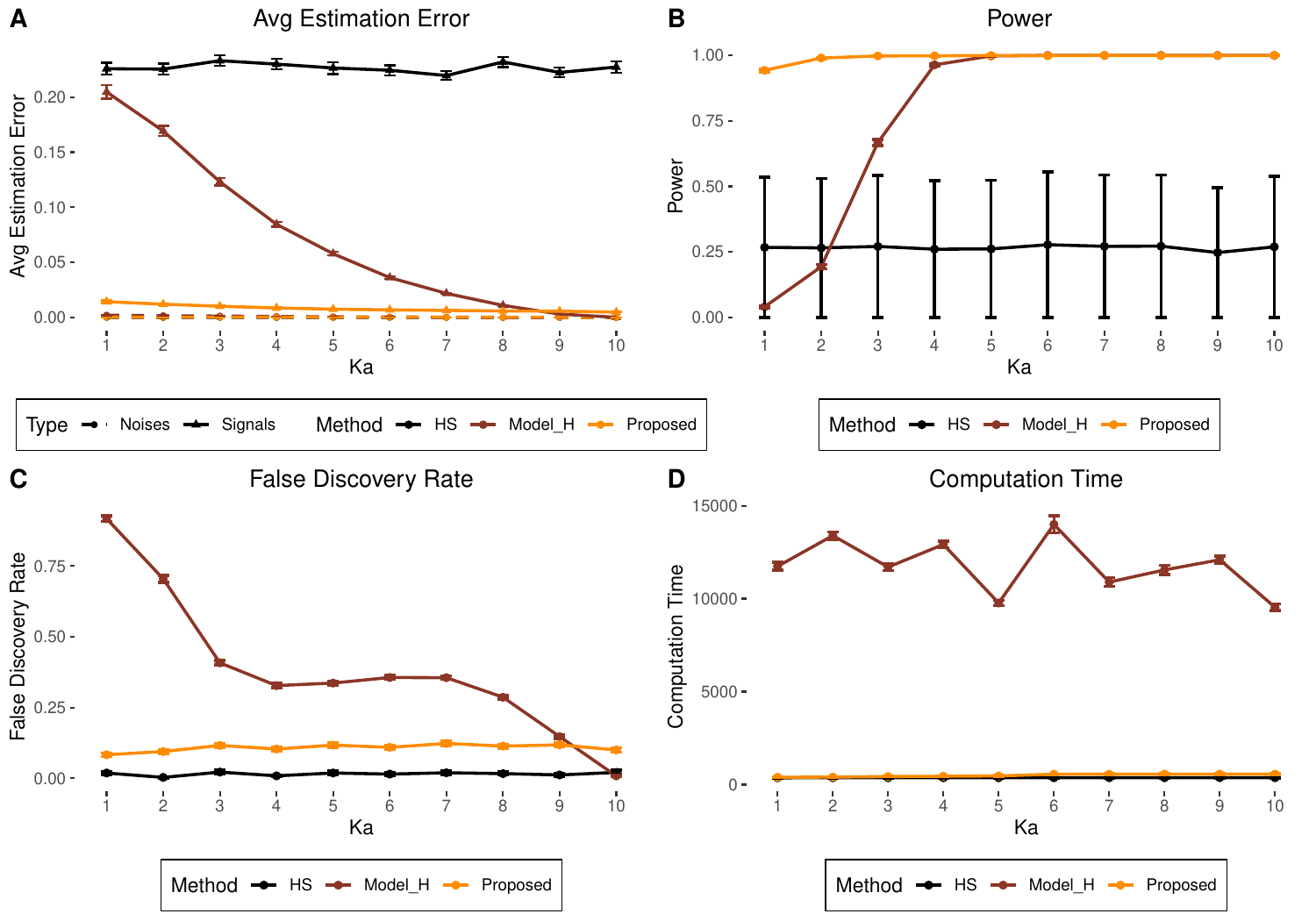}
        \caption{(With Uninformative Sources) Comparison of average estimation error, statistical power, false discovery rate, and computation time between the proposed TRADER (orange), Model $  \mathcal{H}  $ (red), and target-only horseshoe (black) over 100 simulations under a high dimensional setting of $p = 2,000$, with $K = 10$ sources and a varying number $K_a$ of informative sources; $h = 10$, $n = 100$, $n_k = 500$ for all $k = 1, \ldots, 10$, and $s = 20$. The error bars denote the standard deviations.}
    \label{S-fig:AE-cov-type2}
\end{figure}

Figure \ref{S-fig:AE-cov-type2} shows that when the number of informative sources \(K_a\) is small (i.e., many sources are uninformative), Model \(\mathcal{H}\) performs notably worse than TRADER in both estimation and inference metrics. Specifically, the average estimation error for signal coefficients is substantially higher under Model 
\(\mathcal{H}\), while the error for noise coefficients is similar across methods  (Figure~\ref{S-fig:AE-cov-type2}A)

Model \(\mathcal{H}\) also exhibits low power and high FDR, often exceeding 75\% when \(K_a=1\), whereas TRADER maintains stable and controlled FDR across all \(K_a\) (Figures~\ref{S-fig:AE-cov-type2}B--C). 

As \(K_a\) increases and more sources become informative, Model \(\mathcal{H}\)'s performance improves and gradually approaches that of TRADER in estimation error, power, and FDR. However, even in these favorable cases, Model \(\mathcal{H}\) incurs significantly higher computational cost (Figure~\ref{S-fig:AE-cov-type2}D). This difference reflects TRADER's scalability and efficient use of summary-level information.

Overall, these findings highlight two key advantages of TRADER: (1) Robustness against uninformative sources: TRADER's adaptive weighting automatically downweights misleading sources, while Model \(\mathcal{H}\), which implicitly assumes all sources are informative, suffers from degraded inference; and (2) Computational efficiency: TRADER avoids large-scale matrix inversion and achieves over 10-fold speedup compared to Model \(\mathcal{H}\) even without parallelization. 

\begin{figure}[ht]
    \centering
        \includegraphics[width=0.8\linewidth]{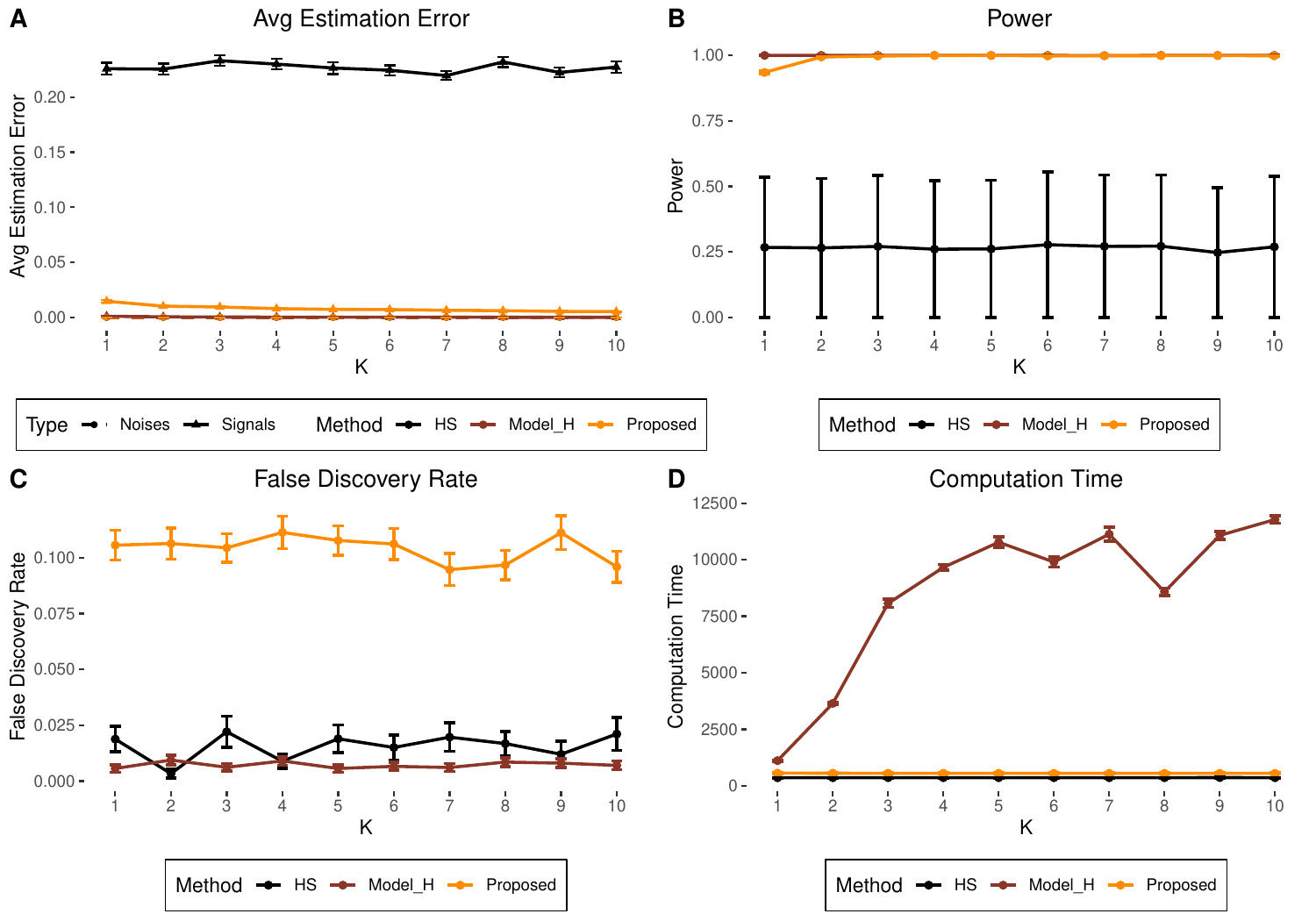}
        \caption{(All Informative Sources) Comparison of average estimation error, statistical power, false discovery rate, and computation time between the proposed TRADER (orange), Model $  \mathcal{H}  $ (red), and target-only horseshoe (black) across 100 simulations in a high-dimensional setting with $  p = 2,000  $, $  K  $ sources ($  K  $ ranging from 1 to 10), $  h = 10  $, $  n = 100  $, $  n_k = 500  $ for all $  k = 1, \ldots, 10  $, and $  s = 20  $. Error bars represent standard deviations.}
    \label{S-fig:AE-cov-type1}
\end{figure}

The suboptimal performance of Model \(\mathcal{H}\) can be explained by its inability to distinguish informative from uninformative sources, a challenge that TRADER overcomes via its similarity-guided prior structure. To further assess the performance of TRADER and to provide a fairer comparison with Model \(\mathcal{H}\), we conducted an additional simulation study under an oracle setting where all $K=10$ sources are informative. The target data follow simulation setting III, and the source datasets are generated as informative sources from that setting. We vary 
$K$ from 1 to 10.

Figure~\ref{S-fig:AE-cov-type1} presents results under this oracle setting. In this case, TRADER and Model \(\mathcal{H}\) achieve comparable estimation error (Figure~\ref{S-fig:AE-cov-type1}A) and power (Figure~\ref{S-fig:AE-cov-type1}B). Model \(\mathcal{H}\) exhibits slightly lower FDR, reflecting its direct access to individual-level source data under ideal conditions (Figure~\ref{S-fig:AE-cov-type1}C). However, this performance comes at substantially higher computational cost (Figure~\ref{S-fig:AE-cov-type1}D). Moreover, this oracle scenario assumes all sources are informative, an assumption rarely satisfied in practice. As shown in Figure~\ref{S-fig:AE-cov-type2}, Model \(\mathcal{H}\) deteriorates sharply when uninformative sources are present, whereas TRADER remains stable. Together, these results highlight TRADER’s robust and computationally efficient performance across realistic heterogeneous settings.

\newpage
\section{Detail on the real data analysis}
\label{S-sec:realdata}
\subsection{Data preprocessing}
First, we excluded the following genetic variants from the GTEx genotype data ($\Xb$):  genetic variants with ambiguous alleles, minor allele frequency (MAF) less than 0.05, or Hardy-Weinberg equilibrium (HWE) $p$-values less than 0.05. The gene-expression data ($\yb$) were adjusted for potential confounding effects, including sex, sequencing platform, the top three principal components of genotype data, and the top probabilistic estimation of expression residuals (PEER) factors. The number of PEER factors included in the adjustment was determined by tissue sample size: 15 (less than 150 samples), 30 (150--250 samples), and 35 (larger than 250 samples). Covariates were sourced from the GTEx portal, and biallelic genetic variants within a 1Mb region of the target gene were selected as features.
 \begin{figure}[ht]
	\centering
	\includegraphics[width=0.6\textwidth]{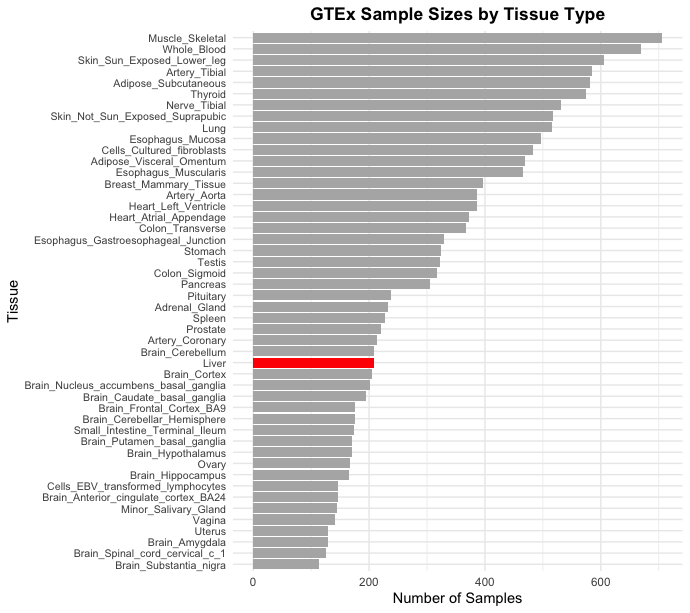}
	\caption{Sample size distribution across datasets in the GTEx dataset. The red bar represents the sample size of the target dataset, while grey bars indicate the sample sizes of the source datasets.}
	\label{S-fig:gtexsamplesize}
\end{figure}

\newpage

\subsection{Complete plot}
\label{S-supp:finemapping-example}
 \begin{figure}[ht]
	\centering
	\includegraphics[width=\textwidth]{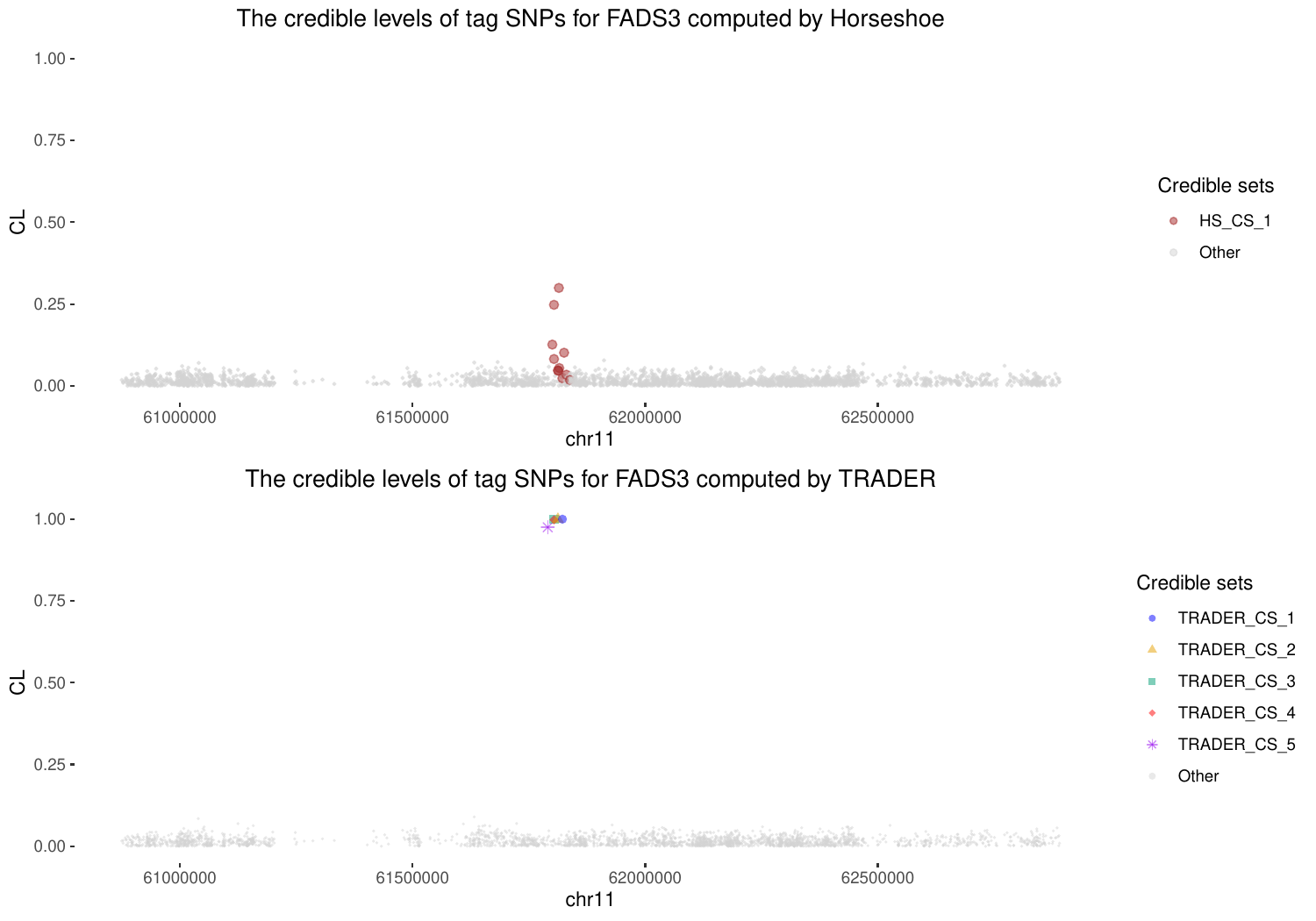}
	\caption{Full plot of the fine-mapping results of \textit{FADS3}. The semitransparent points represent pruned variants in fine mapping. The first and second panel
    illustrates the CLs of tag genetic variants computed by standard HS prior and TRADER. In the each panel, each color represents a 95\% CS. Each CS is named in the format $\{\textbf{method name}\}\textbf{_CS_}\{\textbf{index}\}: \text{chr_}\{\text{chromosome ID}\}\text{_}\{\text{chromosome position}\}.$}
	\label{S-fig:FADS3}
\end{figure}

\newpage
\begin{spacing}{0.9}
\bibliography{ref}
\end{spacing}